\documentclass[hyper,letterpaper]{JHEP3}
\usepackage{amsmath,amssymb,amsfonts,bm}
\usepackage{graphicx}
\usepackage{multirow}
\usepackage{verbatim}
\usepackage{appendix}

\usepackage{url}
\usepackage{float}

\newcommand{\bea}{\begin{eqnarray}}
\newcommand{\eea}{\end{eqnarray}}
\newcommand{\be}{\begin{equation}}
\newcommand{\ee}{\end{equation}}

\newcommand{\unknot}{{\,\raisebox{-.08cm}{\includegraphics[width=.37cm]{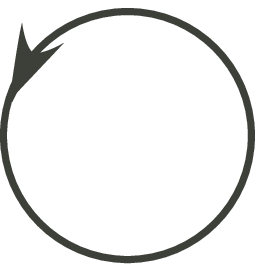}}\,}}

\newcommand{\Z}{{\mathbb Z}}
\newcommand{\R}{{\mathbb R}}
\newcommand{\C}{{\mathbb C}}

\newcommand{\Q}{{\mathbb Q}}

\newcommand{\cF}{{\mathcal{F}}}
\newcommand{\cN}{{\mathcal{N}}}

\def\Tr{{\rm Tr \,}}

\newcommand{\cC}{{\cal C }}

\newcommand{\cO}{{\cal O }}
\newcommand{\cH}{{\cal H }}

\renewcommand{\P}{{\cal P}}
\newcommand{\cp}{{\mathbb{C}}{\mathbf{P}}}
\renewcommand{\S}{{\bf S}}
\newcommand{\B}{{\bf B}}
\renewcommand{\bar}{\overline}
\renewcommand{\hat}{\widehat}

\title{Volume Conjecture: Refined and Categorified}

\author{Hiroyuki Fuji$^{1,2}$, Sergei Gukov$^{1,3}$ and Piotr Su{\l}kowski$^{1,4,5}$
\hspace*{8cm} $\quad\;$ {\it with an appendix by}  Hidetoshi Awata$^{6}$
\\
$^1$ California Institute of Technology, Pasadena, CA 91125, USA \\
$^2$ Nagoya University, Dept. of Physics, Graduate School of Science, \\
$\ $ Furo-cho, Chikusa-ku, Nagoya 464-8602, Japan \\
$^3$ Max-Planck-Institut f\"ur Mathematik, Vivatsgasse 7, D-53111 Bonn, Germany \\
$^4$ Institute for Theoretical Physics, University of Amsterdam, \\
$\ $ Science Park 904, 1090 GL, Amsterdam, The Netherlands \\
$^5$ Faculty of Physics, University of Warsaw, ul. Ho{\.z}a 69, 00-681
Warsaw, Poland \\
$^6$ Nagoya University, Graduate School of Mathematics, Nagoya 464-8602, Japan}

\abstract{The {\it generalized volume conjecture} relates asymptotic behavior of the colored Jones polynomials
to objects naturally defined on an algebraic curve, the zero locus of the $A$-polynomial $A(x,y)$.
Another ``family version'' of the volume conjecture depends on a quantization parameter, usually denoted $q$ or $\hbar$;
this {\it quantum volume conjecture} (also known as the AJ-conjecture) can be stated in a form of
a $q$-difference equation that annihilates the colored Jones polynomials and $SL(2,\C)$ Chern-Simons partition functions.
We propose refinements / categorifications of both conjectures that include an extra deformation parameter $t$
and describe similar properties of homological knot invariants and refined BPS invariants.
Much like their unrefined / decategorified predecessors, that correspond to $t=-1$,
the new volume conjectures involve objects naturally defined on an algebraic curve $A^{\text{ref}} (x,y; t)$
obtained by a particular deformation of the $A$-polynomial, and its quantization $\hat A^{\text{ref}} (\hat x, \hat y; q, t)$.
We compute both classical and quantum $t$-deformed curves in a number of examples coming from
colored knot homologies and refined BPS invariants.
\\
\\
\\
\\
\\
\\
{\tt CALT-68-2866}}

\begin{document}

%%%%%%%%%%%%%%%%%%%%%%%%%%%%%%%%%%%%%%%%%%%%%%%%%%%%%%%%%%%%%%%%%%%%%

\section{Introduction}    \label{sec-intro}

The story of the ``volume conjecture'' started with the crucial observation \cite{Kashaev} that the so-called Kashaev invariant of a knot $K$
defined at the $n$-th root of unity $q = e^{2 \pi i / n}$ in the classical limit has a nice asymptotic behavior
determined by the hyperbolic volume $\text{Vol} (M)$ of the knot complement $M = S^3 \setminus K$.
Shortly after, it was realized \cite{MurMur} that the Kashaev invariant is equal to the $n$-colored Jones polynomial of a knot $K$
evaluated at $q = e^{2 \pi i / n}$, so that the volume conjecture could be stated simply as
\be
\lim_{n \to \infty} \frac{2 \pi \log |J_n (K; q = e^{2 \pi i / n})|}{n} \; = \; \text{Vol} (M) \,.
\label{VCbasic}
\ee

The physical interpretation of the volume conjecture was proposed in \cite{Apol}.
Besides explaining the original observation \eqref{VCbasic} it immediately led to a number
of generalizations, in which the right-hand side is replaced by a function of various parameters (see \cite{DGreview} for a review).
Below we state two such generalizations -- associated, respectively, with the parameters $\hbar$ and $u$ -- that
in the rest of the paper will be ``refined'' or, morally speaking, ``categorified.''

\subsection{Generalized volume conjecture}

Once the volume conjecture is put in the context of analytically continued Chern-Simons theory,
it becomes clear that the right-hand side is simply the value of the classical $SL(2,\C)$ Chern-Simons action functional on a knot complement $M$.
Since classical solutions in Chern-Simons theory ({\it i.e.} flat connections on $M$) come in families,
parametrized by the holonomy of the gauge connection on a small loop around the knot,
this physical interpretation immediately leads to a ``family version'' of the volume conjecture \cite{Apol}:
\be
J_n (K; q = e^{\hbar}) \;\overset{{n \to \infty \atop \hbar \to 0}}{\sim}\;
\exp \left( \frac{1}{\hbar} S_0 (u) \,+\, \ldots %\sum_{n=0}^\infty S_{n+1} (u,t) \, \hbar^{n}
\right)
\label{VCparam}
\ee
parametrized by a complex variable $u$. Here, the limit on the left-hand side is slightly more interesting
than in \eqref{VCbasic} and, in particular, also depends on the value of the parameter $u$:
\be
q = e^{\hbar} \to 1 \,, \qquad
n \to \infty \,, \qquad
q^n = e^u \equiv x  ~~~\text{(fixed)}
\label{VClimit}
\ee
In fact, Chern-Simons theory predicts all of the subleading terms in the $\hbar$-expansion denoted by ellipsis in \eqref{VCparam}.
These terms are the familiar perturbative coefficients of the $SL(2,\C)$ Chern-Simons partition function on $M$.

\subsection{Quantum volume conjecture}

Classical solutions in Chern-Simons theory ({\it i.e.} flat connections on $M$) are labeled
by the holonomy eigenvalue $x = e^u$ or, to be more precise, by a point on the algebraic curve
\be
\cC: \quad \left\{(x,y)\in\mathbb{C}^*\times \mathbb{C}^*\Big|
A(x,y) \; = \; 0 \right\}\,,
\label{Acurve}
\ee
defined by the zero locus of the $A$-polynomial, a certain classical invariant of a knot.
In quantum theory, $A(x,y)$ becomes an operator $\hat A (\hat x, \hat y; q)$
and the classical condition \eqref{Acurve} turns into a statement that the Chern-Simons partition function
is annihilated by $\hat A (\hat x, \hat y; q)$. This statement applies equally well to Chern-Simons theory
with the compact gauge group $SU(2)$ that computes the colored Jones polynomial $J_n (K;q)$ as well as to
its analytic continuation that localizes on $SL(2,\C)$ flat connections.
In the former case, one arrives at the ``quantum version'' of the volume conjecture \cite{Apol}:
\be
\hat A \; J_* (K;q) \; \simeq \; 0 \,,
\label{VCquant}
\ee
which in the mathematical literature was independently proposed around the same time~\cite{Garoufalidis} and is know as the AJ-conjecture.
The action of the operators $\hat x$ and $\hat y$ follows from quantization of Chern-Simons theory.
With the standard choice of polarization\footnote{Although different choices of polarization will not play an important role
in the present paper, the interested reader may consult {\it e.g.} \cite{GMtorsion,Tudor} for further details.}
one finds that $\hat x$ acts as a multiplication by $q^n$, whereas $\hat y$ shifts the value of~$n$:
\begin{align}
& \hat x J_n \; = \; q^n J_n \label{xyactionJ} \\
& \hat y J_n \; = \; J_{n+1} \notag
\end{align}
In particular, one can easily verify that these operations obey the commutation relation
\be
\hat y \hat x \; = \; q \hat x \hat y
\label{xycomm}
\ee
that follows from the symplectic structure on the phase space of Chern-Simons theory.
Therefore, upon quantization a classical polynomial relation of the form \eqref{Acurve} becomes
a $q$-difference equation for the colored Jones polynomial or Chern-Simons partition function.
Further details, generalizations, and references can be found in \cite{DGreview}.\\

One of the goals of the present paper is to propose a ``refinement'' or ``categorification''
of the generalized and quantum volume conjectures \eqref{VCparam} and \eqref{VCquant}.

\subsection{Quantization and deformation of algebraic curves}

The above structure -- and its ``refinement'' that we are going to construct -- is not limited to applications in knot theory.
Similar mathematical structure appears in matrix models \cite{BIPZ,Dijkgraaf:1990rs,Fukuma:1990jw},
in four-dimensional $\cN=2$ supersymmetric gauge theories \cite{SW-I,Nekrasov,NS,Braverman:2004cr},
and in topological string theory \cite{ADKMV,DHSV,DHS,ACDKV}.

In all of these problems, the semiclassical limit is described by a certain ``spectral curve'' \eqref{Acurve}
defined by the zero locus of $A(x,y)$.
Motivated by application to knots, we shall refer to the function $A(x,y)$ as the $A$-polynomial even in situations
where its form is not at all polynomial.

\bigskip
\begin{figure}[ht]
\centering
\includegraphics[width=4.0in]{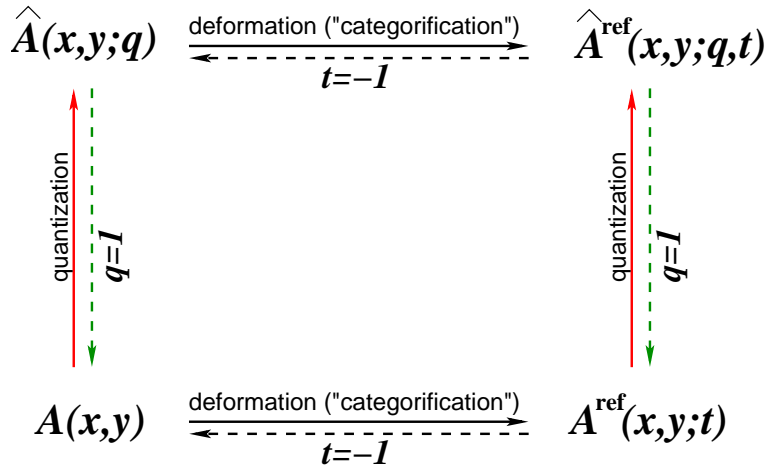}
\caption{Deformation and quantization of the $A$-polynomial. The horizontal arrows describe a deformation / refinement,
such that the unrefined case corresponds to $t=-1$. The vertical arrows represent quantization, {\it i.e.} a lift
of classical polynomials $A(x,y)$ and $A(x,y;t)$ to quantum operators.}
\label{fig:AAAA}
\end{figure}

Furthermore, in most of these problems the classical curve \eqref{Acurve} admits two canonical deformations
with the corresponding parameters $q$ and $t$.
The first one is the deformation quantization in which $x$ and $y$
turn into non-commutative generators of the Weyl algebra, {\it cf.} \eqref{xycomm}.
The second deformation, parametrized by $t$, does not affect commutativity of $x$ and $y$,
{\it i.e.} it is an ordinary deformation.
How these deformations affect the classical curve defined by the zero locus of $A(x,y)$ is illustrated in Figure \ref{fig:AAAA}.

A systematic procedure for lifting the polynomial $A(x,y)$ to the quantum operator $\hat A (\hat x, \hat y; q)$
is described in \cite{Tudor} for curves coming from triangulated 3-manifolds and, more generally, in \cite{abmodel}
for abstract curves defined by the equation $A(x,y)=0$. The general approach of \cite{abmodel} is based
on the topological recursion, which allows to compute $\hat A (\hat x, \hat y; q)$
order by order in the $q$-expansion following the steps of \cite{Mironov,Marino:2006hs,eyn-or,BKMP,DijkgraafFuji-2}
where similar computations of the partition function were discussed.
Thus, under favorable conditions described in \cite{abmodel}, the quantum operator
\be
\hat A (\hat x, \hat y; q) \; = \; \sum_{m,n}\, a_{m,n}\, q^{c_{m,n}}\, \hat x^m\, \hat y^n
\label{Aqnice}
\ee
can be obtained simply from the data $\{ a_{m,n} \}$ of the original polynomial $A(x,y) = \sum a_{m,n} x^m y^n$
and from the data $\{ c_{m,n} \}$ of the Bergman kernel $B(u_1,u_2)$ which, for curves of arbitrary genus,
is given by a derivative of the logarithm of the odd characteristic theta function.
Specifically,
given the Bergman kernel $B(u_1,u_2)$ for the classical curve \eqref{Acurve},
one can first compute the ``torsion'' $T(u)$,
\be
\log T (u) \; = \; \lim_{u_1 \to u_2=u} \int
\left( \frac{du_1 \, du_2}{(u_1 - u_2)^2} - B(u_1,u_2) \right) \,,
\ee
and then find the exponents $\{ c_{m,n} \}$ by solving
\be
\sum_{m,n}\, a_{m,n}\, c_{m,n}\, x^m y^n \; = \;
\frac{1}{2} \left( \frac{\partial_u A}{\partial_v A} \partial_v^2 + \frac{\partial_u T}{T} \partial_v \right) \; A
\ee
together with $A(x,y) = 0$. 
Substituting the resulting data $\{ c_{m,n} \}$ into \eqref{Aqnice} gives the quantization of $A(x,y)$.
In the above equations we used the relations
\be
x = e^u, \qquad  \qquad y = e^v \,.  \label{xuyv}
\ee

In this paper, our main focus will be the deformation in the other direction, associated with the parameter $t$.
Some prominent examples of classical $A$-polynomials and their
$t$-deformations which we find are given in table \ref{table}. Important
examples of quantum refined $A$-polynomials, which we will derive in this paper, are revealed in table \ref{table-qt}.

\begin{table}[h]
\centering
\begin{tabular}{|@{$\Bigm|$}c|c@{$\Bigm|$}c|c@{$\Bigm|$}|}
\hline
\rule{0pt}{5mm}
\textbf{Model} & $A(x,y)$ & $A^{\text{ref}} (x,y;t)$  \\[3pt]
\hline
\hline
\rule{0pt}{5mm}
unknot & $(1-x)(1-y)$ & $(1 + t^3 x) ( - t^{-3})^{1/2} - (1 - x) y$  \\[3pt]
\hline
\rule{0pt}{5mm}
trefoil & $(y-1) (y + x^3)$ & ~~~$y^2 -\frac{1 - x t^2 + x^3 t^5 + x^4 t^6 + 2 x^2 t^2 (t+1)}{1
+ x t^3} y + \frac{(x-1) x^3 t^4}{1 + x t^3}$~~~    \\[3pt]
\hline
\rule{0pt}{5mm}
tetrahedron & $1 - y + x(-y)^f$ & $1 + t y - t x (-y)^f$   \\[3pt]
\hline
\rule{0pt}{5mm}
conifold & $1 - y + x(-y)^f + Q x (-y)^{f+1}$ & $1 + t y - t x (-y)^f + Q \sqrt{-t} \, x (-y)^{f+1}$  \\[3pt]
\hline
\end{tabular}
\caption{Classical curves in prominent examples, in unrefined limit (which are well known, left column) and for general $t$ (derived in this paper, right column). In this paper we also derive refined $A$-polynomials for general $T^{2,2p+1}$ torus knots; for explicit examples for low values of $p$ see table \protect\ref{table_A-poly4}. Note that in section \protect\ref{sec:Bmodel} the deformation parameter $t$ is identified with $-q_1$ (when $q_2 = q = 1$). \label{table} }
\end{table}

\section{The new conjectures: incorporating $t$}

In this section we describe general aspects of the mathematical structure shared by a wide variety of
examples, ranging from the counting of refined BPS invariants to categorification of quantum group invariants.
Then, in later sections we focus on each class of examples separately.

In particular, one of our goals is to promote the volume conjectures \eqref{VCparam}
and \eqref{VCquant} to the corresponding refined / categorified versions, both for knots and for the refined BPS invariants.

\subsection{Quantum volume conjecture: refined}

The fact that a $q$-difference operator $\hat A (\hat x, \hat y; q)$ annihilates the partition function
of Chern-Simons theory / matrix model / B-model / instanton partition function is easy to refine.
Just like each of these partition functions becomes $t$-dependent,
so does the operator $\hat A (\hat x, \hat y; q,t)$ that annihilates it.
The commutation relations \eqref{xycomm} do not change and, therefore,
our proposal for the refinement of \eqref{VCquant} is easy to state:
\begin{subequations}\label{VCquantref}
\be
\hat A^{\text{ref}} (\hat x, \hat y; q, t) \; P_* (K;q,t) \; \simeq \; 0 \qquad \qquad \text{(knots)}
\ee
or
\be
\hat A^{\text{ref}} (\hat x, \hat y; q, t) \; Z_{\text{BPS}}^{\text{open}} (u,q,t) \; \simeq \; 0 \qquad \qquad \text{(BPS states)}
\ee
\end{subequations}
While the formulation of this refined / homological version is simple and follows the lines
of the ordinary quantum volume conjecture \eqref{VCquant}, its interpretation is rather deep and non-trivial.
It involves details of the framework in which \eqref{VCquantref} arises and will be given a proper treatment in the following sections.
Here, we only remark that polynomials $P_n (K;q,t)$ which appear in (\ref{VCquantref}a) as $t$-dependent analogs
of the colored Jones polynomials $J_n (K;q)$ are Poincar\'e polynomials of the $n$-colored $sl(2)$ knot homology groups $\cH^{sl(2),V_n} (K)$:
\be
P_n (K;q,t) \; = \; \sum_{i,j} q^i t^j \dim \cH_{i,j}^{sl(2),V_n} (K) \,,
\label{Pnoptimistic}
\ee
such that
\be
J_n (K;q) \; = \; P_n (K;q,t=-1) \,.
\ee
Because the $t$-deformation does not affect the commutation relation \eqref{xycomm},
the operators $\hat x$ and $\hat y$ act on $P_n$ exactly as in \eqref{xyactionJ}:
\begin{align}
& \hat x P_n \; = \; q^n P_n \label{xyactionP} \\
& \hat y P_n \; = \; P_{n+1} \notag
\end{align}

\begin{table}[h]
\centering
\begin{tabular}{|@{$\Bigm|$}c|c@{$\Bigm|$}|}
\hline
\rule{0pt}{5mm}
\textbf{Model} &  $\widehat{A}^{\text{ref}} (\hat{x},\hat{y};q,t)$ \\[3pt]
\hline
\hline
\rule{0pt}{5mm}
unknot &  $(1 + t^3 q \hat{x}) ( - q^{-1} t^{-3})^{1/2} - (1 - \hat{x})\hat y$ \\[3pt]
\hline
\rule{0pt}{5mm}
\multirow{2}{*}{trefoil}   &   $  \frac{1}{q + \hat{x}^2 q^{2} t^3} \; \hat{y}
+ \frac{\hat{x}^3 (x - q) t^4}{q (q + \hat{x}^2 t^3) (1 + \hat{x} q t^3)} \; \hat{y}^{-1} +   $    \\[3pt]
\rule{0pt}{5mm}
        &    $  -   \frac{t^2 \hat{x}^2}{1 + \hat{x}^2 q t^3} -
\frac{ q - \hat{x} q t^2 + \hat{x}^{4} t^6 +
  \hat{x}^{2} t^2 (1 + t + q t)}{(q + \hat{x}^{2} t^3) (1 + \hat{x} q t^3)}     $         \\[3pt]
\hline
\rule{0pt}{5mm}
tetrahedron & $1 + t \hat{y} - t \sqrt{q} \, \hat{x}(-\hat{y})^f  \; \simeq  \; 1 + t$ \\[3pt]
\hline
\rule{0pt}{5mm}
conifold &  $1 + t \hat{y} - t \sqrt{q} \, \hat{x}(-\hat{y})^f  + Q \sqrt{-tq} \, \hat{x}(-\hat{y})^{f+1} \; \simeq \; 1 + t$~ \\[3pt]
\hline
\end{tabular}
\caption{Refined quantum curves derived in this paper. The notation ``$\ldots \simeq 1 + t$'' should be understood as $\widehat{A}^{\text{ref}} Z_{\text{BPS}}^{\text{open}} = 1 + t$. Note that in section \protect\ref{sec:Bmodel} the quantization parameter $q$ is identified with $q_2$,
and the deformation parameter is $t= - \frac{q_1}{q_2}$. \label{table-qt} }
\end{table}

\subsection{Generalized volume conjecture: refined}

The refinement / categorification of the generalized volume conjecture \eqref{VCparam} involves
taking the limit \eqref{VClimit} while keeping the extra parameter $t$ fixed:
\be
q = e^{\hbar} \to 1 \,, \qquad t = \text{fixed} \,, \qquad x \equiv e^u = q^n = \text{fixed}
\label{reflimit}
\ee
We conjecture that, in this limit, the homological (resp. refined) knot (resp. BPS) invariants
have the following asymptotic behavior:
\begin{subequations}\label{VCparamref}
\be
P_n  \; \simeq \;
\exp\left( \frac{1}{\hbar} S_0 (u,t) \,+\,\sum_{n=0}^\infty S_{n+1} (u,t) \, \hbar^{n} \right)
\qquad \qquad \text{(knots)}
\ee
or
\be
Z_{\text{BPS}}^{\text{open}} (u,q,t) \; \simeq \;
\exp\left( \frac{1}{\hbar} S_0 (u,t) \,+\,\sum_{n=0}^\infty S_{n+1} (u,t) \, \hbar^{n} \right)
\qquad \qquad \text{(BPS states)}
\ee
\end{subequations}
with the leading term (``classical action'')
\be
S_0 (u,t) \; = \; \int v du \; = \; \int \log y \frac{dx}{x}
\label{S0ref}
\ee
defined as an integral on a classical curve
\be
\cC^{\text{ref}}: \quad
\left\{(x,y)\in\mathbb{C}^*\times \mathbb{C}^*\Big|
A^{\text{ref}} (x,y;t) \; = \; 0
\right\}\,,
\ee
which is a deformation of the classical A-polynomial curve $A(x,y)=0$.
In writing \eqref{S0ref} we used the same convention as in (\ref{xuyv}).

%%%%%%%%%%%%%%%%%%%%%%%%%%%%%%%%%%%%%%%%%%%%%%%%%%%%%%%%%%%%%%%%%%%%%%%%%%%
%%%%%%%%%%%%%%%%%%%%%%%%%%%%%%%%%%%%%%%%%%%%%%%%%%%%%%%%%%%%%%%%%%%%%%%%%%%

\section{Examples coming from knots}
\label{sec:knots}

In this section, we illustrate the refined / categorified volume conjectures stated in the general form
in \eqref{VCquantref} and \eqref{VCparamref} for a large class of examples coming from knots.
In these examples, the invariants $P_n (q,t)$ whose recursive behavior is captured by the conjectures
encode the graded dimensions of the homological knot invariants.

\begin{figure}[ht]
\centering
\includegraphics[width=4.0in]{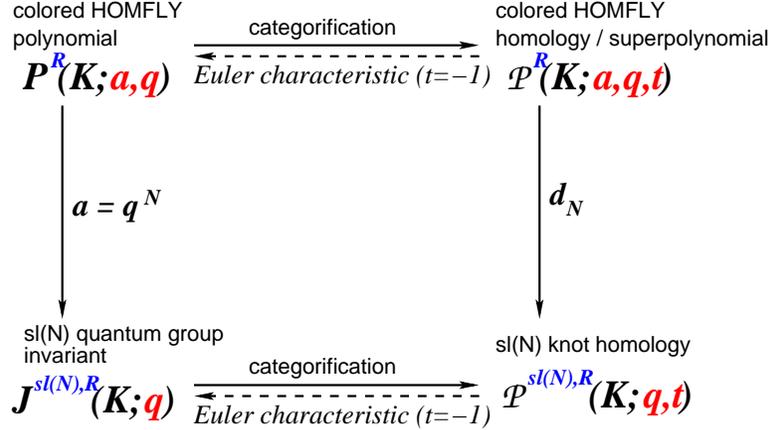}
\caption{Categorification of quantum knot invariants.
To help the reader navigate through this picture we suggest to keep track of the variables $a$, $q$, $t$,
as well as the rank of $sl(N)$. Note, the polynomial (resp. homological) knot invariants which have $a$-dependence
(resp. $a$-grading) are {\it not} labeled by $sl(N)$.}
\label{fig:superpolunmial_hier}
\end{figure}

There are many different kinds of homological knot invariants: doubly-graded and triply-graded,
reduced and unreduced, with different choices of framing and grading conventions.
And, in our discussion we will need at least a basic understanding of these concepts
in order to have most fun with the conjectures (\ref{VCquantref}a) and (\ref{VCparamref}a).
In other words, we will need to have at least a rough understanding of the relation between
different types of knot invariants shown in Figure \ref{fig:superpolunmial_hier}.
Luckily, much of this picture can be explained building on the relations between polynomial
knot invariants, which hopefully are more familiar to the reader.
\bigskip

We start with the lower left corner of the Figure \ref{fig:superpolunmial_hier} that describes
the simplest family of polynomial knot invariants $J^{\frak{g}, R} (K;q)$
labeled by a representation $R$ of a Lie algebra $\frak{g} = \text{Lie} (G)$.
In the special case when $R = V_n$ is the $n$-dimensional representation of $\frak{g} = sl(2)$,
the quantum group invariant $J_n (K;q) := J^{sl(2),V_n} (K;q)$ is the $n$-colored Jones polynomial of a knot/link $K$
that we already encountered in the review of the generalized volume conjectures \eqref{VCparam} and \eqref{VCquant}.
In general, a mathematical definition of $J^{\frak{g}, R} (K;q)$ involves associating a quantum R-matrix to
every crossing in a plane diagram of a knot $K$.
Physically, these quantum group invariants are simply the normalized expressions for
the partition function of Chern-Simons gauge theory \cite{Witten_Jones}:
\be
Z_{G}^{\text{CS}}(M,K_R;q)
\; := \; \int [dA] \, W_R(K)[A] \, e^{ik S_{\text{CS}}[A;M]}
\label{ZCSdef}
\ee
with a Wilson loop operator $W_R(K)[A]:={\rm Tr}_R \text{P} \exp\left[\oint_K A\right]$
supported on a knot $K$ and decorated by a representation $R$.
Here, $S_{\rm CS}[A;M]$ is the famous Chern-Simons action functional on a 3-manifold $M$,
\begin{eqnarray}
S_{\text{CS}}[A;M] \; = \; \frac{1}{4\pi}\int_M {\rm Tr}_{\text{adj}}
\left( A\wedge dA+\frac{2}{3}A\wedge A\wedge A\right) \,.
\label{SCSdef}
\end{eqnarray}
In general, the partition function \eqref{ZCSdef}
is a rather complicated function of the coupling constant $k$ or, equivalently, the ``quantum'' parameter $q=e^{\frac{2\pi i}{k+h}}$.
However, once normalized by that of the unknot, it magically becomes a polynomial in $q$ with integer coefficients,
at least when $M=\S^3$:
\begin{eqnarray}
J^{\frak{g}, R} (K;q) \; =\; \frac{Z_{G}^{\rm CS}(\S^3,K_R;q)}{Z_{G}^{\rm CS}(\S^3,\unknot_R;q)} \,.
\label{JasZZratio}
\end{eqnarray}
The fact that the final result turns out to be a polynomial, let alone integer coefficients,
is not at all obvious in either R-matrix or path integral formulation of $J^{\frak{g}, R} (K;q)$.
This nice property, however, is a precursor of knot homologies, which beautifully explain it.

Another nice property comes from a closer look at $\frak{g} = sl(N)$, which will be the focus our paper.
(Although there are straightforward analogs for other classical groups, we will not consider them here.)
Then, not only $J^{sl(N), R} (K;q)$ turn out to be polynomials in $q$, they exhibit a very simple dependence on $N$.
Namely, for each $R$ (= Young tableaux) there exists a polynomial invariant $P^R (K;a,q)$ of a knot $K$, such that
\be
J^{sl(N), R} (K;q) \; = \; P^R (K; a = q^N, q) \,.
\label{JfromP}
\ee
This relation is represented by a vertical arrow on the left in Figure \ref{fig:superpolunmial_hier}.
The polynomial $P^R (K;a,q)$ is called the colored HOMFLY polynomial of $K$.
To be more precise, it is the {\it normalized} HOMFLY polynomial, meaning that $P^R (\unknot) = 1$.

Once we explained the left side of Figure \ref{fig:superpolunmial_hier}, we can easily
describe its categorification shown on the right. To categorify $J^{sl(N), R} (K;q)$ means to construct
a doubly-graded homology theory $\cH^{sl(N),R}_{i,j} (K)$, with gradings $i$ and $j$, such that
the polynomial $J^{sl(N), R} (K;q)$ is its $q$-graded Euler characteristic:
\be
J^{sl(N), R} (K;q) \; = \; P^R (K; a = q^N, q) \; = \; \sum_{i,j} (-1)^j q^i \dim \cH^{sl(N),R}_{i,j} (K) \,.
\label{JfromH}
\ee
Similarly, a categorification of $P^R (K;a,q)$ is a triply-graded homology
theory $\cH^{R}_{i,j,k} (K)$, with gradings $i$, $j$ and $k$, whose graded Euler characteristic is
\be
P^R (K;a,q) \; = \; \sum_{i,j,k} (-1)^k a^i q^j \dim \cH^{R}_{i,j,k} (K) \,.
\label{PfromH}
\ee
The relations \eqref{JfromH} and \eqref{PfromH} are represented by horizontal arrows in Figure \ref{fig:superpolunmial_hier}.
Sometimes, it is convenient to express these relations as specializations
of the corresponding Poincar\'e polynomials $\P^{sl(N),R} (q,t)$ and $\P^R (a,q,t)$ at $t=-1$.
For example, the Poincar\'e polynomial of the triply-graded homology theory $\cH^{R}_{i,j,k} (K)$
is defined as follows
\be
\text{``superpolynomial''} \qquad
\P^R (K; a,q,t) \; := \; \sum_{i,j,k} \, a^i q^j t^k \, \dim \cH^{R}_{i,j,k} (K)
\label{superPdef}
\ee
and often is called the colored superpolynomial.
To be more precise, just like below eq.~\eqref{JfromP} we pointed out that
$P^R (a,q) = \P^R (a,q,t=-1)$ is the {\it normalized} colored HOMFLY polynomial,
it is important to emphasize that $\P^R (K; a,q,t)$ is the Poincar\'e polynomial
of the {\it reduced} homology theory, in a sense that $\P^R (\unknot) = 1$.

If one naively combines \eqref{JfromH} and \eqref{PfromH} one may be tempted to conclude that
$\P^{sl(N),R} (q,t)$ is a specialization of the superpolynomial at $a=q^N$,
as was stated {\it e.g.} in \cite{AS} and in some other recent papers.
We emphasize that, in general, this is {\it not} the case:
\be
\P^R (a=q^N,q,t) \; \ne \;
\P^{sl(N),R} (q,t) \equiv \sum_{i,j} q^i t^j \dim \cH^{sl(N),R}_{i,j} (K) \,.
\label{superwrong}
\ee
The reason is very simple and becomes crystal clear if one attempts to test \eqref{superwrong}
even in the basic case of $N=1$ and, say, $R = \Box$.
Indeed, the $sl(1)$ theory is trivial in any approach to knot polynomials / homologies.
In other words, for any knot the $sl(1)$ homology $\cH^{sl(1),\Box}_{i,j} (K)$
is one-dimensional and the corresponding quantum group invariant $J^{sl(1),\Box} (K;q)$ consists of a single monomial.

This fact has a nice manifestation in the relation \eqref{JfromP}, which says that almost all terms
in the HOMFLY polynomial $P^{\Box} (a,q)$ cancel out in the specialization $a=q$, leaving behind a single term.
This remarkable property of the HOMFLY polynomial %reflects the fact that $sl(1)$ theory is ``trivial'' and
can be viewed as a non-trivial constraint on the coefficients of the polynomial $P^{\Box} (a,q)$.
Since the coefficients of $P^{\Box} (a,q)$ can be positive and negative, this condition indeed
can be satisfied if the total number of ``minuses'' (counted with multiplicity) is balanced by the total number of ``pluses.''
This, of course, can not work for \eqref{superwrong} where the superpolynomial $\P^{\Box} (a,q,t)$
has only positive coefficients, due to \eqref{superPdef}. Therefore, setting $a=q$ will never reduce
the total sum of the coefficients in this polynomial, and there is no way it can be equal to
the Poincar\'e polynomial $\P^{sl(1),R} (q,t)$ of a one-dimensional homology $\cH^{sl(1),R}_{i,j} (K)$.

A more conceptual reason why \eqref{superwrong} can not be true is that, while a specialization to $a=q^N$ is
perfectly acceptable at the polynomial level (the left side of Figure \ref{fig:superpolunmial_hier}),
it has to be replaced by a suitable operation from homological algebra in order to
make sense at the higher categorical level (the right side of Figure \ref{fig:superpolunmial_hier}).
As explained in \cite{DGR}, the suitable operation, which categorifies the specializations $a=q^N$,
involves taking homology in the triply-graded theory with respect to differentials $d_N$, $N \in \Z$.
Indeed, the ``extra terms'' in the superpolynomial $\P^R (a,q,t)$ that are not part of
$\P^{sl(N),R} (q,t)$ and otherwise would cancel upon setting $t=-1$ always come in pairs,
so that a more proper version of \eqref{superwrong} reads
\be
\P^R (a,q,t) \; = \;
R^{sl(N),R} (a,q,t) + (1 + a^{\alpha} q^{\beta} t^{\gamma}) Q^{sl(N),R} (a,q,t) \,,
\label{superright}
\ee
where $R^{sl(N),R} (a,q,t)$ and $Q^{sl(N),R} (a,q,t)$ are polynomials with non-negative coefficients,
such that the $sl(N)$ homological invariant is a specialization of the ``remainder'' (not the full superpolynomial as in \eqref{superwrong}):
\be
R^{sl(N),R} (a=q^N,q,t) \; = \; \P^{sl(N),R} (q,t) \,,
\label{PRright}
\ee
whereas the extra pairs of terms in \eqref{superright} that come from $Q^{sl(N),R} (a,q,t)$
are killed by the differential $d_N$ of $(a,q,t)$-degree $(\alpha,\beta,\gamma)$.

Now, once we introduced the cast of characters and explained the relations between them,
we can get straight down to business.
We start in the next subsection with the calculation of colored superpolynomials for $(2,2p+1)$ torus.
Then, in section \ref{sec:knotrecursions} we use the results of these calculations to derive and study
the recursion relations, {\it a.k.a.} the quantum volume conjecture (\ref{VCquantref}a).
Starting in section \ref{sec:knotrecursions}, we focus on the $a=q^2$ specializations
of $S^r$-colored superpolynomial, which we denote as
\be
P_n (q,t) \; := \; \P^{S^{n-1}} (a=q^2,q,t) \,.
\label{Pq2special}
\ee
In section \ref{sec:homological} we present some evidence that in the refined volume conjectures \eqref{VCquantref} and \eqref{VCparamref}
one can replace this definition of $P_n (q,t)$ with a true $sl(2)$ homological knot invariant \eqref{Pnoptimistic}.
Finally, in section \ref{sec:knotS0} we consider the classical limit and discuss
the leading ``volume'' term $S_0 (u,t)$ that dominates the asymptotic behavior (\ref{VCparamref}a)
of these homological knot invariants.
According to \eqref{S0ref}, this semi-classical limit is controlled by the $t$-deformed
algebraic curve, $A^{\text{ref}} (x,y;t) = 0$, whose quantization will be revisited in section \ref{sec:quantizability}.

In a physical realization of knot homologies \cite{GSV}, the superpolynomial $\P^{S^{n-1}} (a,q,t)$
and its specialization \eqref{Pq2special} are certain indices that count refined BPS invariants,
essentially identical to $Z_{\text{BPS}}^{\text{open}} (u,q,t)$ where $u = q^n$.
Motivated by this, in the later section \ref{sec:Bmodel} we perform a similar analysis
of more general refined open BPS partition functions and find many similar patterns.

\subsection{$S^r$ homological invariants of $(2,2p+1)$ torus knots}

Our goal in this section is to compute the colored superpolynomials $\P^{R} (a,q,t)$
for $(2,2p+1)$ torus knots and symmetric and anti-symmetric representations, $R=S^r$ and $R=\Lambda^r$.
In performing this calculation, we first review the analogous computation of the polynomial knot
invariants (on the left side of Figure \ref{fig:superpolunmial_hier}) in Chern-Simons gauge theory and then ``refine'' it.
We should stress right away, however, that the refined calculation is {\it not} done in a topological 3d gauge theory
or, at least, such a 3d gauge theory interpretation of the formal steps that we are going to take is not known at present.

In Chern-Simons gauge theory, one can efficiently compute \eqref{ZCSdef}
using the topological invariance of the theory.
Dividing a 3-manifold $M$ into two pieces $M_1$ and $M_2$ along a surface $\Sigma$,
one can express the partition function $Z_{SU(N)}^{\rm CS}(M,K_R;q)$
as a pairing between the two elements
$|\psi_{M_1;K_R}\rangle$ and $|\psi_{M_2;K_R}\rangle$
in the physical Hilbert space ${\cal H}_{\Sigma}$:
\begin{eqnarray}
Z_{SU(N)}^{\rm CS}(M,K_R;q)={}_{\Sigma}\langle \psi_{M_2;K_R}|\psi_{M_1;K_R}\rangle_{\Sigma} \,.
\label{ZHilbert}
\end{eqnarray}
The physical Hilbert space ${\cal H}_{\Sigma}$ is obtained by the canonical quantization
of the Chern-Simons gauge theory on $\Sigma\times \mathbb{R}$, and the following correspondence is found in \cite{Witten_Jones,EMSS}:
\begin{center}
${\cal H}_{\Sigma} \; \simeq \; \bigl\{$
conformal blocks for $G/G$ WZW model on $\Sigma~\bigr\}$,
\end{center}
where $G=SU(N)$.
If the knot $K_R$ meets the surface $\Sigma$ at $m$ points,
the physical Hilbert space ${\cal H}_{\Sigma;R_1\cdots R_m}$
consists of conformal blocks for $m$-point functions which carry representations $R_a = R$ or $\bar{R}$ ($a=1,\cdots ,m$)
depending on the orientation at the intersection.

\begin{figure}[h]
\begin{center}
\includegraphics[width=8cm,keepaspectratio,clip]{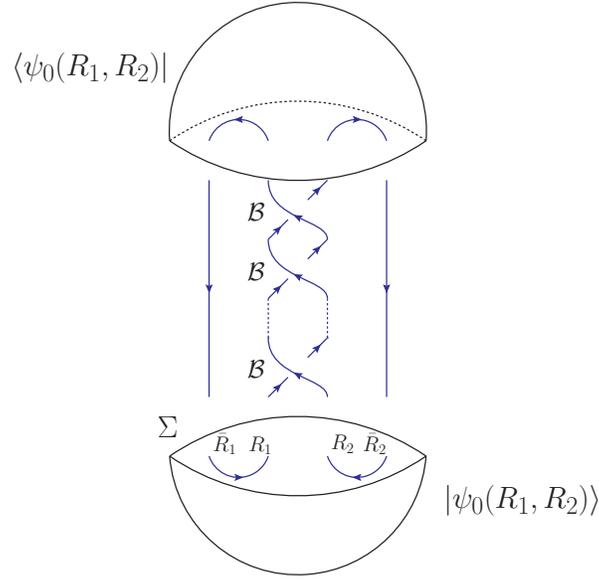}
\end{center}
\caption{Slicing and braiding.}
\label{Braid_Slice}
\end{figure}

Here we will focus on the torus knots $T^{2,2p+1}$ in a 3-sphere and follow the steps of~\cite{RGK}.
Among various choices of the surface $\Sigma$, we will consider a slicing of $\S^3$ into
a pair of 3-dimensional balls $\B^3$ connected by a cylinder, as in Figure \ref{Braid_Slice}.
The physical Hilbert space ${\cal H}_{\S^2; R\, R \, \bar{R}\,\bar{R}}$ for this slice
consists of conformal blocks for the four point function:
\begin{eqnarray}
\phi_Q(R,R,\bar{R},\bar{R}),\quad Q\in R\otimes R,
\end{eqnarray}
whose the intermediate state carries an irreducible representation $Q \in R\otimes R$.
In general, ${\cal H}_{\S^2; R_1 R_2 R_3 R_4}$ is spanned by orthonormal states:
\begin{eqnarray}
\langle\phi_{Q^{\prime}}(R_1,R_2,R_3,R_4)
|\phi_{Q}(R_1,R_2,R_3,R_4)\rangle=\delta_{QQ^{\prime}},\quad
Q,Q^{\prime}\in R_1\otimes R_2.
\end{eqnarray}

\begin{figure}[h]
\begin{center}
\includegraphics[width=5cm,keepaspectratio,clip]{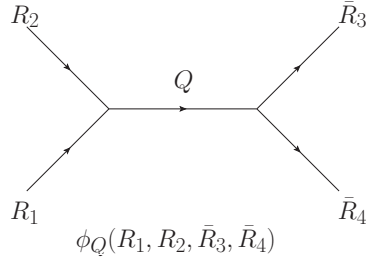}
\end{center}
\label{fig:Conf_block}
\caption{Conformal block $\phi_{Q^{\prime}}(R_1,R_2,\bar{R}_3,\bar{R}_4)$ for the four point function.}
\end{figure}

\begin{figure}[h]
\begin{center}
\includegraphics[width=15cm,keepaspectratio,clip]{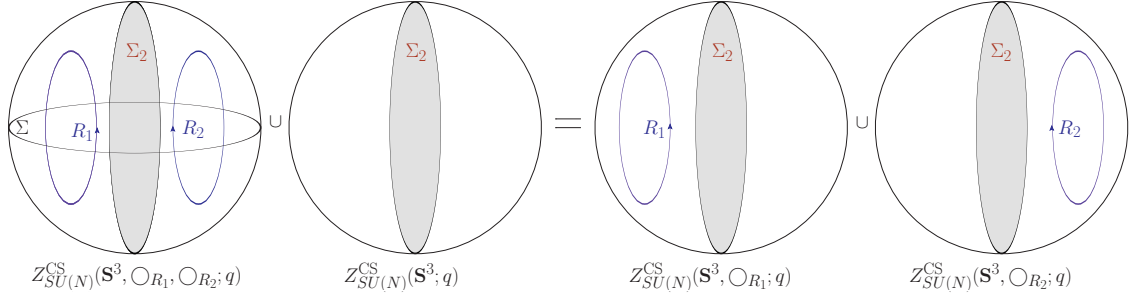}
\end{center}
\caption{Slicing along $\Sigma_2 \simeq \S^2$.}
\label{fig:Another_S2}
\end{figure}

For a state $|\psi_0(R_1,R_2)\rangle\in {\cal H}_{\S^2;R_1R_2\bar{R}_2\bar{R}_1}$
associated with the lower half of the 3-ball in Figure \ref{Braid_Slice},
the following expansion can be considered:
\begin{eqnarray}
|\psi_0(R_1,R_2)\rangle \; = \; \sum_{Q\in R_1\otimes R_2}\mu^Q_{R_1R_2}|\phi_{Q}(R_1,R_2,\bar{R}_2,\bar{R}_1)\rangle \,.
\label{three_ball}
\end{eqnarray}
The coefficient $\mu^Q_{R_1R_2}$ is determined by taking a pairing of this state with itself:
\begin{eqnarray}
\langle \psi_0(R_1,R_2)|\psi_0(R_1,R_2)\rangle &=& \sum_{Q\in R_1\otimes R_2}\left(\mu^Q_{R_1R_2}\right)^2
\nonumber \\
&=& Z_{SU(N)}^{\rm CS}(\S^3,\unknot_{R_1}\unknot_{R_2};q) \,.
\end{eqnarray}
Since the two unknots here are not linked, we can separate them by an application
of another slicing along a surface $\Sigma_2 \simeq \S^2$, as in Figure \ref{fig:Another_S2}:
\be
Z_{SU(N)}^{\rm CS}(\S^3,\unknot_{R_1}\unknot_{R_2};q)
Z_{SU(N)}^{\rm CS}(\S^3;q)
\; = \; Z_{SU(N)}^{\rm CS}(\S^3,\unknot_{R_1};q)
Z_{SU(N)}^{\rm CS}(\S^3,\unknot_{R_2};q).
\ee

The partition function of the unknot $\unknot_R$
in $\S^3$ is given by the quantum dimension:
\be
\frac{Z_{SU(N)}^{\rm CS}(\S^3,\unknot_{R};q)}
{Z_{SU(N)}^{\rm CS}(\S^3;q)} \; = \; \dim_q R \,.
\label{unknotdimqR}
\ee
Here, the quantum dimension $\dim_qR$ is a specialization of the
Schur polynomial $s_R(x)$:
\be
\dim_q R \; = \; s_R(q^{\varrho}),\quad \varrho=(\varrho_1,\cdots,\varrho_N),\quad \varrho_i=\frac{N+1}{2}-i \,,
\label{dimqRsR}
\ee
which enjoys the following identity:
\be
s_{R_1}(x)s_{R_2}(x) \; = \sum_{Q\in R_1\otimes R_2}s_Q(x) \,.
\label{irr_decomp_schur}
\ee
Using these relations, the coefficient $\mu_Q$ can be determined as:
\be
\left(\mu^Q_{R_1R_2}\right)^2
\; = \; s_Q(q^{\varrho})\cdot Z_{SU(N)}^{\rm CS}(\S^3;q) \,.
\label{muCS}
\ee

The state $|\psi_{2p+1}(R_1,R_2)\rangle$ associated with the top part of the picture \ref{Braid_Slice}
is constructed by acting $2p+1$ times with the braid operator ${\cal B}_{R_1R_2}$ on the 3-ball state $|\psi_{0} (R_1,R_2)\rangle$:
\begin{eqnarray}
|\psi_{2p+1}(R_1,R_2)\rangle \; = \; {\cal B}_{R_1R_2}^{2p+1}|\psi_{0}(R_1,R_2) \rangle \,.
\end{eqnarray}
The action of the braid operator ${\cal B}_{R_1R_2}$ on the conformal block
$\phi_Q(R_1,R_2,R_3,R_4)$ obeys a monodromy transformation.
The eigenvalue of the monodromy for the conformal block is determined by the
conformal weights $h_{R_a}$ of primary states in the $G/G$ WZW model \cite{Moore_Seiberg}:
\begin{eqnarray}
\lambda_Q^{\prime}(R_1,R_2) \; = \;
\epsilon_{R_1R_2}^Qe^{\pi i(h_{R_1}+h_{R_2}-h_Q)}
\; = \; \epsilon_{R_1R_2}^Qq^{\frac{1}{2}(C_2(R_1)+C_2(R_2)-C_2(Q))} \,.
\end{eqnarray}
The quadratic Casimir $C_2(R)$ for a representation $R$ of $SU(N)$ is given by
\begin{eqnarray}
C_2(R) \;= \; \frac{1}{2}\left(\kappa_R+N |R|-\frac{|R|^2}{N}\right) \,,
\end{eqnarray}
where $R_i$ denotes the number of boxes in the $i$-th row of the Young diagram for the representation $R$,
and $|R|:=\sum_i R_i$, and $\kappa_R=|R|+\sum_iR_i(R_i-2i)$.
The sign $\epsilon_{R_1R_2}^Q=\pm 1$ is determined by whether $Q$
appears symmetrically or antisymmetrically in $R_1\otimes R_2$.
In order to keep the canonical framing for the Chern-Simons partition function,
it is necessary to make a correction to the eigenvalue for the braid
operator by a factor
$q^{\frac{1}{2}(C_2(R_1)+C_2(R_1)+|C_2(R_1)-C_2(R_2)|)}$
\cite{RGK,Witten_Jones}. Therefore, the resulting eigenvalue of the braid operator is given by
\begin{eqnarray}
\lambda_Q^{(+)}(R_1,R_2)=\epsilon_{R_1R_2}^Qq^{C_2(R_1)+C_2(R_2)+|C_2(R_1)-C_2(R_2)|/2-C_2(Q)/2} \,.
\label{unrefeigl}
\end{eqnarray}

Combining the above formulae, we can evaluate the braid operator ${\cal B}_{RR}^{2p+1}$
sandwiched between two 3-ball states $|\psi_0(R,R)\rangle$:
\begin{eqnarray}
Z_{SU(N)}^{\rm CS}(\S^3,T^{2,2p+1}_R;q)
& =& \langle \psi_0(R,R)|{\cal B}_{RR}^{2p+1}|\psi_0(R,R)\rangle
\label{ZT22p1braid} \\
& = & \sum_{Q\in R\otimes R}\lambda^{(+)}_Q(R,R)^{2p+1}\left(\mu^Q_{RR}\right)^2
\nonumber \\
& = & Z_{SU(N)}^{\rm CS}(\S^3;q)\sum_{Q \in R \otimes R}\lambda^{(+)}_Q(R,R)^{2p+1}{\rm dim}_qQ \,.
\nonumber
\end{eqnarray}
Our next goal is to categorify / refine this computation.

In practice, this will amount to replacing every ingredient with its analog that
depends not only on $q$, but also on the new variable $t$ or, rather,
two variables $q_1$ and $q_2$ (that are related to $q$ and $t$ via a simple change of variables \eqref{GSvarchng}):
\begin{eqnarray}
\underline{\text{~~CS gauge theory~~}} && \underline{\text{~~refined invariants~~}}
\nonumber \\
Z_{SU(N)}^{\rm CS}(\S^3,K_R;q)
&\qquad \leadsto \qquad&
Z^{\text{ref}}_{SU(N)}(\S^3,K_R;q_1,q_2)
\nonumber \\
\dim_q R = s_R(q^{\varrho})
&\qquad \leadsto \qquad&
M_R(q_2^{\varrho};q_1,q_2)
\nonumber \\
q^{C_2(R)}
&\qquad \leadsto \qquad&
q_1^{\frac{1}{2}||R||^2}q_2^{-\frac{1}{2}||R^{t}||^2}q_2^{\frac{N}{2}|R|}q_1^{-\frac{1}{2N}|R|^2}
\label{Casimir_ref} \\
& \vdots & \nonumber
\end{eqnarray}

\subsubsection{Refined braid operators and gamma factors}
\label{sec:refinedbraid}

A physical framework for knot homologies was first proposed in \cite{GSV} and later studied from various viewpoints
and in a number of closely related systems in \cite{Gknothom,Wknothom,DGH,AS}.
Regardless of the details and duality frames used, the basic idea is that a graded vector space $\cH^{R} (K)$
associated to a knot $K$ colored by a representation $R$ is identified with the space of refined BPS invariants
that carry information not only about the charge of the BPS state but also about the spin content:
\be
\cH^R (K) \; = \; \cH^{\text{ref}}_{\text{BPS}} \, .
\label{HHBPS}
\ee
This interpretation can be used for performing concrete computations
\cite{GIKV,AS,IK11} (see also
\cite{Taki:2008hb,Awata:2009yc,Awata:2009sz} and \cite{Carqueville:2011sj,ORSG,Aganagic:2012au})
as well as for studying the structure of $\cH^{\frak g, R}_{i,j} (K)$ for various $\frak g$ and $R$ \cite{GWalcher,GS}.
In this section, we will use both methods -- based on concrete formulas for torus knots and on structural properties for arbitrary knots --
to compute colored HOMFLY homology and colored superpolynomials of $(2,2p+1)$ torus knots.
Furthermore, the physical interpretation of the homological knot invariants in terms of the refined ({\it a.k.a.} motivic) BPS invariants
is what will ultimately allow us to treat the latter in more general systems on the same footing, {\it cf.} section \ref{sec:Bmodel}.

First, let us recall the five-brane configuration relevant to the physical description of the $sl(N)$ knot homologies \cite{GSV,Gknothom,Wknothom}:
\bea
\text{space-time} & : & \qquad \R \times T^* \S^3 \times M_4 \nonumber \\
N~\text{M5-branes} & : & \qquad \R \; \times \; \S^3 \; \times \; D \label{theoryA} \\
|R|~\text{M5-branes} & : &  \qquad \R \; \times \; L_K \; \times \; D \nonumber
\eea
where $L_K$ is the total space of the conormal bundle to $K \subset \S^3$ in the Calabi-Yau space $T^* \S^3$,
and in most of applications one usually takes $D \cong \R^2$ and $M_4 \cong \R^4$.
See {\it e.g.} \cite{GS} for further details, references,
and the outline of the relation between different ways of looking at this physical system.

The precise form of the 4-manifold $M_4$ and the surface $D \subset M_4$ is not important,
as long as they enjoy $U(1)_F \times U(1)_P$ symmetry action, where the first (resp. second) factor
is a rotation symmetry of the normal (resp. tangent) bundle of $D \subset M_4$.
Following \cite{Wknothom}, let us denote the corresponding quantum numbers by $F$ and $P$.
These quantum numbers were denoted, respectively, by $2 S_1$ and $2(S_1 - S_2)$ in \cite{AS}
and by $2 j_3$ and $n$ in \cite{GS}.

Something special happens when $K = T^{m,n}$ is a torus knot. Then, as pointed out in \cite{AS},
the five-brane theory in \eqref{theoryA} has an extra $R$-symmetry $U(1)_R$ that acts on $\S^3$
leaving the knot $K = T^{p,q}$ and, hence, the Lagrangian $L_K \subset T^* \S^3$ invariant.
Following \cite{AS}, we denote the quantum number corresponding to this symmetry by $S_R$,
and also introduce the partition function
\be
Z^{\text{ref}}_{SU(N)}(\S^3, T^{m,n}_R; q, t) \; := \; \Tr_{\cH^{\text{ref}}_{\text{BPS}}} \; (-1)^{S_R} q^{P} t^{F - S_R}
\label{Zrefdef}
\ee
that ``counts'' refined BPS states in the setup \eqref{theoryA}.
A priori, this partition function is different from the Poincar\'e polynomial
of the $sl(N)$ knot homology, which in these notations reads
\be
\P^{sl(N), R} (K;q,t) \; = \; \Tr_{\cH^{\text{ref}}_{\text{BPS}}} \; q^{P} t^{F} \,.
\label{ZrefdefP}
\ee
However, it was argued in \cite{AS} that for some torus knots all refined BPS states \eqref{HHBPS}
have $S_R=0$ and the two expressions actually agree.

Similarly, in a dual description after the geometric transition the setup \eqref{theoryA} turns into a system
\bea
\text{space-time} & : & \qquad \R \times X \times M_4 \label{theoryB} \\
\text{M5-branes} & : &  \qquad \R \times L_K \times D \nonumber
\eea
where $X$ is the total space of the $\cO (-1) \oplus \cO (-1)$ bundle over $\cp^1$,
and BPS states carry a new quantum number which becomes the $a$-grading of $\cH^{\text{ref}}_{\text{BPS}} = \cH^R_{i,j,k} (K)$.
One of the main results in \cite{AS} is that this space has {\it four} gradings: in addition to
the $a$-, $q$- and $t$-grading that in the physics setup correspond to the ``winding number'' $\beta \in H_2 (X,L_K) \cong \Z$,
and to the quantum numbers $P$ and $F$, the space $\cH_{BPS} = \cH^R_{i,j,k} (K)$ has the fourth grading, by $S_R \in \Z$.

Therefore, one of the interesting features of the refined Chern-Simons theory is that it predicts
a new grading on the homology of torus knots and links, thereby upgrading $\cH^{sl(N),R}_{i,j} (K)$
to a triply-graded theory (labeled by $\frak g$ and $R$) and similarly upgrading $\cH^{R}_{i,j,k} (K)$
to a homology theory with as much as {\it four} gradings!
It would be very interesting to study these new extra gradings in other formulations of knot homologies.
Some hints for the extra gradings of torus knot homologies seem to appear in \cite{Cherednik}.

After the geometric transition, the partition function analogous to \eqref{Zrefdef}
``counts'' refined BPS states in the setup \eqref{theoryB}:
\be
Z^{\text{ref}} (\S^3, T^{m,n}_R; a, q, t) \; = \; \Tr_{\cH^{\text{ref}}_{\text{BPS}}} \; (-1)^{S_R} a^{\beta} q^{P} t^{F - S_R} \,.
\label{Zrefdefa}
\ee
When all refined BPS states have $S_R=0$ -- which, following \cite{AS}, will be our working assumption here --
this expression coincides with the colored superpolynomial \eqref{superPdef}, which in these
notations reads $\P^{R} (K;a,q,t) = \Tr_{\cH^{\text{ref}}_{\text{BPS}}} \; a^{\beta} q^{P} t^{F}$.
This will be our strategy for obtaining the colored superpolynomials of $(2,2p+1)$ torus knots.
In fact, via the relation with refined BPS invariants, we will essentially do the computation twice: first, via direct
calculation of the refined Chern-Simons partition function \eqref{Zrefdef} and its large $N$ version \eqref{Zrefdefa},
and then, in section \ref{sec:homological}, by using the structural properties of \eqref{HHBPS} that follow from physics.\\

Although refined Chern-Simons theory is {\it not} a gauge theory,\footnote{At least, such formulation is not
known at present.} its partition function can be evaluated by mimicking the steps in the ordinary Chern-Simons theory.
In particular, one can define ``refined analogs'' of the $S$ and $T$ modular matrices
and the braid operator ${\cal B}_{R_1 R_2}$ used in \eqref{ZT22p1braid}.
In addition to these ingredients, we will also need a modification of the refined braid operator
by the so-called {\it gamma factor} proposed in
\cite{DMMSS,Shakirov,Mironov:2011ym,MMSS,Mironov:2012hz}).
Modulo this modification, the physical meaning of which is still unclear at present, we need the refined variants of

\begin{enumerate}

\item
the partition function of the unknot $Z_{SU(N)}^{\text{CS}}(\S^3,\unknot_R;q)$
in order to determine the coefficient $\mu_Q$, and

\item
the quadratic Casimir factor $q^{C_R}$ in the monodromy by the action of the braid operator.

\end{enumerate}

\noindent
The former, {\it viz.} the refined partition function of the unknot, is given by the partition
function of the refined BPS invariants of the conifold with a D-brane inserted at the appropriate
leg of the toric diagram \cite{IK11}. The resulting partition function is the refined analogue of \eqref{unknotdimqR}
and is given simply by the Macdonald polynomial $M_R(x;q_1,q_2)$:
\be
\frac{Z^{\text{ref}}_{SU(N)}(\S^3,\unknot_R;q_1,q_2)}{Z^{\text{ref}}_{SU(N)}(\S^3;q_1,q_2)}
\; = \; M_R(q_2^{\varrho};q_1,q_2) \, ,
\label{unknotMqtR}
\ee
see also appendix (\ref{sec:macdonald}). The combinatorial expression for the Macdonald polynomial is
\be
M_R(q_2^{\varrho};q_1,q_2) \; = \; \prod_{(i,j)\in R}
\frac{q_2^{\frac{N-i+1}{2}}q_1^{\frac{j-1}{2}}-q_2^{-\frac{N-i+1}{2}}q_1^{-\frac{j-1}{2}}}{q_2^{\frac{R^{t}_j-i+1}{2}}q_1^{\frac{R_i-j}{2}}-q_2^{-\frac{R^{t}_j-i+1}{2}}q_1^{-\frac{R_i-j}{2}}} \,,
\label{Mac_sp}
\ee
where $R^{t}$ denotes the transposition of the Young diagram $R$.
Furthermore, the Macdonald polynomial satisfies the analogue of (\ref{irr_decomp_schur}):
\be
M_{R_1}(x;q_1,q_2) \, M_{R_2}(x;q_1,q_2) \; = \; \sum_{Q\in R_1\otimes R_2}N^Q_{R_1R_2} M_{Q}(x;q_1,q_2) \,,
\ee
where $N^{Q}_{R_1R_2}$ is a certain rational function of $q_1$ and $q_2$,
namely the Littlewood-Richardson coefficient \cite{McD}.
Therefore, as in \eqref{muCS}, we can use these relations to determine
the refined analogue of the coefficient $\mu^Q_{R_1R_2}$
that enters the expression \eqref{three_ball} for the 3-ball partition function:
\be
\left( \mu^Q_{R_1R_2} \right)^2 \; = \; N^{Q}_{R_1R_2} \cdot M_Q(q_2^{\varrho};q_1,q_2)
\cdot Z^{\text{ref}}_{SU(N)}(\S^3;q_1,q_2) \,.
\label{mu_refined}
\ee
Here, following \cite{AS}, we tacitly assumed that generating functions of the refined BPS invariants,
such as \eqref{Zrefdef} and \eqref{Zrefdefa}, can be expressed in the form \eqref{ZHilbert},
as in a local quantum field theory, with a Hilbert space whose states are labeled by conformal blocks.
It would be interesting to understand better the physical basis for this assumption and to study
the unitary structure on the ``Hilbert space'' of the refined Chern-Simons theory.
In particular, in the case of $\Sigma\simeq \mathbf{T}^2$,
one would hope to understand better the identification of the basis of orthonormal states with integrable representations.

As for the second ingredient, the quadratic Casimir factor $q^{C_2(R)}=e^{2\pi ih_R}$,
it is related to the modular transformation
$T = \begin{pmatrix}
1 & 1 \cr
0 & 1
\end{pmatrix}$ which acts in the standard way on the homology cycles of $\Sigma=\mathbf{T}^2$ in the WZW model.
The modular matrix for the action of the $T$-transformation on the characters of $\hat{su}(N)_k$ is
\be
T_{RQ} \; = \; T_{\emptyset\emptyset} \cdot q^{C_2(R)}\cdot \delta_{RQ} \,.
\ee
In \cite{AS} the $T$-matrix for the refined Chern-Simons theory is proposed:
\be
T^{\text{ref}}_{RQ} \; = \; T_{\emptyset\emptyset}\cdot
\frac{q_1^{\frac{1}{2}||R||^2}}{q_2^{\frac{1}{2}||R^{t}||^2}} \frac{q_2^{\frac{N}{2}|R|}}{q_1^{\frac{1}{2N}|R|^2}}\cdot\delta_{RQ} \,,
\ee
where $||R||^2:=\sum_iR_i^2$.
Hence, we adopt a refinement \eqref{Casimir_ref}
and the eigenvalue $\lambda^{(+)}_Q(R,R)$
of the braid operator ${\cal B}_{RR}$ for the refined theory:
\be
\lambda^{(+)}_Q(R,R) \; = \; \epsilon_{RR}^{Q}
\frac{q_1^{||R||^2}}{q_2^{||R^{t}||^2}}\frac{q_2^{N|R|}}{q_1^{\frac{1}{N}|R|^2}}
\frac{q_1^{-\frac{1}{4}||Q||^2}}{q_2^{-\frac{1}{4}||Q^{2}||^2}}
\frac{q_2^{-\frac{N}{4}|Q|}}{q_1^{-\frac{1}{4N}|Q|^2}} \,.
\label{braid_ref}
\ee
Similar framing factors were considered in \cite{DMMSS} and also deduced from the physics of refined BPS invariants in \cite{IK11}.
Here we use slightly modified expressions to match \eqref{unrefeigl} in the unrefined limit $q_1 = q_2$.

Using these ingredients, we find the partition function
$Z_{SU(N)}^{\text{ref}}(\S^3,T^{2,2p+1}_R;q_1,q_2)$ for the $T^{2,2p+1}$ torus knot:
\begin{eqnarray}
&& Z_{SU(N)}^{\text{ref}} (\S^3,T^{2,2p+1}_R;q_1,q_2) \nonumber \\
&& \quad = \; Z_{SU(N)}^{\text{ref}}(\S^3;q_1,q_2) \sum_{Q\in R\otimes R}\gamma^Q_{RR}\cdot
\lambda^{(+)}_Q(R,R)^{2p+1}\cdot N^{Q}_{RR}\cdot M_Q(q_2^{\varrho};q_1,q_2) \,, \label{ZrefTpqq}
\end{eqnarray}
where, following \cite{DMMSS}, we introduced a gamma factor $\gamma^Q_{RR}$.
This factor is needed to make the partition function for the torus knot
invariant under the obvious symmetry $T^{n,m}\leftrightarrow T^{m,n}$:
\be
Z_{SU(N)}^{\text{ref}}(\S^3,T^{n,m}_R;q_1,q_2) \; = \; Z_{SU(N)}^{\text{ref}}(\S^3,T^{m,n}_R;q_1,q_2) \,.
\label{consistency_gen}
\ee
While the proper physical understanding of the gamma factors is lacking, it does not prevent one from doing calculations.
Indeed, the gamma factors can be determined by recursively solving the consistency conditions \eqref{consistency_gen}
for $m=1,\cdots ,n-1$ ($m<n$).
In particular, for torus knots $T^{2,2p+1}$, the gamma factors can be found from a single consistency condition for $p=0$,
{\it i.e.} for $T^{2,1} \simeq \unknot$:
\begin{eqnarray}
\frac{Z_{SU(N)}^{\text{ref}}(\S^3,T^{2,1}_R;q_1,q_2)}{Z_{SU(N)}^{\text{ref}}(\S^3;q_1,q_2)}
&=& \frac{Z_{SU(N)}^{\text{ref}}(\S^3,\unknot_R;q_1,q_2)}{Z_{SU(N)}^{\text{ref}}(\S^3;q_1,q_2)} \label{consistency_gamma} \\
&=&\sum_{Q \in R \otimes R} \gamma^Q_{RR}\cdot \lambda^{(+)}_Q(R,R) \cdot N^{Q}_{RR}
\cdot M_Q(q_2^{\varrho};q_1,q_2) = M_R(q_2^{\varrho};q_1,q_2) \,.
\nonumber
\end{eqnarray}

\subsubsection*{The gamma factors for symmetric and anti-symmetric representations}

In order to determine the gamma factors from the consistency condition (\ref{consistency_gamma}),
one needs the explicit form of the Littlewood-Richardson coefficients $N_{RR}^Q$.
For the symmetric representation $R=S^r$ and the anti-symmetric representation $R=\Lambda^r$,
the explicit expression for the Littlewood-Richardson coefficients can be obtained from the Pieri formula:
\begin{eqnarray}
N_{S^rS^r}^{S^{r+\ell,r-\ell}}
& = & \prod_{j=1}^{r-\ell} { 1-q_1^{j-1} q_2 \over 1-q_1^j }
\cdot \prod_{j=2\ell+1}^{r+\ell} { 1-q_1^{j-1} q_2^2 \over 1-q_1^j q_2 } \cdot \prod_{j=\ell+1}^{r}
\left( { 1-q_1^{j}  \over 1-q_1^{j-1} q_2 } \right)^2 \,, \label{LRsymm}
\\
N_{\Lambda^r\Lambda^r}^{\Lambda^{r+\ell,r-\ell}}
& = & \prod_{i=1}^{\ell} { 1-q_1 q_2^{i-1} \over 1-q_2^i } \cdot \prod_{i=\ell+1}^{2\ell} {1-q_2^{i} \over 1-q_1q_2^{i-1}} \,.
\label{LRantisymm}
\end{eqnarray}

\begin{figure}[h]
\begin{center}
\includegraphics[width=10cm,keepaspectratio,clip]{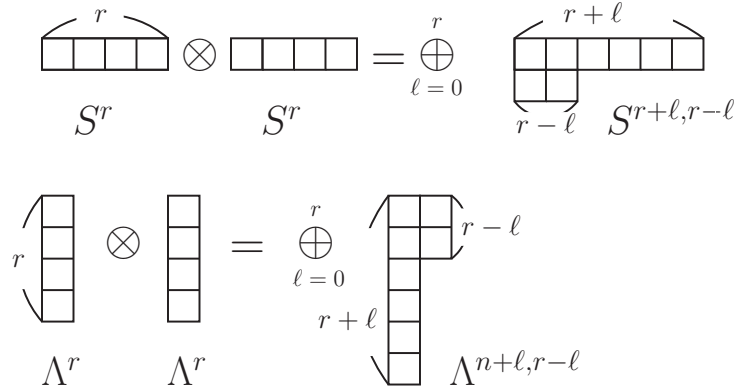}
\end{center}
\caption{Young diagrams for tensor products of symmetric and anti-symmetric representations.}
\end{figure}

For the symmetric and anti-symmetric representations,
the sign factors $\epsilon^{Q}_{RR}$ in $\lambda^{(+)}_Q(R,R)$ look like \cite{RGK,Ramadevi_Sarkar}:
\be
\epsilon_{S^rS^r}^{S^{r+\ell,r-\ell}} = (-1)^{r-\ell} \,, \qquad
\epsilon_{\Lambda^r\Lambda^r}^{\Lambda^{r+\ell,r-\ell}} = (-1)^{\ell} \,.
\ee
%The Macdonald polynomials with the specialization
%$P_{R}(q_2^{\rho};q_1,q_2)$
%for $R=S^{r+\ell,r-\ell}$ and $R=\Lambda^{r+\ell,r-\ell}$ are
%\begin{eqnarray}
%&&P_{S^{r+\ell,r-\ell}}(q_2^{\rho};q_1,q_2)
%=\frac{(A;q_1)_{r+\ell}(q_2^{-1}A;q_1)_{r-\ell}(q_2^2;q_1)_{2\ell}}{(q_2;q_1)_{%2\ell}(q_2;q_1)_{r-\ell}(q_2^2;q_1)_{r+\ell}}A^{-r}t^{2r-\ell},
%\end{eqnarray}
%where $(x;q)_n:=\prod_{i=1}^n(1-xq^{i-1})$ and  we defined
With the explicit expression for the Macdonald polynomials $M_{R} (q_2^{\varrho};q_1,q_2)$ given in (\ref{Mac_sp}),
we can solve the constraint (\ref{consistency_gamma}) and find the following gamma factors (see Appendix \ref{sec:gamma_proof} for details):
\begin{eqnarray}
\gamma_{S^rS^r}^{S^{r+\ell,r-\ell}}
& = & \prod_{i=1}^{r-\ell}\frac{q_1^{\frac{r-i}{2}}q_2^{\frac{1}{2}}-q_1^{-\frac{r-i}{2}}q_2^{-\frac{1}{2}}}{q_1^{\frac{r-i+1}{2}}-q_1^{-\frac{r-i+1}{2}}} \,,
\label{gamma_sym}
\\
\gamma_{\Lambda^r\Lambda^r}^{\Lambda^{r+\ell,r-\ell}}
& = & \prod_{i=1}^{\ell}\frac{q_2^{\frac{i}{2}}-q_2^{-\frac{i}{2}}}{q_2^{\frac{i-1}{2}}q_1^{\frac{1}{2}}-q_2^{-\frac{i-1}{2}}q_1^{-\frac{1}{2}}} \,.
\label{gamma_anti-sym}
\end{eqnarray}
It would be very interesting to understand the physical meaning / origin
of the gamma factors.

Collecting all the ingredients, \eqref{Mac_sp}, \eqref{braid_ref}, and \eqref{LRsymm}--\eqref{gamma_anti-sym},
we obtain a final expression for the partition function \eqref{ZrefTpqq} with $R=S^r$ and $R=\Lambda^r$:
\begin{eqnarray}
R = S^r ~:\quad &&Z_{SU(N)}^{\text{ref}}(\S^3,T^{2,2p+1}_{S^r};q_1,q_2) / Z_{SU(N)}^{\text{ref}}(\S^3;q_1,q_2)
\nonumber \\
&& \quad = \; \sum_{\ell=0}^r
\frac{(q_2;q_1)_{\ell}(q_1;q_1)_r(A;q_1)_{r+\ell}(q_2^{-1}A;q_1)_{r-\ell}}
{(q_1;q_1)_{\ell}(q_2;q_1)_r(q_2;q_1)_{r+\ell}(q_1;q_1)_{r-\ell}}\frac{(1-q_2q_1^{2\ell})}{(1-q_2)}
\nonumber \\
&&\quad \quad\quad\quad\times
A^{-r}
q_1^{\frac{r-\ell}{2}}
q_2^{\frac{3r-\ell}{2}}
\left[
(-1)^{r-\ell}A^{\frac{r}{2}}q_1^{\frac{r^2-\ell^2}{2}}q_2^{-\frac{\ell}{2}}\right]^{2p+1} \,,
\label{Braid_symmetric}
\\
R = \Lambda^r ~:\quad && Z_{SU(N)}^{\text{ref}}(\S^3,T^{2,2p+1}_{\Lambda^r};q_1,q_2) / Z_{SU(N)}^{\text{ref}}(\S^3;q_1,q_2)
\nonumber \\
&&\quad = \; \sum_{\ell=0}^r\frac{(q_1;q_2)_{\ell}(q_2;q_2)_{r+\ell}(A^{-1};q_2)_{r+\ell}(q_1^{-1}A^{-1};q_2)_{r-\ell}}
{(q_2;q_2)_{\ell}(q_1q_2;q_2)_{r+\ell}(q_2;q_2)_{r+\ell}(q_2;q_2)_{r-\ell}}\frac{(1-q_1q_2^{2\ell})}{(1-q_2)}
\nonumber \\
&&\quad\quad\quad\quad \times
A^{r}q_1^{r-\frac{\ell}{2}}q_2^{r-\frac{\ell}{2}}
\left[(-1)^{\ell}A^{\frac{r}{2}}q_1^{\frac{\ell}{2}}q_2^{\frac{\ell^2-r^2}{2}}
\right]^{2p+1},
\label{Braid_anti-symmetric}
\end{eqnarray}
where we used the standard notation for the $q$-Pochhammer symbol
\be
(x;q)_n \; = \; \prod_{k=0}^{n-1} (1 - x q^k) = (1-x) (1-xq) (1-xq^2) \ldots (1-x q^{n-1}) \,,
\ee
and where we introduced
\be
A \; := \; q_2^N \,.
\ee
Now we have all the relevant formulas at our fingertips that one needs to write down the superpolynomials
of $(2,2p+1)$ torus knots colored by the symmetric and anti-symmetric representations, $R = S^r$ and $R = \Lambda^r$.

\subsubsection{From partition functions to superpolynomials}

As we already explained around \eqref{superPdef}, the reduced colored superpolynomial $\P^R (K; a,q,t)$
is defined as the Poincar\'e polynomial of the triply graded homology $\cH^R_{i,j,k} (K)$
that categorifies the colored HOMFLY polynomial $P^R (K;a,q)$.
Here, the word ``reduced'' means that the normalization is such that $P^R (\unknot) = 1$.
Although normalization is one of the delicate points one has to worry about,
luckily it will not be a major issue for us here.

Besides normalization, there are several other choices that affect the explicit form
of the answer and, therefore, need to be explained, especially for comparison with other approaches.
Thus, earlier we already mentioned a very important choice of framing.
Another important choice is a choice of grading conventions.
In order to understand its important role,
let us recall that the triply-graded homology $\cH^R_{i,j,k} (K)$
is related to the doubly-graded $sl(N)$ theory $\cH^{sl(N), R}_{i,j} (K)$
by means of the differentials $d_N$, illustrated in Figure \ref{fig:superpolunmial_hier}:
\be
\cH^{sl(N), R}_{*,*} (K) \; \cong \; \left( \cH^R_{*,*,*} (K) , d_N \right) \,.
\label{specN}
\ee
Taking the Poincar\'e polynomials on both sides gives \eqref{superright}--\eqref{PRright}.

In order to be consistent with the specialization $a=q^N$ (also illustrated on the left side of Figure \ref{fig:superpolunmial_hier}),
the $q$-degree of $d_N$ should be $N$ times greater than its $a$-degree and of opposite sign.
The standard convention for the homological $t$-grading of all differentials $d_N$ with $N>0$ is~$-1$.
Modulo trivial\footnote{Such rescalings are much more elementary choices of notation,
rather than interesting choices of grading convention that affect the explicit form of the results in a more delicate way.
A typical example of such harmless rescaling is a doubling of all $a$- and $q$-gradings
in the last column of \eqref{gradingtabl} that gives $\deg (d_{N>0}) = (-2,2N,-1)$,
as in the middle column, $\deg (d_{N<0}) = (-2,2N,-3)$, {\it etc.}}
rescalings, such as $a \mapsto a^2$ and $q \mapsto q^2$,
there are two sets of conventions consistent with these rules used in the literature:
\be
\begin{array}{l@{\;}|@{\;}c@{\;}|@{\;}c@{\;}c}
\multicolumn{3}{c}{\textbf{(a, q, t) grading conventions:}} \\[.1cm]
\text{differentials} & \text{conventions of } \cite{DGR,GWalcher,AS,DMMSS} & \text{conventions of } \cite{GS} \\\hline
d_{N>0} & (-2,2N,-1) & (-1,N,-1) \\
d_{N<0} & (-2,2N,2N-1) & (-1,N,-3) \\
d_{\text{colored}} & (0,2,2) & (0,1,0) \\
& (-2,0,-3) & (-1,0,-1) \\
& \vdots &
\end{array}
\label{gradingtabl}
\ee
Here, we mostly follow the latter conventions and occasionally, for comparison, state the results in the former conventions.

Another choice of grading conventions comes from a somewhat surprising direction.
A special feature of the colored knot homology is the {\it mirror symmetry} conjectured in \cite{GS}:
\be
\cH^{R}_{i,j,*} (K) \; \simeq \; \cH^{R^t}_{i,-j,*} (K) \,.
\label{mirror}
\ee
It relates triply-graded HOMFLY homologies colored by representations (Young diagrams) $R$ and $R^t$ related by transposition.
Although this nice property is also present even in the basic uncolored case of $R = R^t = \Box$
as a generalization of the $q \leftrightarrow q^{-1}$ symmetry \cite{DGR},
its significance is fully revealed in the colored theory with $R \ne R^t$.
In particular, for our applications it means that
the triply-graded homologies $\cH^{S^r} (K)$ and $\cH^{\Lambda^r} (K)$
are essentially the same, and so are the colored superpolynomials $\P^{S^r}$ and $\P^{\Lambda^r}$.

Put differently, the mirror symmetry \eqref{mirror} implies that instead of two different triply-graded homology theories
$\cH^{S^r} (K)$ and $\cH^{\Lambda^r} (K)$ one really has only one theory, $\cH^r (K)$, labeled by $r$,
such that passing from $R = S^r$ to $R = \Lambda^r$ is achieved by flipping the sign
of the $q$-grading accompanied by a suitable $t$-regrading.
On the other hand, according to \eqref{specN} the sign of the $q$-grading
is correlated with the sign of $N$ in the specialization to $sl(N)$ doubly-graded homology.
Therefore, $\cH^r (K)$ is related to the $sl(N)$ colored knot homology via
\be
\left( \cH^r (K) , d_N \right) =
\begin{cases}
\cH^{sl(N), S^r} (K) \,, & N>0 \\
\cH^{sl(-N), \Lambda^r} (K) \,, & N<0
\end{cases}
\label{gradchoicea}
\ee
or
\be
\left( \cH^r (K) , d_N \right) =
\begin{cases}
\cH^{sl(N), \Lambda^r} (K) \,, & N>0 \\
\cH^{sl(-N), S^r} (K) \,, & N<0
\end{cases}
\label{gradchoiceb}
\ee
The choice between \eqref{gradchoicea} and \eqref{gradchoiceb} is a matter of convention.
But it is an important choice since it certainly affects the form of $\cH^r (K)$
and the corresponding superpolynomial \eqref{superPdef}.

To summarize, it seems that in our class of examples we have to deal with at least two choices,
between grading conventions in \eqref{gradingtabl} and between \eqref{gradchoicea} and \eqref{gradchoiceb}.
A nice surprise is that these two choices are actually related \cite{GS}: switching from \eqref{gradchoicea} to \eqref{gradchoiceb}
has the same effect as switching from one set of grading conventions in \eqref{gradingtabl} to another.
In other words, to quickly go from one set of grading conventions in \eqref{gradingtabl} to another
one can simply exchange the role of symmetric and anti-symmetric representations or, equivalently,
switch the Young tableaux $R$ and its transposed $R^t$.
We shall use this trick in what follows, where our default grading conventions will be that of \cite{GS} and \eqref{gradchoicea}.\\

Keeping in mind the relations between different convention choices,
now we are ready to convert \eqref{Braid_symmetric} and \eqref{Braid_anti-symmetric}
into the colored superpolynomials $\P^{R} (T^{2,2p+1}; a,q,t)$ for symmetric and anti-symmetric representations.
Starting with the symmetric representations $R = S^r$, we can use the following change of variables
\bea
a & = & A \left( \frac{q_1}{q_2} \right)^{3/2} \,, \nonumber \\
q & = & \frac{1}{q_2} \,, \label{GSvarchng} \\
t & = & - \sqrt{\frac{q_2}{q_1}} \,, \nonumber
\eea
to write the refined partition function $Z_{SU(N)}^{\text{ref}}(\S^3,T^{2,2p+1}_{\Lambda^r};q_1,q_2)$
in terms of $a$, $q$, and $t$. For example, from this identification one easily finds
the unreduced superpolynomial $\bar{\P}^{S^r} (\unknot; a,q,t)$ of the unknot, {\it cf.} \eqref{unknotMqtR}:
\begin{eqnarray}
\bar{\P}^{S^r} (\unknot; a,q,t)
&=& \frac{Z_{SU(N)}^{\text{ref}}(\S^3,\unknot_{\Lambda^r};q_1,q_2)}{Z_{SU(N)}^{\text{ref}}(\S^3;q_1,q_2)}
=M_{\Lambda^r}(q_2^{\varrho};q_1,q_2) \nonumber \\
&=& (-1)^r A^{r/2} q_2^{r/2} \frac{(A^{-1};q_2)_r}{(q_2;q_2)_r}
%\nonumber \\
%&&
=(-1)^{\frac{r}{2}}a^{-\frac{r}{2}}q^{\frac{r}{2}}t^{-\frac{3r}{2}} \frac{(-at^3;q)_r}{(q;q)_r} \,.
\label{unknotsuperSr}
\end{eqnarray}
Similarly, the reduced colored superpolynomial of a more general $(2,2p+1)$ torus knot
is related to the partition function $Z_{SU(N)}^{\text{ref}}(\S^3,T^{2,2p+1}_{\Lambda^r};q_1,q_2)$
via a refined analogue of \eqref{JasZZratio},
\be
\P^{S^r} (T^{2,2p+1}; a,q,t) \; = \;
\left( \frac{q_1}{q_2} \right)^{\tfrac{pr}{2}}
\frac{Z_{SU(N)}^{\text{ref}}(\S^3,T^{2,2p+1}_{\Lambda^r};q_1,q_2)}{Z_{SU(N)}^{\text{ref}}(\S^3,\unknot_{\Lambda^r};q_1,q_2)} \,,
\label{superPfromZZ}
\ee
with the same identification of the parameters \eqref{GSvarchng}.
Note how the role of $R = S^r$ and $R = \Lambda^r$ is exchanged in this relation, in line with the above discussion.
Explicitly, from \eqref{Braid_anti-symmetric} we find
\begin{eqnarray}
\P^{S^r} (T^{2,2p+1}; a,q,t)&=&
\sum_{\ell=0}^r\frac{(qt^2;q)_{\ell}(-at^3;q)_{r+\ell}(-aq^{-1}t;q)_{r-\ell}(q;q)_r}
{(q;q)_{\ell}(q^2t^2;q)_{r+\ell}(q;q)_{r-\ell}(-at^3;q)_r}\frac{(1-q^{2\ell+1}t^2)}{(1-qt^2)}
\nonumber \\
&&\quad
\times (-1)^ra^{-\frac{r}{2}}q^{\frac{3r}{2}-\ell}t^{-rp-\ell+\frac{r}{2}}
\left[
(-1)^{\ell}a^{\frac{r}{2}}q^{\frac{r^2-\ell(\ell+1)}{2}}t^{\frac{3r}{2}-\ell}
\right]^{2p+1} \,.
\label{Superpolynomial_sym_refined}
\end{eqnarray}
This is one of the main results, with these choices of conventions, that we will use
for testing the refined / categorified volume conjectures \eqref{VCquantref} and \eqref{VCparamref}.

For completeness, and to illustrate how it is done in general, we also write down the colored
superpolynomial of $(2,2p+1)$ torus knots obtained from \eqref{Braid_symmetric}.
In the same grading conventions as in the relation \eqref{superPfromZZ},
the role of symmetric and anti-symmetric representations is reversed
and we obtain the $\Lambda^r$-colored superpolynomial:
\be
\P^{\Lambda^r} (T^{2,2p+1}; a,q,t) \; = \; \left( \frac{q_2}{q_1} \right)^{\tfrac{pr}{2}}
\frac{Z_{SU(N)}^{\text{ref}}(\S^3,T^{2,2p+1}_{S^r};q_1,q_2)}{Z_{SU(N)}^{\text{ref}}(\S^3,\unknot_{S^r};q_1,q_2)} \,,
\ee
where the parameter identification is essentially the same as in \eqref{GSvarchng} with $q_1$ and $q_2$ interchanged:
\bea
a & = & A \sqrt{\frac{q_2}{q_1}} \,, \nonumber \\
q & = & \frac{1}{q_1} \,, \\
t & = & - \sqrt{\frac{q_1}{q_2}} \,. \nonumber
\eea
The exchange $q_1 \leftrightarrow q_2$ that accompanies $R \leftrightarrow R^t$ is familiar
in the context of refined BPS invariants as well as in the equivariant instanton counting which, of course, are not unrelated.
Hence, from the partition function $Z^{\text{ref}}_{SU(N)}(\S^3,\unknot_{S^r};q_1,q_2)$,
we find the unreduced superpolynomial $\bar{\P}^{\Lambda^r} (\unknot; a,q,t)$ of the unknot:
\begin{eqnarray}
\bar{\P}^{\Lambda^r} (\unknot; a,q,t)
&=& \frac{Z_{SU(N)}^{\text{ref}}(\S^3,\unknot_{S^r};q_1,q_2)}{Z_{SU(N)}^{\text{ref}}(\S^3;q_1,q_2)}
= M_{S^r}(q_2^{\varrho};q_1,q_2)
\nonumber \\
&=& A^{-\frac{r}{2}} q_2^{\frac{r}{2}} \frac{(A;q_1)_r}{(q_2;q_1)_r}
%\nonumber \\
%&&
=(-1)^{-\frac{r}{2}}a^{\frac{r}{2}}q^{\frac{r}{2}}t^{\frac{3r}{2}}
\frac{(-a^{-1}t^{-1};q)_r}{(qt^2;q)_r} \,,
\end{eqnarray}
and the reduced colored superpolynomial of the $(2,2p+1)$ torus knot:
\begin{eqnarray}
\P^{\Lambda^r} (T^{2,2p+1}; a,q,t) &=& \sum_{\ell=0}^r
\frac{(qt^2;q)_{\ell}(q;q)_r(-a^{-1}t^{-1};q)_{r+\ell}(-a^{-1}q^{-1}t^{-3};q)_{r-\ell}(qt^2;q)_r}
{(q;q)_{\ell}(qt^2;q)_r(q^2t^2;q)_{r+\ell}(q;q)_{r-\ell}(-a^{-1}t^{-1};q)_r}
\nonumber \\
&&\quad
\times \frac{(1-q^{2\ell+1}t^2)}{(1-qt^2)}a^{\frac{r}{2}}q^{\frac{3r}{2}-\ell}t^{3r-\ell}
\left[(-1)^{\ell}a^{\frac{r}{2}}q^{\frac{\ell(\ell+1)-r^2}{2}}t^{\ell}\right]^{2p+1} \,.
\label{Superpolynomial_anti-sym_refined}
\end{eqnarray}
These results agrees with the earlier calculations of the colored superpolynomials in \cite{GS,Cherednik}
for small values of $p$ and $r$, and provide a generalization to arbitrary $p$ and $r$.\\

Finally, we complete this part by writing the same formulas for the colored superpolynomials
\eqref{unknotsuperSr}--\eqref{Superpolynomial_anti-sym_refined}
in a different set of grading conventions that we dub ``DGR,'' {\it cf.} \eqref{gradingtabl}.
As we pointed out earlier, a simple way to implement this change of conventions,
based on the ``mirror symmetry'' \eqref{mirror}, is to change $R \mapsto R^t$ on one side
of the relations \eqref{unknotsuperSr}, \eqref{superPfromZZ}, {\it etc.}
As a result, we obtain a nice relation between the $R$-colored superpolynomial of $T^{2,2p+1}$
and the refined partition function of a line operator colored by the same representation $R$
(not $R^t$ as {\it e.g.} in \eqref{superPfromZZ}):
\be
\P^{R}_{\text{DGR}} (T^{2,2p+1}; a,q,t) \; = \; \left( \frac{q_2}{q_1} \right)^{\tfrac{pr}{2}}
\frac{Z^{\text{ref}}_{SU(N)}(\S^3,T^{2,2p+1}_{R};q_1,q_2)}{Z_{SU(N)}(\S^3,\unknot_{R};q_1,q_2)} \,,
\ee
where $R=S^r$ or $\Lambda^r$.
Using \eqref{Braid_symmetric}--\eqref{Braid_anti-symmetric} and the identification of variables \cite{GIKV}:
\begin{eqnarray}
a^2 &=& A \sqrt{\frac{q_2}{q_1}} \,, \nonumber \\
q &=& \sqrt{q_2} \,, \label{DGRvarchng} \\
t &=& - \sqrt{\frac{q_1}{q_2}} \,. \nonumber
\end{eqnarray}
we find explicit expressions for $\P^{R}_{\text{DGR}}(T^{2,2p+1})$ with $R=S^r$ and $R=\Lambda^r$:
\begin{eqnarray}
\P^{S^r}_{\text{DGR}}(T^{2,2p+1}; a,q,t)
&=&\sum_{\ell=0}^r
\frac{(q^2;q^2t^2)_{\ell}(q^2t^2;q^2t^2)_r(-a^2t;q^2t^2)_{r+\ell}(-a^2q^{-2}t;q^2t^2)_{r-\ell}}
{(q^2t^2;q^2t^2)_{\ell}(-a^2t;q^2t^2)_r(q^4t^2;q^2t^2)_{r+\ell}(q^2t^2;q^2t^2)_{r-\ell}}
\nonumber \\
&\times& \frac{(1-q^{4\ell+2}t^{4\ell})}{(1-q^2)}a^{-r}q^{3r-2\ell}t^{r-\ell}
\left[ (-1)^{r-\ell}a^rq^{r^2-\ell(\ell+1)}t^{r^2-\ell^2} \right]^{2p+1}
\label{Symmetric_DGR_refined}
\end{eqnarray}
\begin{eqnarray}
\P^{\Lambda^r}_{\text{DGR}}(T^{2,2p+1};a,q,t)
&=& \sum_{\ell=0}^r
\frac{(q^2t^2;q^2)_{\ell}(-a^{-2}t^{-1};q^2)_{r+\ell}(-a^{-2}q^{-2}t^{-3};q^2)_{r-\ell}(q^2;q^2)_r}{(q^2;q^2)_{\ell}(q^4t^2;q^2)_{r+\ell}(q^2;q^2)_{r-\ell}(-a^{-2}t^{-1};q^2)_r}
\nonumber \\
&\times& \frac{(1-q^{4\ell+2}t^2)}{(1-q^2t^2)}a^r q^{3r-2\ell}t^{3r-\ell}\left[(-1)^{\ell}a^{r}q^{-r^2+\ell(\ell+1)}t^{\ell}\right]^{2p+1}
\label{Anti-Symmetric_DGR_refined}
\end{eqnarray}
as well as the unreduced superpolynomials of the unknot $\bar{\P}^{S^r}_{\text{DGR}}(\unknot)$:
\be
\bar{\P}^{S^r}_{\text{DGR}}(\unknot; a,q,t)
= \frac{Z_{SU(N)}^{\text{ref}}(\S^3,\unknot_{S^r};q_1,q_2)}{Z_{SU(N)}^{\text{ref}}(\S^3;q_1,q_2)}
=(-1)^{\frac{r}{2}}a^{-r}q^{r}t^{-\frac{r}{2}}\frac{(-a^2t;q^2t^2)_r}{(q^2;q^2t^2)_r} \,.
\ee
\be
\bar{\P}^{\Lambda^r}_{\text{DGR}}(\unknot; a,q,t)
= \frac{Z_{SU(N)}^{\text{ref}}(\S^3,\unknot_{\Lambda^r};q_1,q_2)}{Z_{SU(N)}^{\text{ref}}(\S^3;q_1,q_2)}
=(-1)^{\frac{r}{2}}a^rq^rt^{\frac{r}{2}}\frac{(-a^{-2}t^{-1};q^2)_r}{(q^2;q^2)_r} \,.
\ee

%*******************************************************

\subsubsection{Homological algebra of colored knot invariants}
\label{sec:homological}

Our next goal is to describe a very rich structure of the colored superpolynomials
that can be used either as an alternative way to compute them or as a tool to verify their correctness.
As in the previous discussion and in most of the literature on this subject \cite{KhPatterns,KhR1,DGR},
$\P^R (K;a,q,t)$ stands for the {\it reduced} superpolynomial of a knot $K$ colored by $R$,
and its unreduced version is denoted with a bar.

Suppose, for example, that we wish to compute the $S^2$-colored HOMFLY homology $\cH^{S^2}_{i,j,k} (T^{2,5})$
and the corresponding superpolynomial $\P^{S^2} (T^{2,5}; a,q,t)$ of the $(2,5)$ torus knot, also known as the knot ${\bf 5}_1$.
By definition \eqref{superPdef}, at $t=-1$ the colored superpolynomial reduces to the colored HOMFLY polynomial \eqref{PfromH},
which for the knot $T^{2,5}$ has 25 terms, see {\it e.g.} \cite{LinZheng}:
\begin{eqnarray}
P^{S^2} (T^{2,5}; a,q)
&=& a^6 \left( q + q^4 + q^5 + q^7 \right) \label{P25S2} \\
&& + a^5 \left( - q^{-2} - q^{-1} - q - 2 q^2 - q^3 - q^4 - 2 q^5 - q^6 - q^7 - q^8 \right) \nonumber \\
&& + a^4 \left( q^{-4} + q^{-1} + 1 + q^2 + q^3 + q^4 + q^5 + q^6 + q^8 \right) \,. \nonumber
\end{eqnarray}
Each terms in this expression comes from a certain generator
of the triply-graded colored HOMFLY homology $\cH^{S^2}_{i,j,k} (T^{2,5})$, {\it cf.} Figure \ref{fig:superpolunmial_hier}.
Therefore, we conclude that $\cH^{S^2}_{i,j,k} (T^{2,5})$ must be at least 25-dimensional.
How can we restore the homological $t$-grading?

There are many ways to do that, based on the structure of the commuting differentials \eqref{gradingtabl}.
For instance, one way is to pick a relation in the infinite set \eqref{specN} labeled by $N \in \Z$, and to study its implications.
Thus, in our present example, the relation \eqref{specN} with $N=2$ says that the homology of $\cH^{S^2}_{i,j,k} (T^{2,5})$
with respect to the differential $d_2$ should be isomorphic to the $S^2$ colored $sl(2)$ homology $\cH^{sl(2),S^2}_{i,j} (T^{2,5})$,
\be
\left( \cH^{S^2}_{*,*,*} (K) , d_2 \right)
\; \cong \;
\cH^{sl(2), S^2}_{*,*} (K)
\; \cong \;
\cH^{so(3), V}_{*,*} (K) \,,
\ee
where in the last isomorphism we used the identification between the symmetric representation $R = S^2$ of $sl(2)$
and the 3-dimensional vector representation of $so(3)$. The latter homology was studied in \cite{GWalcher},
where explicit answers were tabulated for all prime knots with up to 7 crossings.\footnote{It is not difficult to
extend these calculations to larger knots,
see {\it e.g.} \cite{GS} for the calculation of the Kauffman homology for the $(3,4)$ torus knot $T^{3,4} = {\bf 8}_{19}$.}
In particular, for the $(2,5)$ torus knot $T^{2,5}$ the homology $\cH^{so(3), V}_{i,j} (K)$ is also 25-dimensional,
and by matching the results of \cite{GWalcher} with the specialization $\P^{S^2} (T^{2,5}; a=q^2,q,t)$
one can restore the $t$-grading of every term in \eqref{P25S2}:
\begin{eqnarray}
\P^{S^2} (T^{2,5}; a,q,t)
&=& a^6 \left( q t^6 + q^4 t^8 + q^5 t^8 + q^7 t^{10} \right) \label{superP25S2} \\
&+& a^5 \left( q^{-2} t^3 + q^{-1} t^3 + q t^5 + 2 q^2 t^5 + q^3 t^5 + q^4 t^7 + 2 q^5 t^7 + q^6 t^7 + q^7 t^9 + q^8 t^9 \right) \nonumber \\
&+& a^4 \left( q^{-4} + q^{-1} t^2 + t^2 + q^2 t^4 + q^3 t^4 + q^4 t^4 + q^5 t^6 + q^6 t^6 + q^8 t^8 \right) \, , \nonumber
\end{eqnarray}
where we tacitly assumed that the entire $S^2$-colored homology $\cH^{S^2}_{i,j,k} (T^{2,5})$
is indeed 25-dimensional, so that $Q^{sl(2),S^2} = 0$ in \eqref{superright} for the present example.
This can be easily justified by looking at the other differentials in \eqref{specN}
or the corresponding Poincar\'e polynomials \eqref{superright}--\eqref{PRright}.

Everything we saw in this simple example can be easily generalized
to other knots (in fact, not just torus knots) and other representations.
As the knot $K$ is getting bigger, the size of the colored HOMFLY homology $\cH^{R}_{i,j,k} (K)$ typically grows as well.
At the same time, for larger knots and larger homology
more differentials act on $\cH^{R}_{i,j,k} (K)$ in an interesting way, thus providing non-trivial constraints.
Among the infinite set of differentials in \eqref{gradingtabl} there are some special ones,
which always act non-trivially, no matter how large or small the knot $K$ is.
These are the so-called {\it canceling} differentials which get their name after the fact that,
when acting on $\cH^{R}_{i,j,k} (K)$, they cancel almost all of the terms, except a single one, {\it i.e.}
\be
\dim \left( \cH^R (K) , d_{\text{canceling}} \right) \; = \; 1 \,.
\label{specNcanc}
\ee
At first, even the very existence of such differentials might seem very surprising.
However, they all usually have a simple origin and interpretation.

Let us consider, for example, the $S^r$-colored HOMFLY homology $\cH^{S^r}_{i,j,k} (K)$
relevant to the present paper.
Then, as we already pointed out in the discussion below \eqref{superwrong},
the $sl(1)$ homology $\cH^{sl(1), S^r}_{i,j} (K)$ should be trivial ({\it i.e.} one-dimensional) for every knot $K$.
When combined with \eqref{specN}, this basic property implies that the differential $d_1$
must be canceling for the $S^r$-colored HOMFLY homology $\cH^{S^r}_{i,j,k} (K)$.

Another canceling differential in the same theory is $d_{-r}$, whose origin is very similar:
the representation $R = \Lambda^r$ of $sl(N)$ is trivial when $N=r$.
As a result, $\cH^{sl(r), \Lambda^r}_{i,j} (K)$ should also be one-dimensional for
every knot $K$, and from \eqref{gradchoicea} it follows that $d_{-r}$ is also
a canceling differential for $\cH^{S^r}_{i,j,k} (K)$.
To summarize, in our standard conventions
the canceling differentials in the $S^r$-colored HOMFLY theory have degree
\be
\deg (d_1) = (-1,1,-1) \qquad \text{and} \qquad \deg (d_{-r}) = (-1,-r,-3) \,.
\label{dcancgrading}
\ee
These canceling differentials pair up almost all of the terms in the Poincar\'e polynomial
of the homology $\cH^{S^r}_{i,j,k} (K)$ which, therefore, has the structure \eqref{superright}:
\begin{eqnarray}
\P^{S^r} (K;a,q,t) &=& a^{rs} q^{-rs} t^{0} + (1+a^{-1}qt^{-1}) Q_1 (K;a,q,t) \nonumber \\
&=&a^{rs}q^{r^2s}t^{2rs}+(1+a^{-1}q^{-r}t^{-3}) Q_{-r} (K;a,q,t) \,,
\label{super_monomial1}
\end{eqnarray}
with a monomial ``remainder'' $R (a,q,t) = a^i q^j t^k$ and $(\alpha, \beta, \gamma) = (-1,1,-1)$
or $(-1,-r,-3)$ for $d_1$ or $d_{-r}$, respectively.
Since \eqref{super_monomial1} is obtained by taking the Poincar\'e polynomial of both sides
in \eqref{specNcanc}, the polynomials $Q_1$ and $Q_{-r}$ have non-negative integer coefficients.
Furthermore, the $(a,q,t)$-degrees of the ``remainder''
are determined by a single integer $s(K)$, the so-called $S$-invariant of the knot $K$ (see \cite{GS} for details).

The structure \eqref{super_monomial1} is a ``colored generalization'' of the familiar
property of the ordinary HOMFLY homology \cite{DGR} that corresponds to $r=1$
and comes equipped with two canceling differentials $d_1$ and $d_{-1}$.
In that context, the $S$-invariant is expected to provide a lower bound on
the slice\footnote{The slice genus of a knot $K$ in $\S^3$ (sometimes also known as the Murasugi genus or four-ball genus)
is the least integer $g_*$, such that $K$ is the boundary of a connected, orientable surface of genus $g_*$
embedded in the 4-ball ${\bf B}^4$ bounded by $\S^3$. The slice genus is also a lower bound for the unknotting number
of the knot $K$, {\it i.e.} the least number of times that the string must be allowed to pass through itself
in order to ``untie'' the knot.} genus $g_* (K)$ of the knot, which in our normalization reads
\be
|s (K)| \; \le \; g_* (K)
\ee
and is often tight.
The slice genus of the $(p,q)$ torus knot is \cite{MilnorSing,Rasmussen}:
$$
g_* (T^{p,q}) \; = \; \frac{(p-1)(q-1)}{2}
\qquad\qquad
\text{(The Milnor Conjecture)}
$$
so that for $(2,2p+1)$ torus knots we expect
\be
s (T^{2,2p+1}) \; = \; p \,.
\label{2forptorknot}
\ee

Substituting \eqref{2forptorknot} into \eqref{super_monomial1}, it is easy to verify that
our result \eqref{superP25S2} indeed has the expected structure, thereby, illustrating how a combination
of the structural properties that follow from action of differentials can determine $\cH^{R}_{i,j,k} (K)$.
In practice, starting with the colored HOMFLY polynomial $P^R (K;a,q)$, usually one needs only a few of
the relations like \eqref{specN} in order to find its homological lift ({\it a.k.a.} ``categorification'').
Then, the rest of the relations can be used as consistency checks.
For simple knots (with less than 10 crossings or so) this typically gives a largely overconstrained system,
that miraculously has a solution.

Just like it is easier to differentiate a function $f(x)$ rather than to integrate it,
it is much easier to use the structure based on the commuting differentials \eqref{gradingtabl}
to verify the correctness of a particular result than to derive it.
Thus, even though in principle we could extend the derivation of \eqref{superP25S2}
to more general torus knots and higher dimensional representation,
verifying the correctness of our results \eqref{Superpolynomial_sym_refined} and \eqref{Superpolynomial_anti-sym_refined}
is much easier.
Indeed, specializing to $a = - q t^{-1}$ and $a = - q^{-r} t^{-3}$ in \eqref{Superpolynomial_sym_refined}
we can verify \eqref{super_monomial1} in no time!

Similarly, we can run a test on the $\Lambda^r$-colored superpolynomials \eqref{Superpolynomial_anti-sym_refined}
of the $(2,2p+1)$ torus knots. In view of the mirror symmetry \eqref{mirror}, this is not really an independent test,
but it is still instructive to see how it works. (In particular, it helps to understand the role of mirror symmetry.)
As we already discussed earlier, passing from $R = S^r$ to $R^t = \Lambda^r$ can be achieved by changing the sign of $N$.
Therefore, if $d_1$ and $d_{-r}$ are canceling differentials in $\cH^{S^r}_{i,j,k} (K)$,
then $d_{-1}$ and $d_r$ must be canceling differentials in the $\Lambda^r$-colored HOMFLY homology $\cH^{\Lambda^r}_{i,j,k} (K)$.
According to \eqref{gradingtabl}, the degrees of these differentials are
\be
\deg (d_{-1}) = (-1,-1,-3) \qquad \text{and} \qquad \deg (d_{r}) = (-1,r,-1) \,,
\ee
so that \eqref{specNcanc} implies the following structure of the $\Lambda^r$-colored superpolynomial,
{\it cf.} \eqref{superright} and \eqref{super_monomial1}:
\begin{eqnarray}
\P^{\Lambda^r} (K; a,q,t) &=& a^{rs} q^{rs} t^{2rs} + (1+a^{-1}q^{-1}t^{-3}) Q_{-1} (K;a,q,t)
\nonumber \\
&=& a^{rs} q^{-r^2s} t^0 + (1+a^{-1}q^{r}t^{-1}) Q_{r} (K;a,q,t) \,.
\label{super_monomial2}
\end{eqnarray}
It is easy to verify that all our results indeed have this remarkable structure.
Indeed, using \eqref{2forptorknot} for $K = T^{2,2p+1}$ and specializing to $a=-q^{-1}t^{-3}$ and $a=-q^{r}t^{-1}$
in (\ref{Superpolynomial_anti-sym_refined}) we find that, respectively, only $\ell=r$ and $\ell=0$ contributions survive.\\

For completeness, let us describe how the same structure looks in the ``DGR conventions''
\eqref{Symmetric_DGR_refined} and \eqref{Anti-Symmetric_DGR_refined}.
The grading of the canceling differentials $d_1$ and $d_{-r}$ for $R = S^r$
can be read off directly from \eqref{gradingtabl}:
\be
\deg_{\text{DGR}} \, (d_1) = (-2,2,-1) \qquad \text{and} \qquad \deg_{\text{DGR}} \, (d_{-r}) = (-2,-2r,-2r-1) \,,
\ee
and, similarly, the gradings of the canceling differentials $d_{-1}$ and $d_{r}$ for $R = \Lambda^r$ are
\be
\deg_{\text{DGR}} \, (d_{-1}) = (-2,-2,-3) \qquad \text{and} \qquad \deg_{\text{DGR}} \, (d_{r}) = (-2,2r,-1) \,.
\ee
Substituting these values of $(a,q,t)$-degrees $(\alpha, \beta, \gamma)$ in the general formula \eqref{superright},
we arrive to the following structure of the colored superpolynomials in the grading conventions of \cite{DGR}:
\begin{eqnarray}
\P^{S^r}_{\text{DGR}} (K; a,q,t)
&=&a^{2rs}q^{-2rs}t^{0} + (1+a^{-2}q^{-2}t^{-1})Q_1^{\text{DGR}}(K;a,q,t)
\nonumber \\
&=&a^{2rs}q^{2r^2 s} t^{2r^2 s} + (1+a^{-2}q^{-2r}t^{-2r-1})Q_{-r}^{\text{DGR}}(K;a,q,t) \,,
\\
\P^{\Lambda^r}_{\text{DGR}} (K; a,q,t)
&=&a^{2rs} q^{2rs} t^{2rs} + (1+a^{-2}q^{-2}t^{-3})Q_{-1}^{\text{DGR}}(K;a,q,t)
\nonumber \\
&=&a^{2rs} q^{-2r^2 s} t^0 + (1+a^{-2}q^{2r}t^{-1})Q_{r}^{\text{DGR}}(K;a,q,t) \,.
\end{eqnarray}
Using $s=p$ for $K=T^{2,2p+1}$, it easy to verify that our results
\eqref{Symmetric_DGR_refined} and \eqref{Anti-Symmetric_DGR_refined}
exhibit this structure, ensuring the validity of some of the steps in section \ref{sec:refinedbraid}.\\

While the above discussion hopefully makes it fairly convincing
how powerful and elegant the structure of differentials is, it is just the tip of an iceberg!
Indeed, besides the canceling differentials that we discussed in detail and that, in many cases,
alone suffice for deducing the superpolynomials from the colored HOMFLY polynomials,
there is yet another class of ``universal'' differentials found in \cite{GS}.
These differentials are called ``colored'' because they relate homology theories
associated with different representations.
The simplest example of a colored differential discussed in \cite{GS}
is a differential that, when acting on the $S^2$-colored HOMFLY homology,
leaves behind the ordinary, $S$-colored HOMFLY homology, modulo a simple re-grading:
\be
\left( \cH^{S^2} (K) , d_{\text{colored}} \right)_{2i,k+2j,2k} \; \cong \; \cH^{\Box}_{i,j,k} (K) \,.
\ee
As in our discussion of \eqref{specNcanc} and \eqref{super_monomial1},
we can take Poincar\'e polynomials of both sides to learn the following structure
of the $S^2$-colored superpolynomial,
\be
\P^{S^2} (K; a,q,t) \; = \; \P (K; a^2,q^2, q t^2) + (1 + q) Q (a,q,t) \,,
\label{coloredPa}
\ee
which corresponds to the first colored differential on our list \eqref{gradingtabl}.
Similarly, the second colored differential in \eqref{gradingtabl} acts a little differently
and leads to a similar, but different relation:
\be
\P^{S^2} (K; a,q,t) \; = \; a^{s} \P (K; a,q^2, t) + (1 + a^{-1} t^{-1}) Q (a,q,t) \,.
\label{coloredPb}
\ee
Both superpolynomials $\P^{S^2} (K; a,q,t)$ and $\P (K; a,q,t)$ that participate
in these relations can be easily determined from the corresponding HOMFLY polynomials
by using the structure of either $d_2$ or canceling differentials.
Indeed, as a yet another illustration of this method,
let us explain how it works for the torus knots we are interested in.

In the case of torus knots, the starting point of this construction -- namely, the colored HOMFLY polynomial --
is available {\it e.g.} from \cite{LinZheng}.
In fact, for $R=S$, the explicit expression for the (uncolored) HOMFLY polynomial
of an arbitrary torus knot was written already by Jones \cite{Jones}.
In our conventions, the answer for $(2,2p+1)$ torus knots reads
\be
P (T^{2,2p+1}; a,q) \; = \; - a^{p+1} \sum_{i=1}^p q^{2i - p - 1} + a^{p} \sum_{i=0}^p q^{2i - p} \,.
\label{T22p1HOMFLY}
\ee
The corresponding (uncolored) superpolynomials of torus knots have the same structure \cite{DGR}.
In fact, for $T^{2,2p+1}$ torus knots, all of the terms in the superpolynomial $\P (T^{2,2p+1}; a,q,t)$
are ``visible'' in the HOMFLY polynomial \eqref{T22p1HOMFLY} and, therefore, can be determined from
the structure \eqref{super_monomial1} associated with two canceling differentials $d_1$ and $d_{-1}$.
In our grading conventions \eqref{dcancgrading}, these canceling differentials for $r=1$
have degree $(-1,1,-1)$ and $(-1,-1,-3)$, respectively.
Hence, restoring the powers of $t$ in \eqref{T22p1HOMFLY} consistent with $d_1$ and $d_{-1}$ we get
\begin{eqnarray}
\P (T^{2,2p+1}; a,q,t) & = & a^{p+1} \sum_{i=1}^p q^{2i - p - 1} t^{2i+1} + a^{p} \sum_{i=0}^p q^{2i - p} t^{2i} \nonumber \\
& = & \; a^{p} q^{p} t^{2p} \, \frac{a q^{-1} t (1 - q^{-2p} t^{-2p}) + 1 - q^{-2p-2} t^{-2p-2} }{1 - q^{-2} t^{-2}} \,. \nonumber
\end{eqnarray}
Similarly, starting with the $S^2$-colored HOMFLY polynomial (see {\it e.g.} \cite{LinZheng}),
one can derive the $S^2$-colored superpolynomial for all $(2,2p+1)$ torus knots,
\begin{eqnarray}
\P^{S^2} (T^{2,2p+1}; a,q,t)
& = & \frac{q^4 t^4 (a q^2 t^2)^{2 p} (1 + a t) (q + a t)}{1 - q^2 (1 + q) t^2 + q^5 t^4} \label{PS2p} \\
& - & \frac{q^2 t^2 (1 + q) (a q t)^{2 p} (q + a t) (1 + a q^2 t^3)}{1 - q^2 (1 + q^2) t^2 + q^6 t^4}
+ \frac{(a q^{-1})^{2 p} (1 + a q^2 t^3) (1 + a q^3 t^3)}{1 - q^3 (1 + q) t^2 + q^7 t^4} \nonumber
\end{eqnarray}
which, of course, is consistent with our earlier result \eqref{Superpolynomial_sym_refined}.
Substituting these colored superpolynomials into relations \eqref{coloredPa} and \eqref{coloredPb},
it is easy to verify that they work like a charm.\\

Finally, after an exciting discussion of colored and canceling differentials that have a universal nature,
let us consider a somewhat more rudimentary differential $d_2$ that, according to \eqref{specN},
controls specialization to the $sl(2)$ knot homology, see also Figure \ref{fig:superpolunmial_hier}.
This differential is actually very important for the subject of our paper since it accompanies
the specialization to $a=q^2$ and, hence, the definition of the invariant $P_n (q,t)$
that appears in the refined volume conjectures \eqref{VCquantref} and \eqref{VCparamref}.
This invariant is defined as a specialization \eqref{Pq2special}:
\be
P_{n=r+1}(K;q,t) \; := \; \P^{S^r}(K;a=q^2,q,t)
\label{Pq2special123}
\ee
and, contrary to \eqref{superwrong},
for $(2,2p+1)$ torus knots appears to coincide with the Poincar\'e polynomial
of the $n$-colored $sl(2)$ knot homology $\cH^{sl(2),V_n} (K)$.
Indeed, as one can see directly from \eqref{PS2p}, for the trefoil knot $T^{2,3} = {\bf 3}_1$
the $S^2$-colored HOMFLY homology simply contains no terms that can be killed by $d_2$.
For the next torus knot $T^{2,5} = {\bf 5}_1$, there are such terms, but according to our
discussion around \eqref{superP25S2}, none of them are canceled by $d_2$,
{\it i.e.} $Q^{sl(2),S^2} = 0$ in \eqref{superright}, {\it etc.}

The fact that $P_{n}(T^{2,2p+1};q,t)$ defined in \eqref{Pq2special123}
appears to coincide with the Poincar\'e polynomial
$\P^{sl(2),V_n} (T^{2,2p+1}; q,t)$ of the $n$-colored $sl(2)$ knot homology has an important implication:
it suggests that the homological volume conjectures \eqref{VCquantref} and \eqref{VCparamref}
can be formulated directly in terms of the $sl(2)$ knot homology, as in \eqref{Pnoptimistic},
rather than in terms of the specialization of the colored superpolynomial \eqref{Pq2special}.
For example, such homological version of the generalized volume conjecture \eqref{VCparamref} has a nice form
\be
\sum_{i,j} q^i t^j \dim \cH^{sl(2),V_n}_{i,j} (K)
\; \simeq \;
\exp\left( \frac{1}{\hbar} S_0 (u,t) \,+\,\sum_{k=0}^\infty S_{k+1} (u,t) \, \hbar^{k} \right)
\ee
and describes the asymptotic growth of the dimensions of the $n$-colored $sl(2)$ homology groups in the limit \eqref{reflimit}:
\be
q = e^{\hbar} \to 1 \,, \qquad t = \text{fixed} \,, \qquad x \equiv e^u = q^n = \text{fixed} \,.
\ee
Similarly, the homological version of the quantum volume conjecture \eqref{VCquantref} presumably
can be formulated in the form of an exact sequence
\be
0~~
\longrightarrow~~ C^n~~
\longrightarrow^{\kern -13pt \frak{a}^{n}} ~~ C^{n+1}~~
\longrightarrow^{\kern -20pt \frak{a}^{n+1}}~~ C^{n+2}~~
\longrightarrow^{\kern -20pt \frak{a}^{n+2}}~~ \ldots ~~
%\longrightarrow~~
C^{n+d}~~ \longrightarrow~~ 0
\label{VCsequence}
\ee
where
\be
C^n (K) = \bigoplus_{i,j \in \Z} \cH^{sl(2),V_n}_{i,j} (K)
\ee
and the maps $\frak{a}^j$ are determined by the coefficients of the quantum operator
\be
\hat A^{\text{ref}} (\hat x, \hat y; q, t) \; = \; \sum_{j=0}^d  \frak{a}^j (\hat x; q, t) \, \hat y^j \,.
\ee
Although we believe that both volume conjectures \eqref{VCquantref} and \eqref{VCparamref}
work equally well for the $sl(2)$ homological knot invariants \eqref{Pnoptimistic}
as well as for the specialization of the colored superpolynomial \eqref{Pq2special},
we leave this question to a future work.\footnote{All examples considered in this paper suggest that this may be the case.
However, a proper understanding of this issue requires a closer look at how the differential $d_2$ acts on the colored HOMFLY homology.}

In what follows we simply adopt \eqref{Pq2special} as a definition of $P_n (q,t)$.
Then, from \eqref{Superpolynomial_sym_refined} we have
$$
P_{n}(T^{2,2p+1};q,t) =
 \sum_{\ell=0}^{n-1}(-1)^{n-1}
\left[(-1)^{\ell}q^{\frac{(n-1)(n+1)-\ell(\ell+1)}{2}} t^{\frac{3}{2}(n-1)-\ell}\right]^{2p+1}
$$
\be
\times
q^{\frac{n-1}{2}-\ell}t^{\frac{1}{2}(1-2p)(n-1)-\ell}
\frac{(q;q)_{n-1}(t^2q;q)_{\ell}(t^2q^2;q)_{2\ell}(-t^3 q^2;q)_{n+\ell-1}(-tq;q)_{n-\ell-1}}
{(-t^3q^2;q)_{n-1}(q;q)_{\ell}(t^2q;q)_{2\ell}(t^2q^2;q)_{n+\ell-1}(q;q)_{n-\ell-1}} \,.
\label{refined_SU(2)}
\ee
%where for later conveninece we also modified the overall normalization by introducing a factor $(-t)^{p(1-n)}$.
Our next goal is to use this result to test the refined / categorified volume conjecture \eqref{VCquantref}.

%*******************************************************
%*******************************************************

\subsection{Recursion relations for homological knot invariants}
\label{sec:knotrecursions}

In this section we find recursion relations which are satisfied by homological knot invariants,
therefore, providing concrete examples for one of the new volume conjectures proposed in section~2.

In other words, we find \emph{refined} quantum $A$-polynomials $\widehat{A}^{\textrm{ref}}(\hat x, \hat y;q,t)$.
These are the objects which generalize the \emph{unrefined} quantum curves considered {\it e.g.} in \cite{Apol,DHS,DGLZ,Tudor,abmodel}.
Even though we consider simple examples of knots -- the unknot and the trefoil -- the fact
that such refined relations exist and can be explicitly written down is already nontrivial.
Having found these examples, we have no doubt that their generalization to the entire family of torus knots, and even more general knots, exists. Moreover, an important hint about the general structure of refined quantum curves for $(2,2p+1)$ torus knots is given in the next section, where the analysis of the asymptotics of their colored superpolynomials reveals the form of the refined classical curves $A^{\textrm{ref}}(x,y;t)=0$.
To find the full quantum curves in this class of examples, one should ``just'' reintroduce the dependence on $q$.

We will derive refined quantum curves from the analysis of the homological knot invariants found in section \ref{sec:knots}.
It would also be interesting to find general methods of deriving refined recursion relations, similar to \cite{Tudor,abmodel}
in the unrefined case, which \emph{a priori} do not rely on the knowledge of homological invariants.
We plan to address this problem in the follow-up work.

\subsubsection{Unknot}

As we already mentioned in \eqref{unknotMqtR} and \eqref{unknotsuperSr},
in every approach to knot homologies based on the refined BPS invariants \cite{GIKV,AS}
the colored superpolynomial of the unknot is given essentially by the Macdonald polynomial \cite{IK11}.
In particular, after the change of variables \eqref{GSvarchng}, the $S^r$-colored superpolynomial reads:
\be
\bar{\P}^{S^r} (\unknot; a,q,t) \; = \;
(-1)^{\frac{r}{2}}a^{-\frac{r}{2}}q^{\frac{r}{2}}t^{-\frac{3r}{2}} \frac{(-at^3;q)_r}{(q;q)_r} \,.
\ee
Note that when we talk about the unknot, only the unreduced superpolynomial (resp. homology) is non-trivial;
the reduced one is trivial by definition (which involves normalizing by the unknot).
Specializing further to $a=q^2$ we find
\be
P_{n=r+1} (\unknot; q,t) \; = \;
(-1)^{\frac{r}{2}} q^{-\frac{r}{2}} t^{-\frac{3r}{2}} \frac{(-q^2 t^3;q)_r}{(q;q)_r} \,.
\label{Pnunkot}
\ee
Note, at $t=-1$ this expression reduces to the $n$-colored Jones polynomial of the unknot
\be
J_{n} (\unknot; q) \; = \;
\frac{q^{\frac{n}{2}} - q^{-\frac{n}{2}}}{q^{\frac{1}{2}} - q^{-\frac{1}{2}}}
\ee
which can be written as the partition function of the $SU(2)$ Chern-Simons theory
on a solid torus $\S^1 \times {\bf D}^2 \cong \S^3 \setminus \text{unknot}$,
\be
Z^{\text{CS}}_{SU(2)} (\S^1 \times {\bf D}^2; q) \;=\;
\sqrt{\frac{- \hbar}{2 \pi i}} \, \big( x - x^{-1} \big)
\ee
normalized by the partition function of the Chern-Simons theory on $\S^3$,
\be
Z^{\text{CS}}_{SU(2)} (\S^3; q) \;=\;
\sqrt{\frac{- \hbar}{2 \pi i}} \, \Big( e^{\hbar} - e^{- \hbar} \Big)
\ee
where we used $x=e^u$ and $u = n \hbar$, {\it cf.} \eqref{unknotdimqR}.

As the homological knot invariant \eqref{Pnunkot} has a product structure, we can immediately write down the recursion relation it satisfies
\be
P_{n+1} (\unknot; q,t) \; = \; \frac{1 + t^3 q^{n+1}}{1 - q^{n}} (-q^{-1} t^{-3})^{1/2} \, P_n (\unknot; q,t).
\ee
This means  that $P_{n} (\unknot; q,t)$ obeys the refined version (\ref{VCquantref}a) of the quantum volume conjecture with
\be
\hat A^{\text{ref}}_{\unknot}(\hat{x},\hat{y};q,t) \; = \;
(1 + t^3 q \hat{x}) ( - q^{-1} t^{-3})^{1/2} - (1 - \hat x) \hat y \,.   \label{qAt-unknot}
\ee
In the classical limit $q \to 1$ this recursion relation
reduces to the refined classical curve $A^{\text{ref}} (x,y;t) = 0$ defined by
\be
A^{\text{ref}}_{\unknot} (x,y;q=1,t) \; = \;
(1 + t^3 x) ( - t^{-3})^{1/2} - (1 - x) y \,.   \label{At-unknot}
\ee
On the other hand, in the unrefined limit $t=-1$ the relation (\ref{qAt-unknot}) takes the form
\be
\hat A^{\text{ref}}_{\unknot}(\hat x, \hat y ;q) \; = \;
q^{-1/2}(1 - q \hat x)  - (1 - \hat x) \hat y \,,
\ee
and specializing further to $q=1$ we get the classical $A$-polynomial
\be
A_{\unknot} (x,y) \; = \; (1-x)(1-y) \,.   \label{Aunknot1x1y}
\ee

It is also interesting to consider the second order equation satisfied by (\ref{Pnunkot}).
Writing the three consecutive colored polynomials it is not hard to see that the following equation is satisfied
\be
P_{n+1} - (-q t^3)^{1/2} \frac{1 - q t^3 + 2 x t^3 - x^2 t^3 + 2 q x t^3 + q x^2 t^6}{q (x - 1) t^3 (1 + x t^3)}  P_n + P_{n-1} =0 ,  \label{Punknot-recursion2}
\ee
with $x=q^n$. Interestingly, in the unrefined limit $t = -1$ the dependence on $x$ cancels and $J_n (\unknot; q)$ satisfies the recursion relation
\be
J_{n+1} - [2]_q J_n + J_{n-1} \; = \; 0
\ee
with $[2]_q = q^{1/2} + q^{-1/2}$. This means that in the quantum volume conjecture \eqref{VCquant} for the unknot we have
\be
\widehat{A}_{\unknot}(\hat{x},\hat{y};q) \; = \; \hat y  - (q^{1/2} + q^{-1/2}) + \hat y^{-1}.
\ee
In the classical limit $q \to 1$ this becomes equivalent to
\be
A_{\unknot} (x,y) \; = \; (y-1)^2 \,.   \label{Aunknotclassical}
\ee
On the other hand, we can also write the classical limit of (\ref{Punknot-recursion2}) in the polynomial form as
\be
A^{\textrm{ref}}_{\unknot}(x,y;t) = (1 - x) (1 + t^3 x) y^2 -
 (-t^{-3})^{1/2} \big(1 + t^6 x^2 - t^3 (1 - 4 x + x^2)\big) y + (1- x) (1 + t^3 x).
\ee
Starting from this form the unrefined limit reads $A_{\unknot}(x,y)=(x-1)^2 (y-1)^2$, which captures the cases (\ref{Aunknot1x1y}) and (\ref{Aunknotclassical}) which we considered above.

%*******************************************************
%*******************************************************

\subsubsection{Trefoil}

Our next task is to derive refined recursion relations for the trefoil.
Similarly as in the unknot case, we will be able to determine these relations from
the structure of the $sl(2)$ specialization $P_{n}(T^{2,3};q,t)$ of the colored superpolynomial.
For general $(2,2p+1)$ torus knots, we found an explicit expression for
this homological invariant in (\ref{refined_SU(2)}), which for the present purpose
of deriving recursion relations we write in the form\footnote{The equivalence
between this expression and (\ref{refined_SU(2)}) can be easily verified to sufficiently large $n$.
Plus, both expressions enjoy the structural properties discussed in \ref{sec:homological}.}
%which is conjectured in (\ref{Pn-torus-another}). Actually, for the trefoil one can check that the form (\ref{Pn-torus-another}) with $p=1$ is completely fixed by the structure of differentials presented in the previous section. For $p=1$ we therefore consider
%\begin{eqnarray}
%P_{n}({\bf 3_1};q,t)=\sum_{k=0}^{n-1}\sum_{j=0}^{k}q^{n-1+nk+\frac{j(j+1)}{2}}t%^{2k+j}
%\frac{[n-1]^{\prime}!}
%{[k-j]^{\prime}![j]^{\prime}![n-1-k]^{\prime}!
%},
%\end{eqnarray}
%where
%$$
%[n]'! = [1]' [2]' \cdots [n-1]' [n]',\qquad\qquad [n]' = \frac{q^n - 1}{q-1} = %1 + q + \ldots + q^{n-1}.
%$$
%Moreover, as we prove in the appendix \ref{sec:proof_factor}, the double summation in the above expression can be reduced to a single summation
%\begin{eqnarray}
%P_{n}({\bf 3_1};q,t)
%=\sum_{k=0}^rq^{n-1+nk}t^{2k}\prod_{i=1}^{k}\frac{(1-q^{n-i})(1+q^{i}t)}{1-q^i}.
%\end{eqnarray}
%Let us finally write this expression in the form
\be
P_{n}(T^{2,3}; q,t) \; = \; \sum_{k=0}^{n-1} P(n,k) \,,    \label{Pn-single}
\ee
where each $P(n,k)$ has a product structure
\be
P(n,k) = q^{n-1+n k} t^{2k} \prod_{i=1}^k \frac{(1 - q^{n-i})(1 + q^i t)}{1 - q^i}.  \label{Pnk}
\ee
In particular
\be
P(n,0) = q^{n-1}, \qquad P(n,n) = 0.
\ee
Note that for $t=-1$ simplifications occur and only one set of products remains in (\ref{Pnk}).

To derive recursion relations let us write down the following ratios, which is immediate due to the product structure of $P(n,k)$:
\bea
\frac{P(n+1,k)}{P(n,k)} & = & q^{k+1} \frac{1-q^n}{1-q^{n-k}}, \label{ratioA}\\
\frac{P(n,k+1)}{P(n,k)} & = & q^{n} t^2 \frac{(1-q^{n-k-1})(1+q^{k+1}t)}{1-q^{k+1}}, \label{ratioB}  \\
\frac{P(n+1,k+1)}{P(n,k)} & = & q^{n+k+2} t^2 \frac{(1-q^n)(1+q^{k+1}t)}{1-q^{k+1}}.
  \label{ratioC}  \eea
These equations are equivalent to the following ones (obtained by clearing the denominators):
\begin{eqnarray}
q^k P(n+1,k) - q^n P(n+1,k) & = & (1-q^n) q^{2k+1} P(n,k),   \label{recA} \\
q^k P(n,k+1) - q^{n+k} t^2 (1 - q^n t) P(n,k) &  & \nonumber \\
= -q^{2n-1} t^2 P(n,k) &  + & q^{2k+1}\Big(P(n,k+1) + t^3 q^n P(n,k)  \Big), \label{recB} \\
P(n+1,k+1) - t^3 q^{n+2k+3} (1 - q^n) P(n,k) &  &  \nonumber \\
= q^{k+1} \Big(P(n+1,k+1) & + & t^2 q^{n+1}(1-q^n) P(n,k)  \Big).    \label{recC}
\end{eqnarray}
These are linear equations in $P(n,k)$ (possibly with shifted arguments
$n$ or $k$). Therefore, ideally, we would like to perform the sum over
$k$ as in (\ref{Pn-single}) to transform them into equations for various
$P_n$ (possibly with shifted $n$'s). However such summation cannot be
directly performed because of $k$-dependent factors $q^k$ and
$q^{2k}$. Nonetheless, we can take these factors into account at the
expense of introducing auxiliary quantities:
\be
R_n = \sum_{k=0}^{n-1} q^k P(n,k),   \quad \qquad S_n  = \sum_{k=0}^{n-1} q^{2k} P(n,k).    \label{RnSn}
\ee
Now resummation of (\ref{recA})-(\ref{recC}) can be performed and the answer written in terms of $P_n$, $R_n$, and $S_n$. Note that because of shifts in $n$ we need to take care of some boundary terms arising in various cases for $k=0$ or $k=n$. However all such boundary terms ultimately cancel and we get the following system of equations:
\bea
R_{n+1} + a P_{n+1} & = & b S_n,   \label{recAbis} \\
c R_n + d P_n & = & e S_n, \label{recBbis} \\
P_{n+1} - R_{n+1} + f R_n & = & g S_n,   \label{recCbis}
\eea
where
\be
a = -q^n, \quad b = q(1-q^n), \quad c = q^{-1} - t^2q^n(1-q^n t),\quad d = t^2 q^{2n-1},
\ee
$$
e = q^{-1} + t^3 q^{n+1},\quad f = -t^2 q^{n+2} (1-q^n), \quad g = t^3 q^{n+3} (1-q^n).
$$
Now we can determine $S_n$ from the first equation and substitute to the remaining two equations.
This gives a system of two equations, which allows to determine $R_n$ and $R_{n+1}$ in terms of $P_n$ and $P_{n+1}$:
\bea
R_n & = & \frac{-d(b+g) P_n + (1 + a) e P_{n+1}}{b c - e f + c g}, \\
R_{n+1}  & = & \frac{-b d f P_ n + (b c  + a e f - a c g) P_{n+1}}{b c - e f + c g}.
\eea
Finally we notice that $R_n$ is related to $R_{n+1}$ simply by a shift of $n$ by one unit.
Therefore, shifting the second equation above and comparing with the first one we get a homogeneous,
3-term recursion relation
\be
\boxed{\phantom{\int^1}
\alpha P_{n-1} + \beta P_n + \gamma P_{n+1} \; = \; 0 \,,
\phantom{\int^1}}
\label{RefineRecursion}
\ee
where
\bea
\alpha & = & \frac{q^{3 n-1} (q^n - q) t^4}{(q + q^{2 n} t^3) (1 +
   q^{n+1} t^3)}, \label{A}  \\
\beta & = & -\frac{t^2}{q^{-2 n} + q t^3} -
\frac{ q - q^{n+1} t^2 + q^{4 n} t^6 +
  q^{2 n} t^2 (1 + t + q t)}{(q + q^{2 n} t^3) (1 + q^{n+1} t^3)},  \label{B}  \\
\gamma & = & \frac{1}{q + q^{2 n+2} t^3} .   \label{C}
\eea
We can also rewrite the above relation in the operator form (\ref{VCquantref}a):
\be
\hat A^{\text{ref}} (\hat x, \hat y; q,t) P_n (T^{2,3};q,t) \; = \; 0 \,.
\ee
%acting on $P_{n}(x)\equiv P_{n}(x=q^n)$.
Here, as in \eqref{xyactionP}, $\hat{y}$ acts as a shift operator $P_n \to P_{n+1}$
and $\hat{x}$ acts by multiplication by $q^n$, so that $\hat x$ and $\hat y$ obey the commutation relation $\hat{y} \hat{x} = q \hat{x}\hat{y}$.
Then, the relation (\ref{RefineRecursion}) can be expressed in terms of the refined quantum $A$-polynomial
\be
\widehat{A}^{\text{ref}}(\hat{x},\hat{y};q,t) \; =  \; \alpha \hat{y}^{-1} + \beta + \gamma\hat{y} \,,   \label{trefoil-Aqt}
\ee
with
\bea
\alpha & = & \frac{x^3 (x - q) t^4}{q (q + x^2 t^3) (1 +
   x q t^3)} \,,   \\
\beta & = & -\frac{t^2 x^2}{1 + x^2 q t^3} -
\frac{ q - x q t^2 + x^{4} t^6 +
  x^{2} t^2 (1 + t + q t)}{(q + x^{2} t^3) (1 + x q t^3)} \,,    \\
\gamma & = & \frac{1}{q + x^2 q^{2} t^3} \,.
\eea
In what follows we analyse various limits of this relation.
In particular, we study the classical limit $q=1$ and the associated asymptotic structure of the $sl_2$ colored polynomial,
and generalize these results to other $(2,2p+1)$ torus knots.
The classical refined $A$-polynomials which we find for other values of $p$
provide an important guidance in generalizing the full quantum relations (\ref{trefoil-Aqt}) to other torus knots.

%*******************************************************************
%*******************************************************************

\subsubsection{Recursion in various limits}

In order to understand better the refined recursion relation (\ref{RefineRecursion}),
let us consider what happens in various special limits, when $t=-1$ or $q=\pm 1$.
We begin with the unrefined limit $t=-1$, which is special for several reasons.
First, a direct substitution $\boxed{t=-1}$ in (\ref{RefineRecursion}) gives rise to the following 3-term homogeneous relation
\be
\alpha_{(t=-1)} J_{n-1} \, + \, \beta_{(t=-1)} J_n \, + \, \gamma_{(t=-1)} J_{n+1} \; = \; 0 \,,    \label{RefineRecursiont1}
\ee
with the coefficients (rescaled by $q^2-q^{1-n}$ compared to (\ref{A})-(\ref{C}))
\bea
\alpha_{(t=-1)} & = & \frac{q^{2 n} (q^n - q)}{q^{2 n} - q}, \\
\beta_{(t=-1)} & = & q \Big(1 + q^{-n} - q^n +
\frac{q - q^n}{q^{2 n} - q} - \frac{q^n - 1}{ q^{2 n+1} - 1}  \Big),  \\
\gamma_{(t=-1)} & = & \frac{q - q^{-n}}{1 - q^{2 n+1}}.
\eea
This is precisely the homogeneous relation found in \cite{Garoufalidis} by hand
and obtained more systematically in the recent work \cite{Tudor,abmodel} by quantizing $A(x,y)$.

Moreover, in the $t=-1$ limit, there is also an inhomogeneous 2-term relation,
which does not exist for other values of $t$.
To derive this inhomogeneous relation we again start from the ratios (\ref{ratioA})-(\ref{ratioC}).
The crucial point is that at $t=-1$ the factors $(1-q^{k+1})$ in the numerator and denominator cancel.
As a result, equations (\ref{ratioB}) and (\ref{ratioC}) take the form
\bea
q^{n-1} q^{-k} J(n,k) & = & J(n,k) - q^{-n} J(n,k+1) \,, \\
q^{-k} J(n+1,k+1) & = & q^{n+2} (1 - q^n) J(n,k) \,,
\eea
and factors $q^{2k}$ do not appear.\footnote{Note that for general $t$ it is not possible to solve for combinations $q^k P(n,k)$ directly, which is why we had to sum over $k$ first using auxiliary $R_n$ and $S_n$, to get 3-term homogeneous relation.}
Moreover if we shift the indices $n$ and $k$ by one unit in the second equation,
we can explicitly solve for the factor $q^{-k} P(n,k)$ to obtain a single relation
$$
J(n,k+1) \; = \; q^n J(n,k) - q^{3n-1} (1 - q^{n-1}) J(n-1,k-1) \,.
$$
Performing the sum over $k$ and using the definition (\ref{Pn-single}), as well as taking care of the boundary terms, we get
\be
J_n = q^{n-1} \frac{1-q^{2n-1}}{1-q^n} - q^{3n-1} \frac{1-q^{n-1}}{1-q^n} J_{n-1}.   \label{trefoil2term}
\ee
This is the same 2-term inhomogeneous relation as presented in \cite{Garoufalidis}.
In fact, the homogeneous relation (\ref{RefineRecursiont1}) also follows from (\ref{trefoil2term}):
if we normalize (\ref{trefoil2term}) so that the inhomogeneous term is an $n$-independent constant, we get
$$
{\rm const} \; = \; \delta_n J_n + \epsilon_n J_{n-1} \; = \; \delta_{n+1} J_{n+1} + \epsilon_{n+1} J_{n} \,,
$$
or, equivalently,
$$
\epsilon_n J_{n-1} + (\delta_n - \epsilon_{n+1} ) J_n - \delta_{n+1} J_{n+1} \; = \; 0 \,.
$$
It is not hard to check that this structure reproduces (\ref{RefineRecursiont1}), with
$\alpha_{(t=-1)} \sim \epsilon_n$,
$\beta_{(t=-1)} \sim (\delta_n - \epsilon_{n+1})$,
and $\gamma_{(t=-1)} \sim - \delta_{n+1}$.

%*******************************************************************
%*******************************************************************

%\subsubsection*{Recursion in the limits $q=\pm 1$}

Other interesting limits of (\ref{RefineRecursion}) arise when $q$ takes special values.
Thus, when $\boxed{q=1}$ the coefficients in (\ref{A})-(\ref{C}) simplify and, up to an overall factor $1+t^3$, take the form
\be
\alpha_{(q=1)} = 0 \,,  \qquad   \beta_{(q=1)} = -(1+t^2+t^3) \,, \qquad  \gamma_{(q=1)} = 1 \,.
\ee
Vanishing of $\alpha_{(q=1)}$ means that the recursion reduces to the 2-term form:
\be
P_{n+1} \; = \; (1+t^2+t^3) P_n \,.
\ee
In fact, in the classical limit $q=1$, even the full-fledged superpolynomial \eqref{Superpolynomial_sym_refined},
without the specialization to $a=q^2$, enjoys a simple and elegant recursion relation, which for $(2,2p+1)$ torus knots looks like:
\be
\P^{S^n} (T^{2,2p+1})
%(T^{2,2p+1}; a,q=1,t)
\; = \; a^p \left[ \frac{1 - t^{2p+2}}{1 - t^2} + a t^3 \frac{1 - t^{2p}}{1 - t^2} \right]  \; \P^{S^{n-1}} (T^{2,2p+1})
%(T^{2,2p+1}; a,q=1,t)
\,.
\ee

In the limit $\boxed{q=-1}$ there are also simplifications.
However, the recursion \eqref{RefineRecursion} still involves 3 terms with
\bea
\alpha_{(q=-1)} & = & \big(1 + (-1)^n  \big)t^4,  \\
\beta_{(q=-1)} & = & (1+t^3)\big(-1+(-1)^nt^2 +t^3\big), \\
\gamma_{(q=-1)} & = & -1+(-1)^n t^3.
\eea
%\be
%P_{n+1} = a^p \Big[ \frac{1 - t^{2p+2}}{1 - t^2}  + a t^3 \frac{1 - t^{2p}}{1 - t^2} \Big]  P_n
%\ee
All these limits have a nice physical interpretation that follows from \eqref{theoryA}
and will be discussed elsewhere.
Basically, setting the parameters $q$ and $t$ to special values means that in the corresponding
generating functions \eqref{ZrefdefP} -- \eqref{Zrefdefa}
one ignores the dependence on either the spin $F$ or the D0-brane charge $P$.

%*******************************************************************
%*******************************************************************

\subsection{Refined $A$-polynomials and the refined ``volume'' $S_0 (u,t)$}
\label{sec:knotS0}

One next goal is to test the second new conjecture of the present paper -- namely,
the refinement of the generalized volume conjecture (\ref{VCparamref}a) --
in a large class of examples associated with $(2,2p+1)$ torus knots.
Specifically, in this section we derive the refined $A$-polynomials $A^{\text{ref}}(x,y;t)$
for $(2,2p+1)$ torus knots and analyze the refined ``volume'' $S_0 (u,t)$ that dominates
the asymptotic behavior (\ref{VCparamref}a), thereby verifying \eqref{S0ref}.

At a very practical level,
the refined $A$-polynomials arise as the $q \to 1$ limit of $\widehat{A}^{\text{ref}}(\hat{x},\hat{y};q,t)$,
as we explained earlier and illustrated in Figure \ref{fig:AAAA}.
Therefore, if one knows the quantum operator $\widehat{A}^{\text{ref}}(\hat{x},\hat{y};q,t)$,
say, as in the case of the unknot or the trefoil knot, then it is trivial to find its classical version, $A^{\text{ref}}(x,y;t)$.
However, our conjecture (\ref{VCparamref}) provides another way to look at the refined $A$-polynomial.
Namely, according to \eqref{S0ref}, it determines the asymptotic behavior of the $n$-colored homological invariants $P_n (q,t)$
in the limit \eqref{reflimit}.
Our conjecture says that, in this limit, the homological invariants $P_n (q,t)$ exhibit exponential growth with the leading term $S_0(u,t)$,
such that
\be
y = e^{S'_0(u,t)} \; = \; e^{\frac{d}{du} S_0(u,t)} \; = \; e^{x \frac{d}{dx} S_0(\log x,t)} \,.   \label{yS0prim}
\ee
This gives another way of expressing the dependence of $y$ on $x$, which is equivalent to the equation $A^{\text{ref}}(x,y;t)=0$.
In other words, the leading order free energy $S_0(u,t)$ computed directly from the asymptotics of $P_n (q,t)$
must agree with the integral (\ref{S0ref}) on an algebraic curve $A^{\text{ref}}(x,y;t) = 0$.
In this section, we will test this equivalence and our conjecture (\ref{VCparamref}a).
We also determine the refined $A$-polynomials even in those examples
where the full quantum curve $\widehat{A}^{\text{ref}}(\hat{x},\hat{y};q,t)$ is not known at present!

In particular, for the unknot and for the trefoil knot, for which we already found
$\widehat{A}^{\text{ref}}(\hat{x},\hat{y};q,t)$ in the previous section,
we will show that their $q \to 1$ limits indeed agree with the refined $A$-polynomials
computed from the asymptotics of the corresponding colored homological invariants.
Furthermore, even for more general $(2,2p+1)$ torus knots, for which the full quantum $A$-polynomials are not known at present,
we will find the refined $A$-polynomials by testing our conjecture (\ref{VCparamref}).
Impatient reader can skip directly to Table \ref{table_A-poly4},
where we list the explicit form of $A^{\text{ref}}_{T^{2,2p+1}} (x,y;t)$ for several small values of $p$.
In section \ref{sec:Bmodel}, we will use analogous techniques to analyze
the refined $A$-polynomials and asymptotic expansions in examples coming from refined BPS state counting.

%*******************************************************************

\subsubsection{Refined $A$-polynomial and $S_0 (u,t)$ for the unknot}

We start our analysis with the example of the unknot.
This example is quite instructive: being relatively simple,
it captures all essential ingredients which arise for more complicated knots.
Recall, that we already determined the quantum refined curve in (\ref{qAt-unknot}),
and its classical limit reads (\ref{At-unknot}):
\be
A^{\text{ref}}_{\unknot}(x,y;t) \; = \;
(1 + t^3 x) ( - t^{-3})^{1/2} - (1 - x) y  = 0 \,.   \label{Atunknot2}
\ee
We would like to verify our conjecture (\ref{VCparamref}a)
and to confirm that the same curve controls the asymptotic expansion of the $n$-colored homological invariants (\ref{Pnunkot}):
\bea
P_{n} (\unknot; q,t) & = &
 q^{-\frac{n-1}{2}} (-t^{-3})^{\frac{n-1}{2}} \frac{(-q^2 t^3;q)_{n-1}}{(q;q)_{n-1}} = \nonumber  \\
& = & (-t^{-3})^{\frac{n-1}{2}} q^{1/2} x^{-1/2} \frac{(-q^2 t^3;q)_{\infty}}{(-q^2 t^3 x;q)_{\infty}}
\frac{(x;q)_{\infty}}{(q;q)_{\infty}},
\eea
where we introduced $x=e^u=q^n$.
Now we can use the expansion of the quantum dilogarithm function, see {\it e.g.} \cite{DGLZ,Kirillov},
\begin{eqnarray}
\log (x;q)_{\infty} \; = \; \frac{1}{\hbar}{\rm Li}_2(x)
+\frac{1}{2}\log(1-x)-\sum_{k=1}^{\infty}\frac{B_{2k}}{(2k)!}
\frac{U_{2k-1}(x)\hbar^{2k-1}}{(1-x)^{2k-1}} \,,
\label{q-dlog_exp}
\end{eqnarray}
where $q=e^{\hbar}$ and $U_k(x)$ is a polynomial of degree $k$
satisfying
\begin{eqnarray}
U_k(x) \; = \; (x-x^2)U^{\prime}_{k-1}(x)+kxU_{k-1}(x) \,, \quad U_0=1 \,.
\end{eqnarray}
{}From this expansion we find the asymptotics
\be
P_n(\unknot; q,t) = \exp \frac{1}{\hbar}\Big(\log x\, \log(-t^{-3})^{1/2} + \textrm{Li}_2(x) - \textrm{Li}_2(-t^3 x)  + \textrm{Li}_2(-t^3) - \frac{\pi^2}{6} + \mathcal{O}(\hbar) \Big),
\ee
so that we identify
\be
S_0(u,t) \; = \;  \log x\, \log(-t^{-3})^{1/2} + \textrm{Li}_2(x) - \textrm{Li}_2(-t^3 x)  + \textrm{Li}_2(-t^3) - \frac{\pi^2}{6} \,.
\label{S0unknot}
\ee
Computing the derivative of this result and using the relation (\ref{yS0prim}) we finally obtain
\be
y = e^{S'_0(u,t)} = (-t^{-3})^{1/2} \frac{1 + t^3 x}{1 - x},
\ee
which is clearly equivalent to the refined $A$-polynomial (\ref{Atunknot2}).

Let us also stress an interesting feature of the leading order free energy (\ref{S0unknot}).
In the unrefined limit, $t=-1$, this free energy vanishes:
\be
S_0(u, t=-1) \; = \; 0 \,,
\ee
which is consistent with the form of the unrefined $A$-polynomial $A_{\unknot}(x,y) = (1-x)(1-y)$ given in (\ref{Aunknot1x1y}).
The factor $(1-y)$ in this classical, unrefined $A$-polynomial is a universal factor associated with abelian flat connections connections.
Usually, it does not lead to an interesting contribution to the free energy, as the integral (\ref{S0ref}) is trivial in this case.
Therefore, our result (\ref{S0unknot}) could be interpreted as a contribution of a ``refined'' abelian flat connection,
which becomes non-trivial once $t \ne -1$.

In what follows we use similar methods to analyze more interesting knots.

%*******************************************************************

\subsubsection{Refined $A$-polynomial and $S_0 (u,t)$ for the trefoil}

The next example is naturally the trefoil knot $T^{2,3} = {\bf 3}_1$.
First, let us discuss the structure of its refined $A$-polynomial
and the associated free energy $S_0(u,t)$ from the viewpoint of the quantum $A$-polynomial given in (\ref{trefoil-Aqt}).
The refined $A$-polynomial can be determined by setting $q = 1$ in (\ref{trefoil-Aqt}):
\be
A_{T^{2,3}}^{\text{ref}}(x,y;t) = y^2 -\frac{1 - x t^2 + x^3 t^5 + x^4 t^6 + 2 x^2 t^2 (t+1)}{1
+ x t^3} y + \frac{(x-1) x^3 t^4}{1 + x t^3} \; = \; 0 \,.
\label{deformed_A-poly_trefoil}
\ee
We stress that for generic values of $t$ this form does not factorize, as the discriminant
$$
\Delta = \Big(\frac{1 + x^2 t^3}{1 + x t^3}\Big)^2 \big(1 - 2 t^2 x + (4 t^2 + 2 t^3 + t^4) x^2 + 2 t^5 x^3 + t^6 x^4\big)
$$
is not a complete square.
This means that the abelian branch -- represented by a factor $y-1$ in the classical, unrefined $A$-polynomial --
is ``mixed'' in together with the non-abelian branch for generic values of $t \ne -1$.
Indeed, the value $t=-1$ is very special because the discriminant factorizes
$$
\Delta_{(t=-1)} \; = \; (1+x^3)^2 \,,
$$
and, as a result, the $A$-polynomial also factorizes
\be
A^{\text{ref}}_{T^{2,3}}(x,y;t=-1) \; = \; (y-1) (y + x^3) \,.
\label{A31nonref}
\ee
In the latter expression, $y-1$ represents the abelian branch, while the ``interesting'' factor $y+x^3$
is what sometimes referred to as the reduced $A$-polynomial for the trefoil knot.
We see that for generic values of $t$ such a factorization does not occur,
and the ``character variety'' $A^{\text{ref}} (x,y;t)=0$ is irreducible, with only one component.
For $t=-1$ it becomes reducible and splits into two (or, in general, several) components.

Once we found the refined curve,
we can easily compute $S_0(u,t)=\int \log y \frac{dx}{x}$ using~\eqref{S0ref}.
For the trefoil knot there is no compact expression for this integral,
but one can write it as a Taylor series in $(t+1)$.
Namely, by solving the quadratic equation $A^{\text{ref}}_{T^{2,3}}(x,y;t)=0$, we find
\bea
y & = & \frac{1 - x t^2 + x^3 t^5 + x^4 t^6 + 2 x^2 t^2 (t+1)}{2(1 + x t^3)} - \frac{1}{2}\sqrt{\Delta} = \nonumber \\
& =& -x^3 + (t+1) x^3
\frac{(x-2) (3 x^2 -2x +2) }{(x-1) (1 - x + x^2)} + \mathcal{O}\big((t+1)^2\big)
\eea
The first term $-x^3$ corresponds to the non-abelian branch in (\ref{A31nonref}).
Then, expanding the integrand $\log y \frac{dx}{x}$ in powers of $(t+1)$
and integrating term by term we find the following structure
\be
S_0(u,t) \; = \; S_0(u) + (t+1) S_0^{(1)}(u) + \sum_{k=2}^{\infty} (t+1)^k S_0^{(k)}(u),
\ee
where $S_0(u)$ is the ordinary Chern-Simons action of the flat
$SL(2,\C)$ connection on the trefoil knot complement:
\bea
S_0(u) & = & \frac{1}{6}\Big(\log (-x^3)\Big)^2, \nonumber\\
S_0^{(1)}(u) & = & \log\frac{(1-x)^3}{x^4 (1-x+x^2)}, \nonumber \\
S_0^{(2)}(u) & = & \frac{1}{2} S_0^{(1)}(u) + \frac{-8x^4 +9x^3 -12x^2 + 5x -3}{2( x-1 ) (1 -x +x^2)^2}, \nonumber \\
& \vdots & \nonumber\\
S_0^{(k)}(u) & = & \frac{1}{k} S_0^{(1)}(u) + \frac{1}{k!} \frac{R_k(x)}{( x-1 )^k (1 -x +x^2)^{2k-2}} \nonumber
\eea
In other words, the $k$-th order term $S_0^{(k)}(u)$ includes $\frac{1}{k} S_0^{(1)}(u)$
and also a rational function with $k! ( x-1 )^k (1 -x +x^2)^{2k}$ in the denominator
and a certain polynomial $R_k(x)$ in the numerator.
In particular, we can sum over all $S_0^{(1)}(u)$ contributions, so that
\be
S_0(u,t) \; = \; \frac{1}{6}\Big(\log (-x^3)\Big)^2 - \log(- t) \log\frac{(1-x)^3}{x^4 (1-x+x^2)} + R(x,t) \,,
\label{S0ut-trefoil}
\ee
where $R(x)$ is certain (complicated) function, whose $(t+1)^k$ coefficient is $\frac{1}{k!} \frac{R_k(x)}{( x-1 )^k (1 -x +x^2)^{2k}}$.
We note that for $t=-1$ all $t$-dependent terms vanish. In the next subsection we will confirm that the same refined $A$-polynomial and the same $S_0(u,t)$ arise from the analysis of the asymptotic expansion of the colored superpolynomials (\ref{refined_SU(2)}).

Let us point out that the curve \eqref{deformed_A-poly_trefoil} also factorizes in the $t=0$ limit:
$$
A^{\text{ref}}_{T^{2,3}}(x,y;t=0) \; = \; y(y-1) \,,
$$
and there is an associated singularity in $S_0(u,t)$ when $t\to 0$.
It would be interesting to understand this singularity better.

\subsubsection{Saddle point analysis of the homological torus knot invariants}
\label{sec:saddle_point}

In this section we analyze general $(2,2p+1)$ torus knots ($p \ge 1$).
For this class of knots, the quantum refined curves $\widehat{A}^{\text{ref}}_{T^{2,2p+1}}(\hat{x},\hat{y};q,t)$
are not known at present (apart from the case of trefoil, \emph{i.e.} $p=1$, with the quantum curve determined in (\ref{RefineRecursion})).
Therefore, we cannot determine refined $A$-polynomials by taking the classical limit of $\widehat{A}^{\text{ref}}$,
as we did for the trefoil knot or for the unknot.
Nevertheless, we can determine refined $A$-polynomials from the asymptotic behavior
of the colored superpolynomials and their specializations (\ref{refined_SU(2)}).
For general values of $p$, we determine parametric representation of such $A$-polynomials,
and for several first values of $p$ we rewrite this parametric form as
a polynomial $A^{\text{ref}}_{T^{2,2p+1}}(x,y;t)$, see Table \ref{table_A-poly4}.

There is one fundamental difference between the specialization $P_{n}(K;q,t)$
of the colored superpolynomial for torus knots found in (\ref{refined_SU(2)}) and that of the unknot (\ref{Pnunkot}).
Namely, the latter is given by an infinite product,
whose asymptotic expansion is obtained simply from the expansion of the quantum dilogarithm (\ref{q-dlog_exp}).
On the other hand, the homological invariants of torus knots are expressed as infinite sums,
with each term in those sums given by infinite products.
Therefore, the analysis of the asymptotic behavior for torus knots is more delicate and requires new methods.

In order to find the ``refined volume'' $S_0(u,t)$ for torus knots,
we apply the saddle point approximation \cite{Hikami_exp,Hikami:2007zz}
to (\ref{refined_SU(2)})
and replace the quantum dilogarithm function by
\begin{eqnarray}
(z;q)_{k}\sim e^{\frac{1}{\hbar}\left({\rm Li}_2(z)-{\rm Li}_2(zq^k)\right)} \,.
\label{q-dilog_approx}
\end{eqnarray}
Furthermore, via an analytic continuation we can approximate the summation in (\ref{refined_SU(2)}),
in the asymptotic limit (\ref{reflimit}), by the following integral:
\be
\boxed{\phantom{\int^1}
P_{n}(T^{2,2p+1};q,t) \; \sim \; \int dz\; e^{\frac{1}{\hbar}\left(V_{(2,2p+1)}(z,x;t)+{\cal O}(\hbar)\right)} \,, \phantom{\int^1}}
\label{Pn-saddle}
\ee
with the ``potential'' function
\begin{eqnarray}
&& V_{(2,2p+1)}(z,x;t)
:=-p\log (-t)\cdot \log x+(p+1)\pi i\log x+\log (x^{\frac{1}{2}}z^{-1})\cdot\log t
\nonumber \\
&&\quad\quad\quad\quad\quad\quad\quad
+(2p+1)\Biggl(
\pi i\log z+\frac{1}{2}\left((\log x)^2-(\log z)^2\right)
+\log (x^{\frac{3}{2}}z^{-1})\cdot \log t\Biggr)
\nonumber \\
&&\quad\quad\quad\quad\quad\quad\quad
+{\rm Li}_2(z)-{\rm Li}_2(x)-{\rm Li}_2(t^2z)+{\rm Li}_2(-t^3x)
+{\rm Li}_2(t^2xz)
\nonumber \\
&&\quad\quad\quad\quad\quad\quad\quad
-{\rm Li}_2(-t^3xz)
+{\rm Li}_2(xz^{-1})
-{\rm Li}_2(-txz^{-1})
+{\rm Li}_2(-t)-{\rm Li}_2(1),
\end{eqnarray}
and where the parameter $z$ is related to $\ell$ in (\ref{refined_SU(2)}) via $\ell=\frac{1}{\hbar}\log z$.
%In the derivation of the function $V(z,x;t)$,
%we again used the expansion of the quantum dilogarithm given in (\ref{q-dlog_exp}).
Now, the dominant contribution to this integral comes from the saddle point\footnote{To be more precise,
there can be subtleties in the treatments of the analytic continuation \cite{Wit-anal},
and the convergence and non-perturbative contributions like ${\cal O}(e^{\hbar})$
of the contour integrals should be discussed more carefully.
Luckily, none of these subtleties affect our derivation of the refined $A$-polynomial
in the asymptotic limit $\hbar\to 0$ around the exponential growth point.
For this reason, we will only discuss the saddle point approximation
in a sense of the {\it optimistic limit}, relegating a more detailed
analysis {\it a la} \cite{Kashaev_Tirkkonen,Murakami_torus,Hikami_Murakami1,Hikami_Murakami2} to future work.}
\begin{eqnarray}
\frac{\partial V_{(2,2p+1)}(z,x;t)}{\partial z}\Bigg|_{z=z_0}=0,
\label{saddle_point}
\end{eqnarray}
and the value of the ``potential'' $V(z,x;t)$ at this saddle point determines $S_0(u,t)$:
\begin{eqnarray}
V_{(2,2p+1)}(z_0,x;t) \; = \; S_0(u,t) \,.
\end{eqnarray}
For the above potential $V_{T^{2,2p+1}}(z,x;t)$,
the critical point condition can simply be expressed as $1=\exp\left(z\partial V_{(2,2p+1)}/\partial z\right)|_{z=z_0}$:
\begin{eqnarray}
1 \; = \; -\frac{t^{-2-2p}(x-z_0)z_0^{-1-2p}(-1+t^2z_0)(1+t^3 xz_0)}{(-1+z_0)(tx+z_0)(-1 +  t^2 x z_0)} \,.
\label{braid_saddle1}
\end{eqnarray}

Moreover, according to (\ref{S0ref}), the solution of $A^{\text{ref}}_{T^{2,2p+1}}(x,y;t)=0$ is related to $S_0(u,t)$ via
\begin{eqnarray}
y(x,t)&=&\exp\left(x\frac{\partial V_{(2,2p+1)}(x,z_0;t)}{\partial x}\right)
\nonumber \\
&=&\frac{t^{2 + 2 p} (-1 + x) x^{1 + 2 p} (t x + z_0) (1 + t^3 x z_0)}{(1 + t^3 x) (x - z_0) (-1 + t^2 x z_0)} \,.
\label{braid_saddle2}
\end{eqnarray}
The equations (\ref{braid_saddle1}) and \eqref{braid_saddle2} constitute our desired result:
they provide an expression for the refined $A$-polynomial, parametrized by $z_0$, for a general $(2,2p+1)$ torus knot.
Moreover, for a fixed value of $p$ it is possible to eliminate $z_0$ from these equations
and write an explicit form of the refined $A$-polynomials $A^{\text{ref}}_{T^{2,2p+1}}(x,y;t)$.
For $p=1,2,3,4$, the explicit expressions of the refined $A$-polynomials are presented in Table~\ref{table_A-poly4}.
%\footnote{
%In the following, we use the notations $T^{2,2p+1}$ and ${\bf 2p+1_1}$
% in \cite{Rolfsen} for $(2,2p+1)$-torus knots simultaneously.}
\begin{table}[h]
\centering
\begin{tabular}{|@{$\Bigm|$}c|@{$\Bigm|$}l|}
\hline
\textbf{Knot}  & $A_K^{\text{ref}} (x,y;t)$  \\
\hline
\hline
$T^{2,3}$ & {\scriptsize $y^2 -$}$\frac{1}{1+ x t^3}${\scriptsize $(1 - x t^2 + x^3 t^5 + x^4 t^6 + 2 x^2 t^2+2 x^2 t^3)y
 +$}$\frac{(x-1) x^3 t^4}{1 + x t^3}$    \\
\hline
$T^{2,5}$ & {\scriptsize $y^3
-$}$\frac{1}{1 + t^3 x}$ {\scriptsize $
(1 - t^2 x + 2 t^2 x^2 + 2 t^3 x^2 - 2 t^4 x^3 - 2 t^5 x^3 +
 3 t^4 x^4 + 4 t^5 x^4 + t^6 x^4
+ t^7 x^5
- t^8 x^5 + 2 t^8 x^6)y^2$
}
\\
&{\scriptsize $+$}$\frac{t^6 (-1 + x) x^5}{(1 + t^3 x)^2}$
{\scriptsize $(
2 - t^2 x + t^3 x + 3 t^2 x^2 + 4 t^3 x^2 + t^4 x^2 + 2 t^5 x^3 +
 2 t^6 x^3 + 2 t^6 x^4
+ 2 t^7 x^4
+ t^9 x^5 + t^{10} x^6)y$
}
\\
&{\scriptsize $-$}
$\frac{t^{12} (-1 + x)^2 x^{10}}{(1 + t^3 x)^2}$  \\
\hline
$T^{2,7}$ &
{\scriptsize $y^4
-$}$
\frac{1}{1+t^3x}
${\scriptsize
$
(
1 - t^2 x + 2 t^2 x^2 + 2 t^3 x^2 - 2 t^4 x^3 - 2 t^5 x^3 +
 3 t^4 x^4
+ 4 t^5 x^4 + t^6 x^4
- 3 t^6 x^5
- 4 t^7 x^5
$}
\\&{\scriptsize $\quad\quad
- t^8 x^5 +
 4 t^6 x^6 + 6 t^7 x^6 + 2 t^8 x^6 + t^9 x^7 - 2 t^{10} x^7 + 3 t^{10} x^8
)y^3
$
}
\\
&{\scriptsize $+$}$\frac{t^8 (-1 + x) x^7}{(1 + t^3 x)^2}
${\scriptsize $(
3 - 2 t^2 x + t^3 x + 6 t^2 x^2 + 8 t^3 x^2 + 2 t^4 x^2 - 3 t^4 x^3 -
 2 t^5 x^3 + t^6 x^3
+ 6 t^4 x^4 + 12 t^5 x^4
$}\\
&{\scriptsize $\quad
+ 10 t^6 x^4 +
 4 t^7 x^4 + 3 t^7 x^5 + 2 t^8 x^5 - t^9 x^5 + 6 t^8 x^6 +
 8 t^9 x^6
+ 2 t^{10} x^6 + 2 t^{11} x^7
- t^{12} x^7 + 3 t^{12} x^8
)y^2$
}
\\
&{\scriptsize $-$}$\frac{t^{16} (-1 + x)^2 x^{14}}{(1 + t^3 x)^3}${\scriptsize
$(
3 - t^2 x + 2 t^3 x + 4 t^2 x^2 + 6 t^3 x^2 + 2 t^4 x^2 + 3 t^5 x^3 +
 4 t^6 x^3
+ t^7 x^3
+ 3 t^6 x^4
+ 4 t^7 x^4
$
}
\\
&{\scriptsize
$\quad + t^8 x^4 + 2 t^9 x^5 +
 2 t^{10} x^5 + 2 t^{10} x^6 + 2 t^{11} x^6 + t^{13} x^7 + t^{14} x^8
)y+$}{\scriptsize $\frac{t^{24} (-1 + x)^3 x^{21}}{(1 + t^3 x)^3}$}  \\
\hline
$T^{2,9}$ & {\scriptsize $y^5
-$}
$\frac{1}{1 + t^3 x} $
{\scriptsize $
(1 - t^2 x + 2 t^2 x^2 + 2 t^3 x^2 - 2 t^4 x^3 - 2 t^5 x^3 +
 3 t^4 x^4 + 4 t^5 x^4 + t^6 x^4
- 3 t^6 x^5
- 4 t^7 x^5
$}
\\
&{\scriptsize $\quad\quad
- t^8 x^5
+  4 t^6 x^6 + 6 t^7 x^6 + 2 t^8 x^6 - 4 t^8 x^7 - 6 t^9 x^7 -
 2 t^{10} x^7
+ 5 t^8 x^8 + 8 t^9 x^8
+ 3 t^{10} x^8 + t^{11} x^9
$}
\\
&{\scriptsize $\quad\quad
-
 3 t^{12} x^9 + 4 t^{12} x^{10})y^4
$
}
\\
&{\scriptsize $+$}$\frac{t^{10} (-1 + x) x^9}{(1 + t^3 x)^2}${\scriptsize
$(4 - 3 t^2 x + t^3 x + 9 t^2 x^2 + 12 t^3 x^2 + 3 t^4 x^2 -
 6 t^4 x^3 - 6 t^5 x^3
+ 12 t^4 x^4
+ 24 t^5 x^4
$}
\\
&{\scriptsize $\quad
+ 18 t^6 x^4 +
 6 t^7 x^4
- 6 t^6 x^5 - 9 t^7 x^5 - 6 t^8 x^5 - 3 t^9 x^5 +
 10 t^6 x^6
+ 24 t^7 x^6
+ 27 t^8 x^6 + 16 t^9 x^6
$}
\\
&{\scriptsize $\quad
+ 3 t^{10} x^6
+  4 t^9 x^7 - 6 t^{11} x^7 - 2 t^{12} x^7 + 12 t^{10} x^8 + 18 t^{11} x^8
+  6 t^{12} x^8 + 3 t^{13} x^9 - 3 t^{14} x^9 + 6 t^{14} x^{10})
y^3
$} \\
&{\scriptsize $-$}
$\frac{t^{20} (-1 + x)^2 x^{18}}{(1 + t^3 x)^3}${\scriptsize $
(6 - 3 t^2 x + 3 t^3 x + 12 t^2 x^2 + 18 t^3 x^2 + 6 t^4 x^2 -
 4 t^4 x^3
+ 6 t^6 x^3
+ 2 t^7 x^3 + 10 t^4 x^4
$}
\\
&{\scriptsize $\quad
+ 24 t^5 x^4 +
 27 t^6 x^4 + 16 t^7 x^4 + 3 t^8 x^4 + 6 t^7 x^5 + 9 t^8 x^5
+
 6 t^9 x^5
+ 3 t^{10} x^5
+ 12 t^8 x^6 + 24 t^9 x^6
$}
\\
&{\scriptsize $\quad
+ 18 t^{10} x^6 +
 6 t^{11} x^6 + 6 t^{11} x^7 + 6 t^{12} x^7
+ 9 t^{12} x^8
+ 12 t^{13} x^8 +
 3 t^{14} x^8
+ 3 t^{15} x^9 - t^{16} x^9 + 4 t^{16} x^{10})
y^2
$}
\\
&{\scriptsize $+$}
$\frac{t^{30} (-1 + x)^3 x^{27}}{(1 + t^3 x)^4}$
{\scriptsize $
(4 - t^2 x + 3 t^3 x + 5 t^2 x^2 + 8 t^3 x^2 + 3 t^4 x^2 + 4 t^5 x^3 +
 6 t^6 x^3 + 2 t^7 x^3
+ 4 t^6 x^4
$}
\\
&{\scriptsize $\quad
+ 6 t^7 x^4 + 2 t^8 x^4 +
 3 t^9 x^5 + 4 t^{10} x^5 + t^{11} x^5 + 3 t^{10} x^6 + 4 t^{11} x^6 +
 t^{12} x^6
+ 2 t^{13} x^7
+ 2 t^{14} x^7 + 2 t^{14} x^8
$}
\\
&{\scriptsize $\quad
+ 2 t^{15} x^8 +
 t^{17} x^9 + t^{18} x^{10})y-$}$\frac{t^{40} (-1 + x)^4 x^{36}}{(1 + t^3 x)^4}
$
\\
\hline
\end{tabular}
\caption{Refined A-polynomials for $p=1,2,3,4$.
\label{table_A-poly4} }
\end{table}
For $p=1$, this result is consistent with the semi-classical limit of
the recursion relation (\ref{deformed_A-poly_trefoil}).
%In the unrefined limit $t\to -1$, we observe
%\begin{eqnarray}
%A_{T^{2,2p+1}}(x,y;t=-1)=(y-1)(y+x^{2p+1})^p.
%\end{eqnarray}
Further examples are also summarized in Appendix \ref{sec:AppendixB}.
Note that for $t=-1$ the first equation above (\ref{braid_saddle1}) simply specifies the value of $z_0$ as $z_0^{2p+1}=-1$, while the second equation (\ref{braid_saddle1}) reduces to the well-known ordinary $A$-polynomial for $T^{2,2p+1}$ torus knot:
$$
y + x^{2p+1} = 0.
$$

Let us also reveal some remarkable properties of the refined $A$-polynomials found above.
We recall that the ordinary $A$-polynomials $A_{K}(x,y)$ are known to be reciprocal \cite{Apol,DGLZ,CCGL}:
\begin{eqnarray}
A_{K} (x,y) \; = \; x^a y^b A_{K} (x^{-1},y^{-1}) \,.
\end{eqnarray}
This property can be understood either as a consequence of the Weyl reflection
in $SL(2,\C)$ Chern-Simons theory or, alternatively, as a symmetry induced by
the orientation-preserving involution on the knot complement, $M = {\bf S}^3 \setminus K$,
which acts as an endomorphism $(-1,-1)$ on $H_1 (\partial M) \cong \Z \times \Z$.
It beautifully generalizes to the refined case: as a careful reader will notice,
the equations (\ref{braid_saddle1}) and \eqref{braid_saddle2} are invariant under the following transformation:
\begin{eqnarray}
t\to t \,, \quad  x\to -t^{-3}x^{-1} \,, \quad z\to t^{-2}z^{-1} \,, \quad  y\to t^{-2p}y^{-1} \,.
\end{eqnarray}
Therefore, the refined $A$-polynomials satisfies a deformed reciprocity:
\begin{eqnarray}
\boxed{\phantom{\int^1}
A^{\text{ref}}_{T^{2,2p+1}}(x,y;t) \; = \; x^a y^b (-t)^c A^{\text{ref}}_{T^{2,2p+1}} (-t^{-3}x^{-1},t^{-2p} y^{-1};t) \,. \phantom{\int^1}}
\label{reciprocity_deform}
\end{eqnarray}

Once we found the refined $A$-polynomials, from (\ref{S0ref}) we can also determine
the classical action $S_0(u,t)$ by computing it iteratively around $t=-1$.
The solutions $z_0 = z_j$ ($j=1,\cdots,2p$) for (\ref{braid_saddle1}) around $t=-1$ are\footnote{
There are also the other solutions:
\begin{eqnarray}
z_0=\pm \frac{1}{t}.
\end{eqnarray}
Plugging these solutions into (\ref{braid_saddle1}), we find
\begin{eqnarray}
y(x,t)=\frac{t^{2 + 2 p} (-1 + x) x^{1 + 2 p} (\pm 1 + t^2 x)^2}{(\mp 1 + t x)^2 (1 + t^3 x)}.
\end{eqnarray}
We assume that these factors correspond to the solutions $z_0$ which are
not the saddle points but the critical points of the potential $V_{(2,2p+1)}(z,x;t)$.
To treat this point in more detail, we need to specify the integration path more carefully.
Since for $p=1$ this solution is not included in (\ref{deformed_A-poly_trefoil}),
we conclude that these solutions do not describe the saddle points relevant to us.}
\begin{eqnarray}
z_j&=&-\xi_j+\frac{(1+t)}{2p+1}\left(-\frac{(1+x)^2}{x}-(2p-1)\xi_j+\frac{2}{1+\xi_j}+\frac{x^2}{x+\xi_j}+\frac{1}{x(1+x\xi_j)}\right)
\nonumber \\
&&
+{\cal
 O}((1+t)^2),
\end{eqnarray}
where $\xi_j=\exp(\frac{2\pi ij}{2p+1})$, $j=1,\cdots,2p$.
Plugging this solution into \eqref{braid_saddle2}, we find the power series expansion for $y_j (x,t)$:
%\footnote{For $p=1$ and $\xi_j=\exp(2\pi ij/3)$,
%this expansion coincides with the previous analysis for ${\bf 3_1}$.}
\begin{eqnarray}
y_j(x,t)&=&-x^{2p+1} \\
&&
+(1+t)\frac{x^{2p+1}}{(-1+x)(x+\xi_j)(1+x\xi_j)}
(-3 x - 2 p x + 2 p x^2 - 2 \xi_j - 2 p \xi_j - x \xi_j +
 2 p x \xi_j
\nonumber \\
&&\quad\quad\quad\quad\quad\quad\quad\quad\quad
- 4 x^2 \xi_j - 2 p x^2 \xi_j + x^3 \xi_j +
 2 p x^3 \xi_j - 3 x \xi_j^2 - 2 p x \xi_j^2 + 2 p x^2 \xi_j^2)
\nonumber \\
&&+{\cal O}((1+t)^2) \,. \nonumber
\end{eqnarray}
In turn, plugging this expansion into (\ref{S0ref}), we obtain
a power series expansion of the classical action $S_0(u,t)$:
\begin{eqnarray}
S_0(u,t)&=&S_0(u)+\sum_{k=1}^{\infty}(t+1)^kS_0^{(k)}(u),
\nonumber \\
S_0(u)&=&\frac{1}{4p+2}\left(\log(-x^{2p+1})\right)^2,
\nonumber \\
S_0^{(1)}(u)&=&\log\left(\frac{(1-x)^3}{x^{2(p+1)}(x+\xi_j)(1+x\xi_j)}\right)
\nonumber \\
S_0^{(2)}(u)&=&\frac{1}{2}S_0^{(1)}(u)
\nonumber \\
&&-\frac{1}{2(2p+1)(-1+x)(x+\xi_j)^2(1+x\xi_j)^2}
\times (9 x^2 + 12 p x^2 + 6 p x^3 + 11 x \xi_j + 18 p x \xi_j
\nonumber \\
&&
+
 3 x^2 \xi_j + 18 p x^2 \xi_j
+ 21 x^3 \xi_j + 30 p x^3 \xi_j +
 x^4 \xi_j + 6 p x^4 \xi_j + 3 \xi_j^2 + 6 p \xi_j^2 + 2 x \xi_j^2
\nonumber \\
&&
+
 12 p x \xi_j^2 + 30 x^2 \xi_j^2 + 48 p x^2 \xi_j^2 + 6 x^3 \xi_j^2 +
 24 p x^3 \xi_j^2 + 13 x^4 \xi_j^2 + 18 p x^4 \xi_j^2 +
 11 x \xi_j^3
\nonumber \\
&&
+ 18 p x \xi_j^3 + 3 x^2 \xi_j^3
+ 18 p x^2 \xi_j^3 +
 21 x^3 \xi_j^3 + 30 p x^3 \xi_j^3 + x^4 \xi_j^3 + 6 p x^4 \xi_j^3 +
 9 x^2 \xi_j^4
\nonumber \\
&&
+ 12 p x^2 \xi_j^4 + 6 p x^3 \xi_j^4),
\nonumber \\
&&\vdots
\nonumber \\
S_0^{(k)}(u)&=&\frac{1}{k}S_0^{(1)}(u)+\frac{1}{k!}\frac{R_k(x)}{(x-1)^k(x+\xi_j)^{2(k-1)}(1+x\xi_j)^{2(k-1)}}.
\end{eqnarray}
Summing the terms $\frac{1}{k} S_0^{(1)}$, we find the following general form of the classical action:
\begin{eqnarray}
S_0(u,t)&=&\frac{1}{2(2p+1)}\left(\log(-x^{2p+1})\right)^2-\log(-t)
 \log\left(\frac{(1-x)^3}{x^{2(p+1)}(x+\xi_j)(1+x\xi_j)}\right)+R(x,t).
\nonumber \\
&&
\end{eqnarray}
For $p=1$, this result agrees with (\ref{S0ut-trefoil}),
which was derived from the refined quantum curve (\ref{trefoil-Aqt}).
Therefore, this agreement proves the consistency of our conjectures in the case of the trefoil knot.

We also note that, because the values of the classical action $S_0(u,t)$ for $z_0=z_j$ and $z_0=z_{j+p}$ coincide,
there are in total $p$ independent solutions.
This indicates that the non-abelian branch of the character variety in the unrefined theory
``splits'' into $p$ independent branches in the refined / categorified theory.
The same splitting and the same number of solutions can be seen more directly
from the form of the refined $A$-polynomials for $(2,2p+1)$ torus knots.

%%%%%%%%%%%%%%%%%%%%%%%%%%%%%%%%%%%%%%%%%%%%%%%%%%%%%%%%%%%%%%%%%%%%%%%%%%%%%%%%%%%%%%%%%%%
\subsubsection{Asymptotic behavior in different grading conventions}
\label{sec:saddle_DGR}

It is instructive to study the asymptotic behavior of the colored superpolynomial and its specialization \eqref{Pq2special}
in different grading conventions. For example, another popular set of grading conventions
is the one
%which we refer to as ``DGR'',
where differentials $d_{N<0}$ have degree $(-2,2N,2N-1)$, see \eqref{gradingtabl}.
One important lesson of this exercise will be the fact that the limit \eqref{reflimit}
has to be slightly modified depending on which grading conventions one chooses.
Conceptually, and also as a simple way to remember which limit to take, one wants
\be
q^{\beta} t^{\gamma}  = e^{\hbar} \; \to \; 1 \,,
\label{gradreflimit}
\ee
where the exponents $\beta$ and $\gamma$ represent the degree\footnote{For simplicity,
here we assume that the $a$-degree of $d_{\text{colored}}$ is equal to zero,
as for the first set of colored differentials in \eqref{gradingtabl}.}
of the colored differential, {\it cf.} \eqref{superright},
\be
\deg (d_{\text{colored}}) \; = \; (0 , \beta, \gamma) \,.
\ee
Indeed, with the refined volume conjectures \eqref{VCquantref} and \eqref{VCparamref}
we wish to probe the ``large volume'' asymptotics of the homological knot invariants $P_n (q,t)$, as $n$ goes to infinity.
On the other hand, as we explained in section \ref{sec:homological}, the dependence of
homological knot invariants on the ``color'' $R = S^{n-1}$ is controlled by
the colored differentials which (in the basic case) change the value of $n$ by one unit, see {\it e.g.} \eqref{coloredPa}.

Therefore, in order to study the asymptotic behavior of the homological knot invariants under $n \to n+1$
for sufficiently large $n$, one needs to take a limit in which discrete values of $n$ are replaced
by a continuous variable $x = e^u$ and different terms in the chain complex \eqref{VCsequence}
are clumped together in a continuous distribution, described by $S_0 (u,t)$.
Therefore, this continuous limit {\it is} precisely the limit in which $d_{\text{colored}}$
changes gradings by a tiny amount, {\it i.e.} the limit \eqref{gradreflimit}.

For example, in the grading conventions of \cite{DGR},
the first colored differential listed in \eqref{gradingtabl} has $(a,q,t)$-degree $(0,2,2)$.
Therefore, the right limit to take in this case is the limit \eqref{gradreflimit} with $\beta = 2$
and $\gamma=2$ or, more precisely,
\be
q^2 t^2 = e^{\hbar} \to 1 \,, \qquad t = \text{fixed} \,, \qquad x \equiv e^u = e^{n \hbar} = \text{fixed} \,.
\label{gradlimit}
\ee

In order to make this a little bit more concrete and to understand the issue better,
let us illustrate how it all works in the large class of examples associated with $(2,2p+1)$ torus knots.
We already computed the colored superpolynomial (\ref{Symmetric_DGR_refined}) for these knots
in the grading conventions of \cite{DGR}, and now to study its asymptotic behavior we will need
a couple of useful identities.
First, applying the Euler-Maclaurin formula \cite{Kirillov}:
\begin{eqnarray}
\sum_{m=M}^Nf(m)&=&\int_M^Nf(t)dt+\frac{1}{2}(f(N)+f(M))
+\sum_{k=1}^{n}\frac{B_{2k}}{(2k)!}\left\{
f^{(2k-1)}(N)
\right\}
\nonumber \\
&&-\int_M^N\frac{\bar{B}_{2n}(t)}{(2n)!}f^{(2n)}(t)dt.
\end{eqnarray}
to the function $f(m)=\log\left(1-X(q^2t^2)^m\right)$, we find
\begin{eqnarray}
\log(X;q^2t^2)_{\infty}=
\frac{1}{\hbar+\epsilon}{\rm Li}_2(X)+\frac{1}{2}\log(1-X)+\sum_{k=1}^{\infty}
\frac{B_{2k}}{(2k)!}\frac{U_{2k-1}(X)}{(1-X)^{2k-1}}(\hbar+\epsilon)^{2k-1}
\end{eqnarray}
where we temporarily introduced $\epsilon:=\log t$.
In particular for $X=e^{\frac{s}{\hbar}}$, one finds the following expansion:
\begin{eqnarray}
\log(X;q^2t^2)_{\infty} \; = \; -\frac{s}{2}\hbar^2-i\pi s\hbar+\frac{\pi^2}{3} +{\cal O}(e^{-\hbar s}) \,,
\end{eqnarray}
which we can apply to our result for the superpolynomial (\ref{Symmetric_DGR_refined})
or, to be more precise, to its specialization \eqref{Pq2special123} at $a = q^2$.

First, let us see what would happen if, instead of the correct limit \eqref{gradlimit} that comes from
the gradings of colored differentials, we naively used the limit \eqref{reflimit}
(suitable for the homological invariants in the grading conventions of \cite{GS}, where $\deg (d_{\text{colored}}) = (0,1,0)$).
We would find that the desired specialization of the colored superpolynomial (\ref{Symmetric_DGR_refined})
can be approximated by an integral, much like in \eqref{Pn-saddle},
\be
{\cal P}^{S^{n-1}}_{\text{DGR}}(T^{2,2p+1};a=q^2,q,t)
\; \sim \; \int dz\; e^{\frac{1}{\hbar^2}V_{(2,2p+1)}^{\text{DGR}}(z,x;t)+{\cal O}(\hbar^{-1})} \,,
\ee
with a very simple potential function $V_{T^{2,2p+1}}^{\text{DGR}} (z,x;t)$:
\be
V_{(2,2p+1)}^{\text{DGR}}(z,x;t)
\; = \; (2p+1)\epsilon\left((\log x)^2-(\log z)^2\right) \,.
\ee
Clearly, the critical point of this potential is $z_0 = 1$
and, therefore, the colored superpolynomial in this grading and in the limit $\hbar\to 0$ behaves as
\begin{eqnarray}
\P^{S^{n-1}}_{\text{DGR}}(T^{2,2p+1};a=q^2,q,t) \; \sim \; e^{\frac{1}{\hbar^2}V_{(2,2p+1)}^{\text{DGR}}(z_0,x;t)} \,.
\end{eqnarray}
This behavior is way too simple to learn anything non-trivial about the ``large color''
behavior of the colored superpolynomial and is not even in the expected form (\ref{VCparamref}a),
which is yet another signal that one needs to be very careful passing from one set of grading conventions to another.

Now, let us consider the asymptotic behavior of the same object in the correct limit \eqref{gradlimit}.
Again, as in \eqref{Pn-saddle}, we can write
\be
\P_{\text{DGR}}^{S^{n-1}}(T^{2,2p+1};a=q^2,q;t)
\; \sim \; \int dz\; e^{\frac{1}{\hbar} \biggl(V_{(2,2p+1)}^{\text{DGR}}(z,x,t) +{\cal O}(\hbar)\biggr)} \,,
\ee
with
\begin{eqnarray}
&&V_{(2,2p+1)}^{\text{DGR}}(z,x;t)=
{\rm Li}_2(-t^{-1}) - {\rm Li}_2(x) + {\rm Li}_2(-xt^{-3}) +
{\rm Li}_2(xz^{-1}) - {\rm Li}_2(-xt^{-1}z^{-1})
\nonumber \\
&&\quad\quad\quad\quad\quad\quad\quad\;
+ {\rm Li}_2(z) - {\rm Li}_2(zt^{-2}) - {\rm Li}_2(-x zt^{-3})
+ {\rm Li}_2(x zt^{-2})-\frac{\pi^2}{6}
\nonumber \\
&&\quad\quad\quad\quad\quad\quad\quad\;
+  (2p+1)\Bigl[ (\log (-t^{-2}))\cdot(\log x) +
    \frac{(\log x)^2 - (\log z)^2}{2}-\pi i\log z\Bigr]
\nonumber \\
&& \quad\quad\quad\quad\quad\quad\quad\;
+(\log t)\cdot  (\log z) \,.
\end{eqnarray}
{}From this potential function, the saddle point condition (\ref{saddle_point}) and the first line of (\ref{braid_saddle2}) we find the equations
for the saddle point that dominates the above integral:

\begin{eqnarray}
&&1=\frac{ (t^2 - z) z^{-1 - 2 p} (-x + z) (t^3 + x z)}{t (-1 + z) (x + t z) (t^2 - x z)},
\nonumber \\
&&y=-\frac{ t^{-1 - 4 p} (-1 + x) x^{1 + 2 p} (x + t z) (t^3 + x z)}
{(t^3 + x) (x - z) (t^2 - x z)}.
\label{Saddle_DGR2}
\end{eqnarray}
Eliminating the variable $z$, we find the refined A-polynomial $A^{\text{DGR}}_{T^{2,2p+1}} (x,y;t)$.
For $p=1$, the refined A-polynomial will be described in Appendix \ref{sec:AppendixB}. For $t=-1$ we find the same behavior as in the analysis of  (\ref{braid_saddle1}) and (\ref{braid_saddle2}): the first equation above only specifies the value $z_0^{2p+1}=1$, and the second one reduces to ordinary unrefined $A$-polynomial equation for $T^{2,2p+1}$ knot $y+x^{2p+1}=0$.

The algebraic equations (\ref{Saddle_DGR2}) can be easily solved around $t=-1$:
\begin{eqnarray}
z_j&=&\xi_j+\frac{\xi_j(1+\xi_j)(-2 x + \xi_j + 2 x \xi_j + x^2 \xi_j - 2 x \xi_j^2)}{(1 + 2 p) (x - \xi_j) (-1 + \xi_j) (-1 + x \xi_j)}(1+t)
\nonumber \\
&&+{\cal O}((1+t)^2),
\end{eqnarray}
with $\xi_j=e^{\frac{2\pi ij}{2p+1}}$.
Plugging this solution into the second equation of (\ref{Saddle_DGR2}),
we find the approximate solution for $y$ (as a function of $x$ and $t$):
\begin{eqnarray}
y&=&-x^{2p+1}\Bigl[
1+(1+t)
(-3 x - 4 p x + 4 p x^2 + 2 \xi_j + 4 p \xi_j + x \xi_j -
 4 p x \xi_j + 4 x^2 \xi_j + 4 p x^2 \xi_j
\nonumber \\
&&\quad\quad\quad\quad\quad\quad\quad
- x^3 \xi_j -
 4 p x^3 \xi_j - 3 x \xi_j^2 - 4 p x \xi_j^2 + 4 p x^2 \xi_j^2)
%\nonumber \\
%&&\quad\quad\quad\quad\quad\quad\quad\quad
/(-1 + x) (x - \xi_j) (-1 + x \xi_j)
\nonumber \\
&&\quad\quad\quad\quad
+{\cal O}(1+t)^2\Bigr].
\end{eqnarray}
{}From (\ref{S0ref}), one can also find the refined classical action $S_0(u,t)$:
\begin{eqnarray}
S_0(u,t)&=&S_0(u)+\sum_{a=1}^{\infty}(1+t)^aS_0^{(a)}(u),
\nonumber \\
S_0(u)&=&\frac{1}{2(2p+1)}\log(-x^{2p+1})^2,
\nonumber \\
S_0^{(1)}(u)&=&\log\frac{x^{4p+2}(x-\xi_j)(1-x\xi_j)}{(1-x)^3},
\nonumber \\
S_0^{(2)}(u)&=&\frac{1}{2}S_0^{(1)}
\nonumber \\
&&
+\frac{1}{2(2p+1)(-1 + x) (x - \xi_j)^2 (-1 + x \xi_j)^2}
\nonumber \\
&&\times\bigl(
-7 x^2 - 8 p x^2 - 2 x^3 - 10 p x^3 + 9 x \xi_j + 14 p x \xi_j +
 5 x^2 \xi_j + 22 p x^2 \xi_j + 19 x^3 \xi_j
\nonumber \\
&&
+ 26 p x^3 \xi_j +
 3 x^4 \xi_j + 10 p x^4 \xi_j - 3 \xi_j^2 - 6 p \xi_j^2 -
 2 x \xi_j^2 - 12 p x \xi_j^2 - 30 x^2 \xi_j^2 - 48 p x^2 \xi_j^2
\nonumber \\
&&
-
 6 x^3 \xi_j^2 - 24 p x^3 \xi_j^2 - 13 x^4 \xi_j^2 -
 18 p x^4 \xi_j^2 + 13 x \xi_j^3 + 22 p x \xi_j^3 + x^2 \xi_j^3 +
 14 p x^2 \xi_j^3
\nonumber \\
&&
+ 23 x^3 \xi_j^3 + 34 p x^3 \xi_j^3 - x^4 \xi_j^3 +
 2 p x^4 \xi_j^3 - 11 x^2 \xi_j^4 - 16 p x^2 \xi_j^4 +
 2 x^3 \xi_j^4 - 2 p x^3 \xi_j^4
\bigr),
\nonumber
\\
&&
\end{eqnarray}
and the general form of the classical action:
\begin{eqnarray}
S_0(u,t)&=&\frac{1}{2(2p+1)}\left(\log(-x^{2p+1})\right)^2-\log(-t)
 \log\left(\frac{x^{4p+2}(x-\xi_j)(1-x\xi_j)}{(1-x)^3}\right)+R(x,t).
\nonumber \\
&&
\end{eqnarray}

%***********************************************************************************
%***********************************************************************************

\subsection{Relation to algebraic K-theory}
\label{sec:quantizability}

So far in this section we have discovered a number of refined $A$-polynomials $A^{\textrm{ref}}(x,y;t)$ for various knots,
including the unknot (\ref{At-unknot}), the trefoil knot (\ref{deformed_A-poly_trefoil}),
and more general $(2,2p+1)$ torus knots discussed in section \ref{sec:saddle_point}.
For the unknot and for the trefoil knot we also found an explicit form of the quantum $A$-polynomial,
given respectively in (\ref{qAt-unknot}) and \eqref{trefoil-Aqt}.

There is no doubt that such refined and quantum $A$-polynomials exist for other knots as well,
and can be determined either by (generalizations of) the techniques we used here or via some other methods.
In any case, when lifting the classical $t$-deformed $A$-polynomial to a quantum operator,
one encounters an important subtlety: not all classical curves are quantizable,
and the existence of a consistent quantum curve depends in a delicate way on the complex
structure\footnote{At first, this may seem a little surprising, because the quantization problem
is about symplectic geometry and not about complex geometry of $\cC$.
(Figuratively speaking, quantization aims to replace all classical objects in symplectic geometry
by the corresponding quantum analogs.) However, our phase space $\C^* \times \C^*$
is very special in a sense that it comes equipped with a whole $\cp^1$ worth of complex and symplectic structures,
so that each aspect of the geometry can be looked at in several different ways, depending on which
complex or symplectic structure we choose. This hyper-K\"ahler nature of our geometry is responsible,
for example, for the fact that a curve $\cC$ ``appears'' to be holomorphic (or algebraic).
We put the word ``appears'' in quotes because this property of $\cC$ is merely an accident,
caused by the hyper-K\"ahler structure on the ambient space, and is completely irrelevant from the viewpoint of quantization.
What is important to the quantization problem is that $\cC$ is Lagrangian with respect
to the symplectic form $\Omega = \frac{i}{\hbar} \frac{dx}{x} \wedge \frac{dy}{y}$.}
({\it i.e.} on the coefficients in the defining equation) of the classical polynomial $A^{\textrm{ref}}(x,y;t)$.
Therefore, one should always verify whether a classical $t$-deformed curve actually admits a consistent quantization.

We should stress that this issue of quantizability is much more delicate for refined $A$-polynomials compared to the ordinary ones.
The coefficients of ordinary $A$-polynomials are merely integer numbers, and quantizability imposes non-trivial constraints on these numbers.
Magically, all $A$-polynomials of knots and 3-manifolds automatically meet these conditions.
In retrospect, this is not too surprising, for otherwise $SL(2,\C)$ Chern-Simons would simply make no sense
on 3-manifolds whose $A$-polynomials fail to meet these constraints \cite{Apol}.

The coefficients of refined $A$-polynomials, on the other hand, are functions of an arbitrary continuous parameter $t$.
Hence, at first it may not be entirely obvious how to reconcile arbitrariness of $t$
with the fact that certain functions of this parameter should satisfy rather strong constraints.
In all examples that we have analyzed, something beautiful happens:
as we explain below, it turns out that $t$ can be any root of unity.
Therefore, it is also natural to think of $\hbar=\log q$ as a purely imaginary number, valued in $i \mathbb{Q}$.
We conjecture that this is a general phenomenon: for any knot, quantizability of the refined $A$-polynomial
$A^{\textrm{ref}}(x,y;t)$ requires $t$ to be a root of unity.
This statement could be verified if new examples of refined $A$-polynomials are found or, conversely,
this property could help in \emph{finding} new examples of refined $A$-polynomials.

Before we explain the condition on $t$, let us recall in more detail the general quantizability criteria for curves \cite{Apol}.
Let us consider a curve \eqref{Acurve}:
\be
\cC:\qquad \left\{(x,y)\in\mathbb{C}^*\times \mathbb{C}^*\Big|
A(x,y) \; = \; 0 \right\}\,,
\ee
and the corresponding partition function
$Z=\exp(\frac{1}{\hbar}S_0+\ldots)=\exp(\frac{1}{\hbar}\int \log y \frac{dx}{x}+\ldots)$, as in (\ref{VCparamref}).
One has to make sure that this partition function is well defined,
which means that all periods of the 1-form ${\rm Im} \, \log y \frac{dx}{x}$ on the curve $\cC$ are trivial,
\be
\oint_{\gamma} \Big( \log |x| d ({\rm arg} \, y) - \log |y| d ({\rm arg} \, x) \Big) \; = \; 0 \,,
\label{qcond0}
\ee
and that the periods of the 1-form ${\rm Re} \, \log y \frac{dx}{x}$ are rational multiples of $2 \pi i$,
so that for all closed paths $\gamma$ on the curve $\cC$
\be
\frac{1}{4 \pi^2} \oint_{\gamma} \Big( \log |x| d \log |y| + ({\rm arg} \, y) d ({\rm arg} \, x) \Big) \; \in \; \mathbb{Q} .
\label{qcondQ}
\ee
The above conditions can be nicely reformulated in terms of algebraic K-theory \cite{abmodel}.
Thus, the integrand $\eta(x,y)=\log |x| d ({\rm arg} \, y) - \log |y| d ({\rm arg} \, x)$
in \eqref{qcond0} is the image of the symbol $\{ x,y \} \in K_2 (\cC)$ under the regulator map \cite{Beilinson,Bloch}.
For curves it is not hard to see that $\eta(x,y)$ is closed, $d\eta=0$.
However, the condition (\ref{qcond0}) means that $\eta(x,y)$ must actually be exact.
In the language of algebraic K-theory, this means that the symbol $\{ x,y \} = 0$ must be trivial in $K_2 (\C (\cC)) \otimes \Q$.
This led two of the authors of the present paper to propose the following criterion for quantizability \cite{abmodel}:
\be
\cC~ \text{is quantizable} \qquad \Longleftrightarrow \qquad \{ x,y \} \in K_2 (\C (\cC)) ~\text{is a torsion class}.
\label{quantcritK2}
\ee
Moreover, it is known that the above condition is equivalent to the existence of a decomposition \cite{Fernando}
\be
x \wedge y \; = \; \sum_i r_i z_i \wedge (1 - z_i) \qquad \qquad {\rm in~} \wedge^2 (\C (\cC)^*) \otimes \Q
\label{qcondzz}
\ee
for some $z_i \in \C (\cC)^*$ and $r_i \in \Q$. In turn, this also means that \cite{Fernando}:
\be
\text{$A(x,y)$ is tempered, {\it i.e.} the roots of all its face polynomials are roots of unity.}   \label{tempered}
\ee
By face polynomials we mean the following. We construct a Newton polygon corresponding to $A(x,y)=\sum_{i,j} c_{(i,j)} x^i y^j$,
and to each point $(i,j)$ of this polygon we associate the coefficient $c_{(i,j)}$.
Then, each face of the polygon consists of several points labeled, in order, by $k=0,1,2,\ldots$,
so that the corresponding monomial coefficients for a given face can be relabeled as $c_k=c_{(i,j)}$.
The face polynomial is defined, then, as $f(z)=\sum_k c_k z^k$.
The condition (\ref{tempered}) states that all roots of face polynomials $f(z)$,
constructed for all boundaries of the Newton polygon, must be roots of unity.
It is this latter condition that we shall consider below in order to test quantizability of the refined $A$-polynomials.

\subsubsection{Quantizability of refined $A$-polynomials}

We can now analyze under what conditions the refined $A$-polynomials which we found earlier are quantizable.
Since we already found the explicit form of such refined polynomials $A^{\textrm{ref}}(x,y;t)$,
in order to test quantizability we can simply apply the condition (\ref{tempered}).
To this end, we need to identify face polynomials associated to the Newton polygons of various knots discussed earlier.
We will analyze separately of the unknot, the trefoil knot, and more general $(2,2p+1)$ torus knots.
We will also discuss an interesting relation between Newton polygons in the refined and unrefined cases.

\begin{figure}[ht]
\begin{center}
\includegraphics[width=0.6\textwidth]{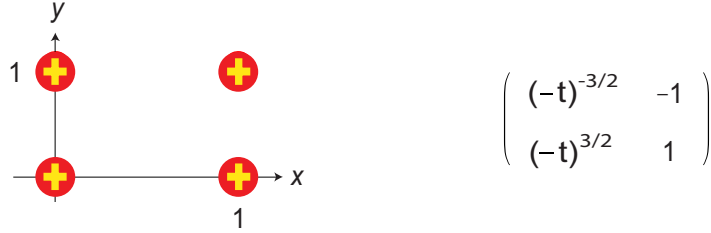}
%\hspace{1in}
%\includegraphics[totalheight=0.12\textwidth]{AtUnknotMatrix}
\end{center}
\caption{Newton polygon for the unknot (left).
Red circles denote monomials of the refined polynomial, and smaller yellow crosses denote monomials of the unrefined polynomial.
In this example both Newton polygons look the same, so that positions of all circles and crosses overlap.
More detailed structure of the refined $A$-polynomial is also shown in a matrix form on the right.
Note, that the role of rows and columns is exchanged in both pictures:
the monomial $c_{i,j}x^i y^j$ is put in the place $(i,j)$ in the Newton polygon,
and it corresponds to the entry $c_{i,j}$ in the $(i+1)^{\text{th}}$ row and in the $(j+1)^{\text{th}}$ column of the matrix on the right.}
\label{fig-unknotNewtonMatrix}
\end{figure}

To start with we consider the unknot; as we will see, general features of this example will also be present for more general torus knot.
The refined $A$-polynomial of the unknot has been found in (\ref{At-unknot}), and in the limit $t=-1$ it reduces as follows:
\be
A^{\text{ref}}_{\unknot} \; = \;
(1 + t^3 x) ( - t^{-3})^{1/2} - (1 - x) y \qquad \xrightarrow[\ t\to -1\ ]{}  \qquad  (1-x)(1-y).
\ee
We present Newton polygons for these two polynomials in Figure \ref{fig-unknotNewtonMatrix} (left).
Bigger red circles denote monomials of the refined polynomial, while smaller yellow crosses denote monomials of the unrefined polynomial.
In the unknot case, both Newton polygons coincide. From the matrix presentation of the refined $A$-polynomial
given in Figure \ref{fig-unknotNewtonMatrix} we can immediately write down face polynomials for all faces of the Newton polygon.
These face polynomials are summarized in table \ref{tab-unknot};
they are all linear in $z$ and their roots are $t^3$, $1$, $(-t)^{-3/2}$, and $-(-t)^{3/2}$.
If we insist that the criterion (\ref{tempered}) should be satisfied, we conclude that the deformation parameter $t$ should be a root of unity.

\begin{table}[ht]
\centering
\begin{tabular}{| c || c | }
\hline
\rule{0pt}{5mm}
face & face polynomial \\[3pt]
\hline
\rule{0pt}{5mm}
first column & $(-t)^{-3/2} z + (-t)^{3/2}$ \\[3pt]
last column & $z-1$  \\[3pt]
first row & $(-t)^{-3/2} - z$    \\[3pt]
last row & $(-t)^{3/2} + z$  \\[3pt]
\hline
\end{tabular}
\caption{Face polynomials for the unknot.}    \label{tab-unknot}
\end{table}

Next, let us consider the class of $(2,2p+1)$ torus knots,
for which the refined $A$-polynomials were found in section \ref{sec:saddle_point}.
When $t=-1$ these polynomials reduce, for general $p$, to the following form\footnote{Here and in what follows,
we multiply $A^{\text{ref}}_{T^{2,2p+1}}$ by a factor $(1+xt^3)^p$ to turn them into a nicer polynomial form.}
\be
A^{\text{ref}}_{T^{2,2p+1}}(x,y;t=-1) \; = \; (x - 1)^p (y - 1)  (y + x^{2p+1})^p \,.
\label{Atorus-nonrefined}
\ee
Interestingly, this form is closely related to the standard $A$-polynomials of $(2,2p+1)$ torus knots $(y-1)(y+x^{2p+1})$,
up to an extra factor $(x-1)^p$ and the power of $p$ in the last factor.
While it is desirable to understand the meaning of the form (\ref{Atorus-nonrefined}) better,
here we note that its Newton polygon is related in an interesting way to the Newton polygon of the refined $A$-polynomial.
Namely, Newton polygons of refined $A$-polynomials for $(2,2p+1)$ knots have hexagonal shape,
as shown in Figure \ref{fig-31NewtonMatrix} (left) for the trefoil, and in appendix \ref{sec:AppendixB} for other torus knots.
(In all of these figures we use the same conventions as in Figure \ref{fig-unknotNewtonMatrix}.)
Then, the Newton polygons for the unrefined $A$-polynomials (\ref{Atorus-nonrefined})
have the same overall shape, with only one small difference: in their center a ``rhomboidal''
collection of points of size $p\times p$ is absent, as clearly seen in the figures in appendix \ref{sec:AppendixB}.

\begin{figure}[ht]
\begin{center}
\includegraphics[width=0.86\textwidth]{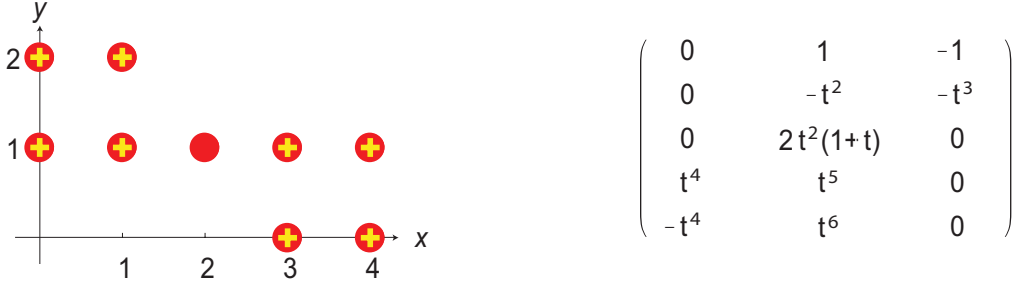}
%\hspace{1in}
%\includegraphics[width=0.2\textwidth]{At31matrix}
\end{center}
\caption{Newton polygons for the trefoil (left). Bigger red circles denote monomials of the refined polynomial, and smaller yellow crosses denote monomials of the unrefined polynomial. More detailed structure of the refined $A$-polynomial is also shown in matrix representation (right). All conventions are the same as in figure \protect\ref{fig-unknotNewtonMatrix}.}
\label{fig-31NewtonMatrix}
\end{figure}

Apart from Newton polygons, in Figure \ref{fig-31NewtonMatrix} (right)
and in appendix \ref{sec:AppendixB} we also present the matrix form of refined $A$-polynomials. From this
presentation it is not hard to read off the face polynomials, and conjecture their form for general $p$,
along each six faces of the hexagonal shape. These face polynomials are listed in table \ref{tab-torus}.
Interestingly, all these polynomials factor into linear factors in $z$,
and the constraint (\ref{tempered}) again leads to the aforementioned conclusion:
the deformation parameter $t$ must be a root of unity.

\begin{table}[ht]
\centering
\begin{tabular}{| c || c | }
\hline
\rule{0pt}{5mm}
face & face polynomial \\[3pt]
\hline
\rule{0pt}{5mm}
first column & $-t^{2p(p+1)} (z-1)^p$ \\[3pt]
last column & $(-1)^p (z+t^3)^p$  \\[3pt]
first row & $z-1$    \\[3pt]
last row & $-t^{2p(p+1)} (z-t^{2p})$  \\[3pt]
lower diagonal & $(-1)^p  t^{3p} (z - t^{2p+1})^p$  \\[3pt]
upper diagonal & $(-1)^{p+1} (z + t^{2p+2})^p$  \\[3pt]
\hline
\end{tabular}
\caption{Face polynomials for $(2,2p+1)$ torus knots.}   \label{tab-torus}
\end{table}

Finally, let us consider the $A$-polynomials for the trefoil knot in
%``DGR''
 grading conventions of \cite{DGR}, {\it cf.} \eqref{gradingtabl}.
Newton polygons for both refined and unrefined $A$-polynomials are shown in Figure \ref{fig-At31dgr-Newton}.
They are also of a hexagonal shape, though much bigger than the Newton polygon for the trefoil grading conventions of \cite{GS}
shown in Figure \ref{fig-31NewtonMatrix}.
The matrix form of the refined $A$-polynomials in grading conventions of \cite{DGR} is presented in Figure \ref{fig-At31dgr-matrix},
and the corresponding face polynomials are listed in table \ref{tab-dgr}.
Again, it is easy to see that the quantizability condition (\ref{tempered}) is satisfied only if $t$ is a root of unity.
Therefore, we conclude that, even though the explicit form of the refined $A$-polynomial $A^{\text{ref}} (x,y;t)$
may be sensitive to the choice of grading, the quantizability is independent of this choice.

\begin{table}[ht]
\centering
\begin{tabular}{| c || c | }
\hline
\rule{0pt}{5mm}
face & face polynomial with $a=q^2$ \\[3pt]
\hline
\rule{0pt}{5mm}
first column & $-t^3 (z - 1)^4 (t^2 z - 1)^2 (t^2 z + 1)^2$ \\[3pt]
last column & $-t^{23} (t z - 1)^2 (t z + 1)^2 (t^3 z + 1)^4$ \\[3pt]
first row & $-t^{34} (z - t) (t^5 z - 1)$ \\[3pt]
last row & $-(z - t^3) (t^3 z - 1)$ \\[3pt]
lower diagonal & $-t^3 (t^5 z - 1)^4$ \\[3pt]
upper diagonal & $-t^{11} (t^6 z + 1)^4$ \\[3pt]
\hline
\end{tabular}
\caption{Face polynomials for the trefoil knot in grading conventions of \cite{DGR}.}  \label{tab-dgr}
\end{table}

%***********************************************************************************
%***********************************************************************************

\section{Examples coming from refined BPS invariants}
\label{sec:Bmodel}

It is known that $\mathcal{N}=2$ gauge theories can be geometrically engineered by considering
type II string theory on appropriate toric Calabi-Yau manifolds \cite{engineering}.
The corresponding mirror manifolds have the form of a hypersurface
\be
z_1 z_2 \; = \; A(x,y)    \label{Amirror}
\ee
in $\C^2 \times \C^{*\, 2}$. Upon the suitable identification of parameters
the mirror curve $A(x,y)=0$ agrees with the Seiberg-Witten curve of the 5-dimensional theory on a circle,
and its appropriate scaling limit reproduces the ordinary Seiberg-Witten curve of the 4-dimensional theory.
While the knowledge of the Seiberg-Witten curve is equivalent to the knowledge of the prepotential of the gauge theory,
in fact much more is known to be true: entire series of gravitational corrections to the prepotential,
which are encoded in the Nekrasov partition function \cite{Nekrasov},
can be determined from topological string theory on the toric Calabi-Yau manifold.
More precisely, topological string computation reproduces Nekrasov partition functions
for 5-dimensional theories in the $\Omega$-background with $\epsilon_1 = - \epsilon_2$,
and the appropriate scaling limit reproduces Nekrasov partition function of the 4-dimensional theory \cite{Iqbal:2003zz}.
The corresponding BPS degeneracies are also encoded in ordinary topological string amplitudes.
For toric manifolds, these topological string amplitudes can be computed using the topological vertex \cite{AKMV}.

Presently, we are interested in refined BPS degeneracies.
It turns out that they are naturally encoded in Nekrasov partition functions in a nontrivial $\Omega$-background,
parametrized by arbitrary values of $q_1 = e^{\epsilon_1}$ and $q_2 = e^{-\epsilon_2}$.
Conjecturally, these amplitudes should be reproduced by a refined version of topological string theory on a toric Calabi-Yau manifold.
So far such genuine formulation of topological strings is not known.
It is however postulated that it should be reproduced via a combinatorial formalism of the refined topological vertex.
While there are various formulations of the refined vertex, see {\it e.g.} \cite{Awata:2005fa},
we use here a combinatorial definition in \cite{Iqbal:2007ii}.
In this case, the refined vertex amplitude can be written in terms of Macdonald polynomials $P_R$ (see section \ref{sec:macdonald})
and skew Schur functions $s_{R/S}$, and it reads
\bea
C_{PQR}(q_1,q_2) & = & \Big( \frac{q_2}{q_1} \Big)^{\frac{||Q||^2 + ||R||^2}{2}} q_1^{\frac{\kappa(Q)}{2}} P_{R^t}(q_1^{-\rho};q_2,q_1)  \\
& & \times
\sum_S \Big( \frac{q_2}{q_1} \Big)^{\frac{|S|+|P|-|Q|}{2}} s_{P^t /S} (q_1^{-\rho} q_2^{-R}) s_{Q/S}(q_1^{-R^t} q_2^{-\rho}).  \nonumber
\eea
In particular
\be
C_{\bullet\bullet R}(q_1^{-1},q_2^{-1}) = (-1)^{|R|} \Big( \frac{q_1}{q_2} \Big)^{|R|/2}  q_1^{||R^t||^2/2} \prod_{(i,j) \in R} \frac{1}{1 - q_1^{a(i,j)+1} q_2^{l(i,j)}}  , \label{C00R}
\ee
where $a(i,j)=R_i-j$ and $l(i,j)=R^t_j-i$ denote the arm-length and leg-length in a diagram $R$.

A few notational remarks are in order. Parameters $t$ and $q$ from the original formulation in \cite{Iqbal:2007ii} are replaced in the above expression respectively by $q_1$ and $q_2$, which agrees with conventions in \cite{GIKV,GS}. The non-refined limit corresponds to $q_1=q_2$. Similarly as in \cite{Iqbal:2007ii}, in computing open string amplitudes we use vertex amplitudes with inverse parameters $C_{PQR}(q_1^{-1},q_2^{-1})$. Moreover, following the convention in \cite{GIKV}, we denote symmetric representation $S^r$ by a Young diagram $R=(r)$. %consisting of one column with $r$ boxes.

The framing factor (associated to the amplitude already expressed in terms of $q_1^{-1}$ and $q_2^{-1}$) is defined as
\be
f_{R}(q_1,q_2) = (-1)^{|R|}\Big( \frac{q_1}{q_2} \Big)^{n(R)} q_2^{-\frac{\kappa_R}{2}},
\ee
where
\begin{eqnarray}
n(R):=\sum_i(i-1)R_i.
\end{eqnarray}
We typically allow general values of framing by including this factor raised to the power $-f$.
So, in particular, for the representation $R=(r)$ the framing factor takes the form
$$
f_{(r)}(q_1,q_2)^{-f} \; = \; (-1)^{f r} q_2^{\frac{f}{2} r(r-1)} \,.
$$

\bigskip
\begin{figure}[ht]
\centering
\includegraphics[width=4.0in]{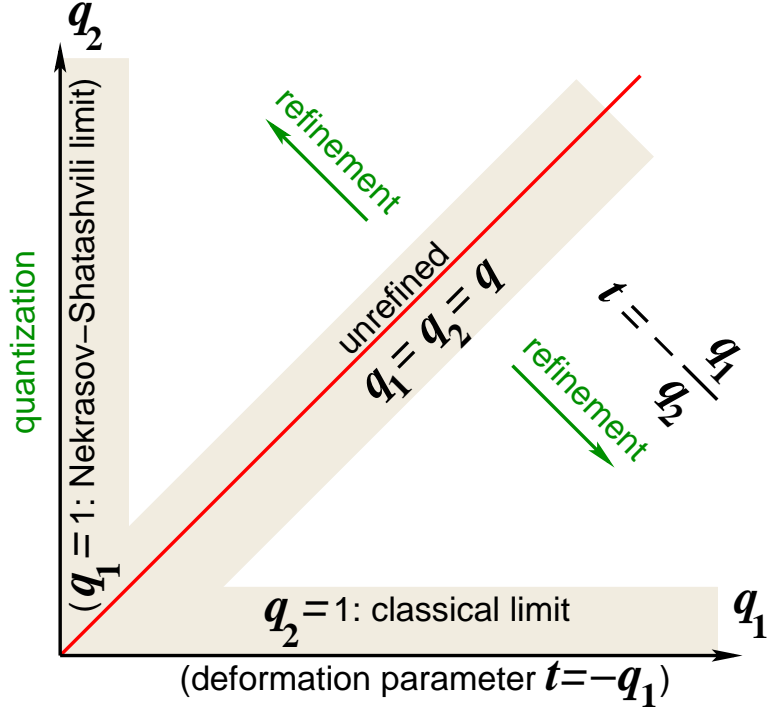}
\caption{With our choice of conventions, the parameter $q_2 = q = e^{\hbar}$ is responsible for
quantization, whereas $t = - \frac{q_1}{q_2}$ is the deformation parameter responsible for the refinement.
Hence, the ``classical limit'' corresponds to $q_2=1$, while $q_1=1$ is the so-called Nekrasov-Shatashvili limit.
Finally, $q_1 = q_2$ defines a locus in the space of parameters where refinement is turned off,
and the problem can be formulated in terms of ordinary topological strings.}
\label{fig:qqqt}
\end{figure}

Using the above conventions we are able to compute refined BPS state generating functions $Z_{\text{BPS}}^{\text{open}} (u,q_2,q_1)$
of toric manifolds in the presence of branes.
However, what we are really interested in are difference operators which annihilate these open refined BPS partition functions
(up to, possibly, some universal inhomogeneous term) and refined mirror curves which arise as classical limits of these difference operators.
As we demonstrate below, from the knowledge of $Z_{\text{BPS}}^{\text{open}} (x,q_2,q_1)$ we are able to determine such difference operators, at least for a class of toric manifolds which do not contain compact 4-cycles. It turns out that the role of parameters in these operators is as follows: $q_2$ is the quantum parameter which enters the commutation relation
\be
\hat{y}\hat{x} \; = \; q_2 \hat{x}\hat{y} \,,
\ee
and, therefore, it plays the role of $q$ in previous sections.\footnote{This will become clearer a little later,
when in \eqref{Zopen-BPS-C3} and \eqref{Zopen-BPS-coni} we start writing the generating functions $Z_{\text{BPS}}^{\text{open}} (x,q_2,q_1)$
in terms of products of the form $\prod_{i=1}^k (1 - q_1 q_2^{i-1})$, where $q_2$ plays the key role.}
Combining this with the fact that $q_1 = q_2 = q$ defines the unrefined limit,
{\it i.e.} the limit $t=-1$ in the notations of the previous sections,
we quickly conclude that, up to a slight redefinition $q_1 \mapsto q_1^2$ and $q_2 \mapsto q_2^2$,
the identification of the parameters is essentially as in \eqref{DGRvarchng}:
\be
\boxed{\phantom{\int^1}
q \; = \; q_2 \,, \qquad
t \; = \; - \frac{q_1}{q_2} \,. \phantom{\int^1}}
\label{qqqt}
\ee
Note, with this identification of the parameters, in the classical limit, {\it i.e.} when $q = q_2 = 1$, we can simply identify $t$ with $-q_1$.
Of course, with different choices of the preferred direction, {\it etc.}, the role of $q_1$ and $q_2$ could be different.

Having found the difference operator, or quantum curve, we can analyze asymptotic behavior of $Z_{\text{BPS}}^{\text{open}} (x,q_2,q_1)$ and determine the leading order amplitude $S_0(x,q_1)=\int \log y \frac{dx}{x}$. Using saddle point analysis we can also determine the form of the refined mirror curve $A^{\text{ref}} (x,y;q_1)$, and show that it agrees with the classical limit $q_2\to 1$ of the difference equation.

Before we present detailed results, let us stress that the form of the open string amplitude depends on various choices, such as preferred direction, framing, and the edge of a toric diagram to which the brane is attached. We will mainly discuss the most interesting case of branes on the external edges, along the preferred direction. This is certainly the most interesting case for $\C^3$ geometry, as in that case $Z_{\text{BPS}}^{\text{open}} (x,q_2,q_1)$ is a nontrivial function of both $q_1$ and $q_2$. For branes associated to either of the two non-preferred directions of $\C^3$, the amplitudes $Z_{\text{BPS}}^{\text{open}} (x,q_2,q_1)$ are given entirely in terms of either $q_1$ or $q_2$, and essentially do not differ from non-refined amplitudes \cite{Iqbal:2007ii}. Moreover, for branes along the preferred direction, an inhomogeneous term arises in the difference equation. Having found such inhomogeneous equation, we show that it implies that a homogeneous equation of a higher order is also satisfied. For more complicated geometries some simplifications occur for branes along non-preferred directions, which is also a reason why the branes along the preferred direction are most interesting in general. In all cases, we will consider branes with arbitrary value of framing.

%***********************************************************************************
%***********************************************************************************

\subsection{Refined quantum mirror curves}

In this section we derive the refined quantum curves relevant in the context of the open BPS state counting (or topological string theory).
They can be regarded as refined and quantum versions of mirror curves $A(x,y)=0$ which appear in the mirror geometry (\ref{Amirror}) for toric three-folds. Specifically, we consider open refined topological string amplitudes for branes along preferred directions, in general framing, for $\C^3$ and conifold geometries. It brings only some technical, but no conceptual challenges to generalize the computations below to the case of ``generalized conifolds'' ({\it i.e.} toric three-folds without compact 4-cycles). To find quantum curves in the present context, in this section we derive the refined open topological string amplitudes $Z_{\text{BPS}}^{\text{open}}$ and find difference equations which they satisfy. Then, in the next section we consider the classical limit $q_2 \to 1$ of the quantum curve, show that it is equivalent to the saddle point analysis of $Z_{\text{BPS}}^{\text{open}}$, and describe properties of the resulting refined mirror curves.

\subsubsection*{$\C^3$ or tetrahedron}

To start with, we consider the simplest toric geometry of $\C^3$, in arbitrary framing $f$.
Its ordinary mirror curve -- which is equivalent to the ``single tetrahedron'' curve from the perspective of hyperbolic geometry and knot theory -- takes the form \cite{abmodel}:
\be
A^{\text{ref}} (x,y; q_1=1) \; = \; 1 - y + x(-y)^f = 0 \,.    \label{mirrorC3f}
\ee
To find the refined and quantum generalization of this curve we consider
the refined BPS partition function in $\C^3$ geometry with a D-brane located along the preferred direction.
In arbitrary framing such an amplitude reads
\be
Z_{\text{BPS}}^{\text{open}} \; = \; \sum_R   f_R(q_1,q_2)^{-f} C_{\bullet\bullet R}(q_1^{-1},q_2^{-1}) s_R(x) \,.
\ee
Performing the summation we get
\be
Z_{\text{BPS}}^{\text{open}} \; = \; \sum_{k=0}^{\infty} (-1)^{(f+1)k} q_2^{\frac{f}{2}k(k-1)} \Big(x \frac{q_1}{\sqrt{q_2}}\Big)^k \prod_{i=1}^k \frac{1}{1 - q_1 q_2^{i-1}} \,.     \label{Zopen-BPS-C3}
\ee
For $f=0$ this result reduces, as it should, to the brane amplitude in $\C^3$ in phase III found in \cite{Iqbal:2007ii} (after identifying $x=-Q$).
If we write now
$$
Z_{\text{BPS}}^{\text{open}} = \sum_{k=0}^{\infty} a_k \,,
$$
we can easily determine the ratio of the consecutive coefficients, $\frac{a_{k+1}}{a_k}$, so that
\be
(1 - q_1 q_2^k) a_{k+1} \; = \; -x (-1)^f q^{fk} \frac{q_1}{\sqrt{q_2}} a_k \,.
\ee
Performing the summation over $k$ on both sides of this equation
we find that the open string amplitude satisfies the following inhomogeneous difference equation
\be
\Big(1 - \frac{q_1}{q_2} \hat{y} + \frac{q_1}{\sqrt{q_2}} \hat{x}(-\hat{y})^f   \Big) Z_{\text{BPS}}^{\text{open}} \; = \; 1 - \frac{q_1}{q_2} \,.
\label{AquantumC3inhom}
\ee
This is the refined and quantum version of the ordinary mirror curve (\ref{mirrorC3f}). Because the inhomogeneous term is independent of $x$, acting on it by $\hat{y}$ leaves it invariant. Therefore, if we multiply both sides of the above equation by $(1-\hat{y})$, the inhomogeneity on the right hand side vanishes, while the degree in $\hat{y}$ of the left hand side increases. Commuting all $\hat{y}$ operators to the right we obtain the following homogeneous equation of a higher degree
\be
\Big(1 - \Big(1 + \frac{q_1}{q_2}\Big) \hat{y} + \frac{q_1}{q_2}\hat{y}^2 +
\frac{q_1}{\sqrt{q_2}} \, \hat{x}(-\hat{y})^f + q_1\sqrt{q_2} \hat{x} (-\hat{y})^{f+1}
  \Big) Z_{\text{BPS}}^{\text{open}} \; = \; 0 \,.
\ee
We note the similarity of the factor $(1-\hat{y})$,
which brings the equation to the homogeneous form,
to the factor representing abelian flat connection in $SL(2,\C)$ Chern-Simons theory.

%***********************************************************************************
%***********************************************************************************

\subsubsection*{Conifold}

We now repeat the above calculation for the conifold, whose ordinary mirror curve in general framing takes the form \cite{abmodel}:
\be
A^{\text{ref}} (x,y;q_1=1) = 1 - y + x(-y)^f + Q x (-y)^{f+1} \; = \; 0 \,.    \label{mirrorConif}
\ee
To find the refined and quantum version of this curve, we again consider the brane located along the preferred direction of the conifold.
More precisely, we wish to consider the brane amplitude normalized by the closed string partition function.
Such an amplitude can be written as
\be
Z_{\text{BPS}}^{\text{open}} \; = \; \sum_{k=0}^{\infty} a_k = \sum_{R=(k)}  s_R(x) f_R(q_1,q_2)^{-f} \frac{b_R}{b_{\bullet}},   \label{Zopen-coni}
\ee
where
\bea
b_R & = & \sum_{P} (-Q)^{|P|} C_{P\bullet R}(q_1^{-1},q_2^{-1})  C_{P^t \bullet \bullet}(q_2^{-1},q_1^{-1})  \nonumber\\
& = & \Big( \frac{q_1}{q_2} \Big)^{||R||^2/2}  P_{R^t}(q_1^{\rho};q_2^{-1},q_1^{-1})
\prod_{i,j=1}^{\infty} (1 - Q q_1^{-i+1/2} q_2^{-j+1/2+R_i})
\eea
and $Q$ denotes the K{\"a}hler parameter of the conifold.
The normalization factor in this case is nothing but
\bea
Z_{\text{BPS}}^{\text{closed}} = b_{\bullet} & = & \sum_{P} (-Q)^{|P|} C_{P\bullet \bullet}(q_1^{-1},q_2^{-1})  C_{P^t \bullet \bullet}(q_2^{-1},q_1^{-1})  \nonumber  \\
& = & \prod_{i,j=1}^{\infty} (1 - Q q_1^{-i+1/2} q_2^{-j+1/2}),
\eea
and it follows that for representation $R=(k)$, {\it cf.} (\ref{C00R}):
$$
\frac{b_{(k)}}{b_{\bullet}} = C_{\bullet \bullet (k)}(q_1^{-1},q_2^{-1}) \prod_{j=1}^k \big(1 - Q q_1^{-1/2} q_2^{-j+k+1/2} \big)
= \Big(- \frac{q_1}{\sqrt{q_2}} \Big)^k \prod_{i=1}^k \frac{1 - Q q_1^{-1/2} q_2^{i-1/2}}{1 - q_1 q_2^{i-1}}.
$$
Collecting the above ingredients, we find the following structure of the open BPS state partition function
\be
Z_{\text{BPS}}^{\text{open}} \; = \; \sum_{k=0}^{\infty} a_k
= \sum_{k=0}^{\infty} (-1)^{(f+1)k} q_2^{\frac{f}{2}k(k-1)} \Big(x \frac{q_1}{\sqrt{q_2}}\Big)^k \prod_{i=1}^k \frac{1 - Q q_1^{-1/2} q_2^{i-1/2}}{1 - q_1 q_2^{i-1}} \,.   \label{Zopen-BPS-coni}
\ee
We also find that the consecutive terms $a_k$ are related as
\be
(1 - q_1 q_2^k) a_{k+1} \; = \; (-1)^{f+1} q^{f k} \frac{q_1}{\sqrt{q_2}} x (1 - Q q_1^{-1/2} q_2^{k+1/2} ) a_k \,.
\ee
Summing both sides over $k$ gives rise to the inhomogeneous equation
\be
\Big(1 - \frac{q_1}{q_2} \hat{y} + \frac{q_1}{\sqrt{q_2}} \hat{x}(-\hat{y})^f  + Q q_1^{1/2} \hat{x}(-\hat{y})^{f+1} \Big) Z_{\text{BPS}}^{\text{open}} \; = \; 1 - \frac{q_1}{q_2} \,.   \label{AquantumConiinhom}
\ee
This is the refined and quantum version of the conifold mirror curve (\ref{mirrorConif}).
We can also bring this equation to the homogeneous form at the expense of increasing its degree in $\hat{y}$,
by multiplying both sides with $(1-\hat{y})$. Commuting all $\hat{y}$ operators to the right gives
\be
\Big(1 - \Big(1 + \frac{q_1}{q_2}\Big) \hat{y} + \frac{q_1}{q_2}\hat{y}^2 +
\frac{q_1}{\sqrt{q_2}} \hat{x}(-\hat{y})^f + \Big(Q\sqrt{q_1} +  q_1\sqrt{q_2} \Big) \hat{x} (-\hat{y})^{f+1}
+ Q\sqrt{q_1} q_2 \hat{x} (-\hat{y})^{f+2}
  \Big) Z_{\text{BPS}}^{\text{open}} = 0.
\ee

%***********************************************************************************
%***********************************************************************************

\subsection{Refined mirror curves, $S_0(u,t)$ and quantizability}

Once we found the refined quantum curves, we can easily determine the classical refined mirror curves.
Namely, much as in the knot theory examples, we can find them in two ways. First, they arise as $q_2\to 1$ limit of the quantum curves. Secondly, they can be determined by the saddle point analysis. As we will see momentarily, both methods consistently give the same result. The limit $q_2\to 1$ of the $\C^3$ quantum curve (\ref{AquantumC3inhom}) leads to the following refined mirror curve
\be
A^{\text{ref}} (x,y;q_1) \; = \; 1 - q_1 y + q_1 x (-y)^f \,.  \label{Aq1-C3}
\ee
Similarly, the refined mirror curve for the conifold can be obtained form (\ref{AquantumConiinhom}) and takes the form
\be
A^{\text{ref}} (x,y;q_1) \; = \; 1 - q_1 y + q_1 x (-y)^f + Q \sqrt{q_1} \, x (-y)^{f+1} \,.    \label{Aq1-coni}
\ee
By looking at the form of this refined $A$-polynomial, a careful reader will recognize
a close relation with the example of the unknot discussed in section \ref{sec:knots}.
Indeed, a simple change of variables
\be
\begin{array}{c@{\quad}c@{\quad}c}
\underline{\text{unknot}} & & \underline{\text{conifold}} \\[.1cm]
x & \mapsto & (-1)^{f+1}t^{-2}xy^f  \\[.1cm]
y & \mapsto & (-t)^{-\frac{1}{2}}y
\end{array}
\label{unknotconi}
\ee
relates refined $A$-polynomials \eqref{Atunknot2} and \eqref{Aq1-coni},
provided that we identify $q_1 = -t$ as in \eqref{qqqt}
that the K\"ahler parameter of the conifold is tuned to a special value:
\be
Q \; = \; it^{-\frac{3}{2}} \,.
\ee
Note, the transformation \eqref{unknotconi}
preserves the holomorphic symplectic form $\Omega = \frac{i}{\hbar} \frac{dx}{x} \wedge \frac{dy}{y}$
on $\C^* \times \C^*$ relevant to quantization.

To confirm that the same refined curves \eqref{Aq1-C3} and \eqref{Aq1-coni}
arise from the saddle point analysis we follow the strategy of section \ref{sec:saddle_point},
where we discussed analogous curves coming from knot theory examples.
Specifically, much like in (\ref{Pn-saddle}), we wish to approximate the BPS partition functions
$Z_{\text{BPS}}^{\text{open}}$ in (\ref{Zopen-BPS-C3}) and (\ref{Zopen-BPS-coni}) by the integral
\be
Z_{\text{BPS}}^{\text{open}} \; \sim \;
\int dz\; e^{\frac{1}{\hbar}\left(V(z,x;q_1)+{\cal
						 O}(\hbar)\right)} \,.     \label{Zopen-BPS-saddle}
\ee
As the computation is analogous for $\C^3$ and the conifold, we present it just in the latter case; the result for $\C^3$ is easily obtained by setting $Q=0$. Therefore, using the conifold amplitude in the form (\ref{Zopen-BPS-coni}) and the expansion of the quantum dilogarithm given in (\ref{q-dlog_exp}), we find
\bea
V(z,x;q_1) & = & (\log z)\big(\pi i (f+1) + \log x q_1   \big) + \frac{f}{2} (\log z)^2 \nonumber \\
& & + \, \textrm{Li}_2(q_1 z) - \textrm{Li}_2(Q q_1^{-1/2} z) - \textrm{Li}_2(q_1) + \textrm{Li}_2(Q q_1^{-1/2}) \,. \nonumber
\eea
The saddle point equation $\partial_z V |_{z=z_0}=0$, written in the exponential form $1=\exp (z \partial_z V |_{z=z_0})$, leads to the condition
\be
1 \; = \; -x q_1 (-z_0)^f  \frac{1 - Q z_0 q_1^{-1/2}}{1 - z_0 q_1} \,.    \label{Coni-saddle1}
\ee
On the other hand, from the relation between $y$ and $S_0\equiv V$ we get
\be
y = e^{S_0'} = e^{x\partial_x V} = z_0 \,.     \label{Coni-saddle2}
\ee
Eliminating $z_0$ from the above two equations we find
\be
A^{\text{ref}} (x,y;q_1) = 1 - q_1 y + q_1 x (-y)^f + Q \sqrt{q_1} \, x (-y)^{f+1}.
\ee
This result indeed reproduces the refined $A$-polynomial for the conifold (\ref{Aq1-coni}),
and setting $Q=0$ gives the refined curve for the $\C^3$ geometry obtained in (\ref{Aq1-C3}).

We also note that the two equations (\ref{Coni-saddle1}) and (\ref{Coni-saddle2}) can be rewritten in the form
\be
\left\{\begin{array}{l} x=x(z_0) = -(-z_0)^{-f} \frac{q_1^{-1}-z_0}{1 - Q z_0 q_1^{-1/2}} \,, \\
y=y(z_0) = z_0  \,. \end{array} \right.
\label{Conif-param}
\ee
This is a parametric form of the refined mirror curve, and for $q_1=1$
it agrees with the parametrization considered in \cite{abmodel},
where such a representation was used to find the (unrefined) quantum curve.
Nonetheless, the framework of \cite{abmodel} cannot be applied to the present case
since it works only if we start from the unrefined classical curve, and upon quantization introduce a single quantum parameter $q=q_1=q_2$.
On the other hand, for the purposes of the present paper, we would like to keep the \emph{classical} parameter $q_1$ fixed in (\ref{Conif-param}),
and perform quantization with respect to another (\emph{quantum}) parameter $q_2$; this is a different quantization problem
compared to the one considered in \cite{abmodel}.

We also note, that in fact it is possible to construct a deformed classical curve, whose associated closed string amplitudes (computed by the ordinary Eynard-Orantin topological recursion \cite{eyn-or}) would coincide with refined topological string amplitudes. Such curves have been constructed in \cite{Sulkowski:2010ux,Sulkowski:2011qs,Eynard:2011vs} for toric manifolds without compact 4-cycles, and in that case a classical deformation parameter was identified as $\beta=-\epsilon_1/\epsilon_2$. For example, for $\C^3$ such a $\beta$-deformed mirror curve can be written as \cite{Eynard:2011vs}:
\be
A(x,y;\beta) \; = \; x^2 y - (1+y)^{1+\beta} \,,
\ee
so that for $\beta=1$ it reduces to the mirror curve of $\C^3$ in framing $\frac{1}{2}$.
However, this curve is completely different than the refined curve which we found now in (\ref{Aq1-C3}). It is desirable to understand the relation between these two deformed classical curves, and between the two corresponding quantization schemes.

%*********************************************

Having found the refined mirror curves, we can finally determine the leading free energy,
given by the integral $S_0(x,q_1)=\int\log y\frac{dx}{x}$ on an appropriate curve.
For $\C^3$ we integrate over the curve (\ref{Aq1-C3}). The explicit expression in terms of the variable $x$
can be written in framing $f=0$, so that
\be
S_0 (x,q_1; f=0) \; = \; - (\log x) ( \log q_1) - \textrm{Li}_2 (-q_1 x) \,.
\label{S0C3}
\ee
In this case, we also clearly see a singularity at $q_1\to 0$. For general framing we can express the exact answer in terms of $y$ variable
\be
S_0(x,q_1; f) \; = \; -\frac{f}{2}  ( \log y)^2 + (\log y) (\log(1-y q_1)) + \textrm{Li}_2(y q_1) \,.
\ee

In our next example, namely the conifold, the free energy $S_0(x,q_1)$ is a $Q$-deformation of the $\C^3$ result.
An explicit expression in terms of the variable $x$ can again be given in framing $f=0$, so that
\be
S_0(x,q_1; f=0) \; = \; - (\log x) ( \log q_1) - \textrm{Li}_2 (-q_1 x) + \textrm{Li}_2 (-\frac{Q x}{\sqrt{q_1}}) \,.
\label{S0conifold}
\ee
In this case we also clearly see a singularity as $q_1\to 0$.
For general framing we can express the exact answer in terms of the variable $y$:
\be
S_0 (x,q_1; f) \; = \; -\frac{f}{2}  ( \log y)^2 + (\log y) \Big(\log\frac{1-y q_1}{1 - \frac{y Q}{\sqrt{q_1}}}\Big) + \textrm{Li}_2(y q_1)
- \textrm{Li}_2 (-\frac{Q y}{\sqrt{q_1}}) \,.
\ee

%***********************************************************************************
%***********************************************************************************

%\subsection{Quantizability}

\begin{figure}[ht]
\centering
\includegraphics[width=0.25\textwidth]{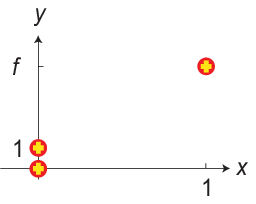} \hspace{1.5in}
\includegraphics[width=0.25\textwidth]{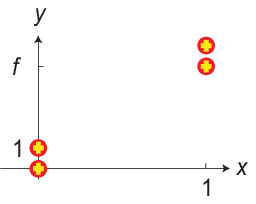}
\caption{Newton polygons for refined (black dots) and unrefined (white dots) $A$-polynomials, for $\C^3$ (left) and conifold (right).}
\label{fig:NewtonC3Coni}
\end{figure}

Finally, let us comment on the quantizability of the refined classical curves which we found in this section.
These curves are, in a sense, more general than the refined $A$-polynomials for knots analysed in section \ref{sec:knots}.
This is so because, apart from the deformation parameter $q_1$, they in general depend on K{\"a}hler parameters $Q_i$.
Therefore, the quantizability conditions should impose some constraints on $q_1$ as well as on $Q_i$, and this is indeed what happens.

Much as in the analysis of the refined $A$-polynomials for knots,
we can consider the quantization condition (\ref{tempered}).
To this end, we need to construct face polynomials $f(z)$ for all faces of Newton polygons for BPS quantum curves.
Such Newton polygons, for both $\C^3$ and the conifold, are shown in figure \ref{fig:NewtonC3Coni}.
These polygons coincide in refined and unrefined cases.
In fact, they are quite simple, as each face contains only two lattice points,
so that all face polynomials $f(z)$ are linear in $z$. For example, for $\C^3$ they look like
\be
z-q_1 \,,\qquad z+(-1)^f q_1 \,,\qquad z - (-1)^f \,.
\ee
Therefore, in the refined $\C^3$ case, we immediately conclude that meeting the constraint (\ref{tempered})
requires $q_1$ to be a root of unity, much as in the knot theory examples, where the deformation parameter $t$ was forced to be a root of unity.
In the conifold case, we find that both $q_1$ and $Q$ must be roots of unity.
This imposes an interesting condition on the K{\"a}hler parameter:
$\log Q$ must be a pure imaginary rational number.
We expect that similar conditions arise for refined mirror curves of general manifolds,
{\it i.e.} $q_1$ and all K{\"a}hler parameters $Q_i$ in general are required to be roots of unity.

%***********************************************************************************
%***********************************************************************************

\subsection{Refinement as a twisted mass parameter}
%\subsection{Surface operators and refinement of the Nekrasov-Shatashvili limit}

Now, let us discuss the physical interpretation of the general setup considered in this section
and the corresponding interpretation of the classical action $S_0 (u,t)$.

As we already reviewed in the beginning of this section, compactification of type II
string theory on a toric Calabi-Yau 3-fold $X$ is a natural arena for a ``geometric engineering''
of $\cN=2$ supersymmetric gauge theories in the remaining four space-time dimensions.
In this setup, BPS states of the ``effective'' four-dimensional gauge theory
are simply the closed BPS states from the vantage point of the Calabi-Yau 3-fold $X$.
Moreover, counting {\it refined} closed BPS states means that, from the viewpoint of the 4d $\cN=2$ theory,
we keep track of their spin $j_3$.

Much of our interest comes from counting {\it open} refined BPS states
in a slight generalization of this setup, with a D4-brane added:
\be
\begin{matrix}
{\mbox{\rm space-time:}} & \qquad & \R^4 & \times & X \\
& \qquad & \cup &  & \cup \\
{\mbox{\rm D4-brane:}} & \qquad & \R^2 & \times & L
\end{matrix}
\label{surfeng}
\ee
Since the extra D4-brane here is supported on a special Lagrangian submanifold $L \subset X$,
it preserves half of the supersymmetry and, as explained in \cite{DGH},
yields a ``geometric engineering'' of a half-BPS surface operator in $\cN=2$ gauge theory on $\R^4$
(see \cite{Ramified,Gknothom,Rigid} for a purely field theoretic definition of surface operators).
In a special case when $X$ is the conifold and $L_K$ is the Lagrangian submanifold associated to a knot $K$,
the system \eqref{surfeng} becomes identical to the physical setup \eqref{theoryB} discussed in section \ref{sec:knots},
after compactification on a circle from M-theory to type IIA string theory.

{}From the viewpoint of the $\cN=2$ gauge theory on $\R^4$,
the generating function of open BPS invariants $Z_{\text{BPS}}^{\text{open}}$
that plays a central role in this section can be interpreted as the instanton partition function
of the gauge theory in the presence of a surface operator \cite{DGH,Braverman,AGGTV,AFKMY},
much like its ``closed'' counterpart $Z_{\text{BPS}}^{\text{closed}}$
is the K-theory version of the equivariant instanton counting on $\R^4$ without surface operators \cite{Nekrasov,NYlectures}.
Let $SO(2)_1 \times SO(2)_2$ be the rotation symmetry of $\R^4$ preserved by the surface operator in \eqref{surfeng},
such that $SO(2)_2$ is the rotation symmetry along the $\R^2 \subset \R^4$ and $SO(2)_1$
is the rotation symmetry of its normal bundle.
Note, from the viewpoint of the Euclidean $\cN=(2,2)$ theory on the surface operator,
$SO(2)_2$ is a rotation symmetry (part of the two-dimensional Poincar\'e group),
while $SO(2)_1$ is a global R-symmetry.
In fact, both symmetries in this brane system were already discussed in section \ref{sec:refinedbraid},
where we gave them a name $U(1)_P$ and $U(1)_F$, respectively.
In particular, the global R-symmetry is:
\be
U(1)_F \; \cong \; SO(2)_1
\label{U1Fso2}
\ee

If, as usual, we denote by $\epsilon_1$ and $\epsilon_2$ the equivariant parameters
for $SO(2)_1$ and $SO(2)_2$, respectively,
then the generating functions $Z_{\text{BPS}}^{\text{closed}}$ and $Z_{\text{BPS}}^{\text{open}}$
are known to have the following form, see {\it e.g.} \cite{KreflWalcher,KlemmHuang}:
\bea
Z_{\text{BPS}}^{\text{closed}} (\epsilon_1, \epsilon_2)
& = & \exp \Bigl(\, \frac{1}{\epsilon_1 \epsilon_2} \cF (\epsilon_1, \epsilon_2) \, \Bigr) \label{ZclosedW} \\
& = & \exp \Bigl(\, \frac{1}{\epsilon_1 \epsilon_2} \cF^{(0)}
+ \frac{\epsilon_1 + \epsilon_2}{\epsilon_1 \epsilon_2} \cF^{(1)} + \ldots \, \Bigr) \nonumber
\eea
and
\be
Z_{\text{BPS}}^{\text{open}} (\epsilon_1, \epsilon_2)
\; = \; \exp \Bigl(\, \frac{1}{\epsilon_2} \widetilde{\cal W} + \ldots  \, \Bigr)
\label{ZopenW}
\ee
where $\cF^{(0)}$ is the prepotential of the four-dimensional $\cN=2$ theory
and $\widetilde{\cal W}$ is the twisted superpotential of the two-dimensional $\cN=(2,2)$ theory on the surface operator.
In particular, as explained in \cite{AGGTV} (see also \cite{AwataYamada,TakiMaruyoshi,AFKMY}),
the twisted superpotential is given by the integral
of the Seiberg-Witten differential $v du = \log y \frac{dx}{x}$ over a path:
\be
\widetilde{\cal W} (u) \; = \; \int^u v du
\ee
on the curve $A(x,y)=0$.

Now, let us explain how our discussion in this paper compares to the Nekrasov-Shatashvili limit
and its relation to quantum integrable systems.
There are some similarities and some differences, and both are important.
First, the limit considered in \cite{NS} in our notations is
$\epsilon_1 = 0$ with $\epsilon_2$ playing the role of the quantization parameter (in the quantum integrable system).
This is very similar to the limit \eqref{reflimit} and the discussion in this section,
where $\hbar = - \epsilon_2$ also plays the role of the quantization parameter
and the only essential difference is that $\epsilon_1$ (equivalently, $q_1 = e^{\epsilon_1}$)
is allowed to take any finite values, so that for small $\hbar$ we have $t = - \frac{q_1}{q_2} \sim - q_1$.
In this respect, what we consider can be viewed as a refinement of the Nekrasov-Shatashvili limit, {\it cf.} Figure \ref{fig:qqqt}.

There is an important difference, however, which has to do with the fact that \cite{NS}
consider the partition function of the 4d gauge theory or, from the vantage point of the Calabi-Yau 3-fold $X$,
the generating function of {\it closed} BPS invariants \eqref{ZclosedW}.
In particular, the Yang-Yang function of the quantum integrable system
is identified with the following limit of the instanton partition function
in the absence of any surface operators \cite{NS}:
\be
\lim_{\epsilon_1 \to 0} \, \epsilon_1 \, \log Z_{\text{BPS}}^{\text{closed}} (\epsilon_1, \epsilon_2)
\; = \; \frac{1}{\epsilon_2} \cF^{(0)} + \cF^{(1)} + \cO (\epsilon_2)
\label{YangYang}
\ee
This expansion should not be confused with a similar-looking expansion in \eqref{ZopenW},
which describes the behavior of the {\it open} BPS partition function or,
from the gauge theory viewpoint, represents the contribution of a surface operator.
In particular, the leading term $\cF^{(0)}$ in \eqref{YangYang} depends only on the closed string moduli,
whereas a similar leading term $\widetilde{\cal W}$ in \eqref{ZopenW} depends on both open and closed string moduli.
Thus, the variable $u$ in our discussion or, equivalently, $x = e^u$ is an open string modulus.

Keeping these remarks in mind, we can express the refined volume conjecture (\ref{VCparamref}b)
as a statement that the free energy $\log Z_{\text{BPS}}^{\text{open}} (q_1, q_2)$
has a first-order pole in the limit \eqref{reflimit}:
\be
\log Z_{\text{BPS}}^{\text{open}} (\epsilon_1, \epsilon_2)
\; = \; \frac{S_0 (u,t)}{\hbar} + \ldots
\qquad \text{as~~} \boxed{ \phantom{I} q_2 = e^{\hbar} \to 1
\text{~and~} \frac{q_1}{q_2} = - t = \text{finite}~}~ \,,
\label{logZopenSt}
\ee
where we used \eqref{qqqt}.
This, at the same time, is both {\it open} and {\it refined} generalization of \eqref{YangYang}.
In particular, in the special case $t = -1$ (or, equivalently, $q_1 = 1$)
we recover \eqref{ZopenW}, with the twisted superpotential
\be
\widetilde{\cal W} (u) \; = \; S_0 (u,-1) \,.
\ee
Therefore, we conclude that $S_0 (u,t)$ which appears in the refined volume conjecture (\ref{VCparamref}b)
is a ``refinement'' of the twisted superpotential in the $\cN=(2,2)$ surface operator theory\footnote{To be more precise,
it is a three-dimensional $\cN=2$ theory on a circle that is relevant to the K-theoretic version of the vortex partition function \cite{DGH}.}
with a twisted mass for the global symmetry
$U(1)_F$ (= rotation in the plane orthogonal to the surface operator), {\it cf.} \eqref{ZrefdefP}:
\be
\tilde m_F \; = \; \log (-t) \,.
\label{mftrel}
\ee
As a simple example, let us consider a two-dimensional $\cN=(2,2)$ theory with a tower of Kaluza-Klein states
obtained by reducing a three-dimensional $\cN=2$ chiral multiplet on a circle.
The twisted superpotential of this theory has the familiar form $\widetilde{\cal W} = \textrm{Li}_2(-t x)$,
where we assumed that the 3d chiral multiplet on $\R^2 \times \S^1$ has charge $+1$
with respect to both $U(1)_F$ as well as the global $U(1)$ symmetry with twisted mass $\log x$.
Therefore, in the twisted superpotential \eqref{S0conifold} we can recognize a contribution of two chiral multiplets.
Similarly, in the expression \eqref{S0C3} for the tetrahedron (or $\C^3$) we recognize contribution of a single chiral multiplet.
In general, the number of chiral multiplets is equal to the number of dilogarithms in $S_0 (u,t)$,
and the charges of chiral multiplets are simply the powers of $t$ and $x$ in the arguments of these dilogarithms.
It is easy to recognize such contributions {\it e.g.} in \eqref{Pn-saddle}.

To summarize, each chiral multiplet in the D4-brane theory \eqref{surfeng}
(or, to be more precise, in the five-brane theory on a circle, {\it cf.} \eqref{theoryB})
contributes to the twisted chiral superpotential a dilogarithm term:
$$
\text{tetrahedron}~\Delta
\qquad \leftrightarrow \qquad
\text{chiral}~\phi
\qquad \leftrightarrow \qquad
\begin{array}{l}
\text{twisted superpotential} \\[.1cm]
S_0 (\Delta; \vec u, t) = \textrm{Li}_2( e^{\vec n \cdot \vec u + n_F \, \tilde m_F} )
\end{array}
$$
where $n_F$ is the charge of the chiral multiplet under the global R-symmetry \eqref{U1Fso2}
and $\vec n$ denotes the charges of the chiral multiplet under all other flavor symmetries (with twisted mass parameters $\vec u$).
Here, we also identified the example of a single chiral multiplet (or $\C^3$) discussed in the present section with
a single tetrahedron in examples coming from 3-manifolds, {\it cf.} \cite{DGH,DGSdual,abmodel,DGG,CCV}.

\begin{table}[h]
\centering
\begin{tabular}{|@{$\Bigm|$}c|c@{$\Bigm|$}|}
\hline
\rule{0pt}{5mm}
\textbf{Model} & values of $n_F^i$  \\[3pt]
\hline
\hline
\rule{0pt}{5mm}
unknot & $0$, $3$, $3$   \\[3pt]
%\hline
%\rule{0pt}{5mm}
%$T^{2,2p+1}$ & $0$, $0$, $2$, $3$, $2$, $3$, $0$, $1$, $1$    \\[3pt]
\hline
\rule{0pt}{5mm}
tetrahedron & $1$   \\[3pt]
\hline
\rule{0pt}{5mm}
conifold & $1$, $- \frac{1}{2}$  \\[3pt]
\hline
\end{tabular}
\caption{Values of $n_F^i$ in prominent examples. \label{tab-nF} }
\end{table}

More generally, one can consider a three-dimensional $\cN=2$ theory $\mathcal{T}$
that contains $N_f$ chiral multiplets $\phi_i$, $i=1, \ldots, N_f$
with charges $n^i_a$ under global flavor symmetries $U(1)_a$, $a = 1, \ldots, N$.
This could be either a low-dimensional effective field theory in brane systems \eqref{theoryB} and \eqref{surfeng},
or a three-dimensional $\cN=2$ theory $T_M$ associated to a triangulation of a 3-manifold $M$,
\be
M \; = \; \bigcup_i \, \Delta_i \,.
\ee
In either case, the refinement is achieved by assigning each chiral multiplet $\phi_i$ (resp. each tetrahedron $\Delta_i$)
one extra charge $n_F^i$ that describes how $\phi_i$ transforms under the R-symmetry $U(1)_F$,
which may already be a part of $\prod_a U(1)_a$.
Then, using \eqref{mftrel}, we conclude that passing from unrefined theory
to refined theory has the effect of introducing the $t$-dependence in
the classical ``volume functional'' (= twisted superpotential) via a simple rule:
\be
\widetilde{\cal W} (x_i) = \sum_i \textrm{Li}_2 \big( \prod_a x_a^{n_a^i} \big)
\quad \leadsto \quad
S_0 (x_i, t) = \sum_i \; \textrm{Li}_2 \; \Big( (-t)^{n_F^i} \prod_a x_a^{n_a^i} \Big)
\ee
modulo logarithmic ambiguities that depend on choices of framing, polarization, {\it etc.}
In other words, at least in such models,
the essential information about the refinement is contained in a set of charges $\{ n_F^i \}$
that need to be assigned to chiral multiplets or, in the language of 3-manifolds, to tetrahedra $\Delta_i$.
While many examples are considered in the present paper (see Table \ref{tab-nF}),
a systematic rule for assigning $\{ n_F^i \}$ will be presented in the follow-up work.

%***********************************************************************************
%***********************************************************************************
%***********************************************************************************
%***********************************************************************************

%%%%%%%%%%%%%%%%%%%%%%%%%%%%%%%%%%%%%%%%%%%%%%%%%%%%%%%%%%%%%%%%%%%%%%%%%%%

\acknowledgments{We thank H.-J.~Chung and R.H.~Dijkgraaf
for useful discussions during the early stages of this work.
We thank M. Aganagic, A. Iqbal, D. Krefl, and  Sh. Shakirov for discussions and comments.
The authors would also like to thank the following institutions for their hospitality:
California Institute of Technology (H.F.),
the Banff International Research Station (H.F., P.S.),
and the Simons Center for Geometry and Physics (H.F., S.G., P.S.).
The work of H.F. is supported by the Grant-in-Aid for Young Scientists
(B) [\# 21740179] from the Japan Ministry of Education, Culture, Sports,
Science and Technology, and the Grant-in-Aid for Nagoya University
Global COE Program, ``Quest for Fundamental Principles in the Universe:
from Particles to the Solar System and the Cosmos.''
The work of S.G. is supported in part by DOE Grant DE-FG03-92-ER40701FG-02 and in part by NSF Grant PHY-0757647.
The research of P.S. is supported by the DOE grant DE-FG03-92-ER40701FG-02,
the European Commission under the Marie-Curie International Outgoing Fellowship Programme, and the Foundation for Polish Science.
Opinions and conclusions expressed here are those of the authors and do not necessarily reflect the views of funding agencies.}

%%%%%%%%%%%%%%%%%%%%%%%%%%%%%%%%%%%%%%%%%%%%%%%%%%%%%%%%%%%%%%%%%%%%%%%%%%%%%%%%%%%%%%%%
\newpage
\appendix
\section{Miscellaneous results on knot invariants}
\label{sec:knot_app}

\subsection{Macdonald polynomials}
\label{sec:macdonald}
Let $p_n$ be the power sum
symmetric function in $x=(x_1,x_2,\ldots)$:
\begin{eqnarray}
p_n:=\sum_{i=1}^{\infty}x_i^n.
\end{eqnarray}
For any symmetric functions $f$ and $g$, a scalar product is
introduced as:
\begin{eqnarray}
\langle f(p),g(p)\rangle_{q_1,q_2}:=f(p^*)g(p)|_{\text constant\;\;part},
\quad p_n^*:=n\frac{1-q_1^n}{1-q_2^n}\frac{\partial}{\partial p_n}.
\end{eqnarray}
with $p=(p_1,p_2, \ldots )$ and $p^* = (p_1^*, p_2^*, \ldots )$.
The Macdonald function $P_R(x;q_1,q_2)$ is uniquely specified by the
following orthogonality condition and normalization:
\begin{eqnarray}
&&\langle P_R(x;q_1,q_2),P_{Q}(x;q_1,q_2)\rangle_{q_1,q_2}=0, \quad (R\ne
 Q)\\
&&P_R(x;q_1,q_2)=\sum_{Q\le R}u_{RQ}(q_1,q_2)m_R(x),\quad u_{RR}(q_1,q_2)=1
\end{eqnarray}
where $m_R(x)$ is the monomial symmetric function:
\begin{eqnarray}
m_R(x):=\sum_{\sigma}x_1^{R_{\sigma(1)}}x_2^{R_{\sigma(2)}}\cdots,
\end{eqnarray}
and $u_{RQ}(q_1,q_2)\in\mathbb{Q}(q_1,q_2)$.
The dominance partial ordering $R>Q$ denotes the condition: $|R|=|Q|$
and $R_1+\cdots +R_i\ge Q_1+\cdots +Q_i$ for all $i$.
The scalar product of the Macdonald functions is given by
\begin{eqnarray}
\langle P_{R}(x;q_1,q_2),P_{Q}(x;q_1,q_2)\rangle_{q_1,q_2}=\prod_{i,j}
\frac{1-q_1^{R_i-j+1}q_2^{R_j^{t}-i}}{1-q_1^{R_i-j}q_2^{R_j^{t}-i+1}}.
\end{eqnarray}

\noindent{\underline{Explicit form of the Macdonald function $P_R(x;q_1,q_2)$}}\\
Using the definitions above, the Macdonald functions are determined explicitly.
Up to $r\le 3$, the Macdonald functions
$P_{S^{r+\ell,r-\ell}}(x;q_1,q_2)$ and $P_{\Lambda^{r+\ell,r-\ell}}(x;q_1,q_2)$ are
\begin{eqnarray}
P_{S^1}(x,q_1,q_2)&=&p_1,
\nonumber \\
P_{S^2}(x,q_1,q_2)&=&\frac{(1+q_1)(1-q_2)}{1-q_1q_2}\frac{p_1^2}{2}+\frac{(1-q_1)(1+q_2)}{1-q_1q_2}\frac{p_2}{2},\quad
P_{S^{1,1}}(x,q_1,q_2)=\frac{p_1^2}{2}-\frac{p_2}{2},
\nonumber \\
P_{S^3}(x;q_1,q_2)
&=&\frac{(1 + q_1) (1 - q_1^3) (1 - q_2)^2}{(1-q_1)(1-q_1q_2)(1-q_1^2q_2)}\frac{p_1^3}{6}+\frac{(1-q_1^3)(1-q_2^2)}{(1-q_1q_2)(1-q_1^2q_2)}\frac{p_1p_2}{2}
\nonumber \\
&&
+\frac{(1-q_1)(1-q_1^2)(1-q_2^3)}{(1-q_2)(1-q_1q_2)(1-q_1^2q_2)}\frac{p_3}{3},
\nonumber \\
P_{S^{2,1}}(x;q_1,q_2)
&=&\frac{(1-q_2)(2 + q_1 + q_2 + 2 q_1 q_2)}{1-q_1q_2^2}\frac{p_1^3}{6}
+\frac{(q_2-q_1)(1+q_2)}{1-q_1q_2^2}\frac{p_1p_2}{2}
\nonumber \\
&&
-\frac{(1-q_1)(1-q_2^3)}{(1-q_2)(1-q_1q_2^2)}\frac{p_3}{3},
\nonumber
\end{eqnarray}
\begin{eqnarray}
P_{\Lambda^1}(x,q_1,q_2)&=&p_1,
\nonumber \\
P_{\Lambda^{2}}(x,q_1,q_2)&=&
P_{S^{1,1}}(x,q_1,q_2),\quad
P_{\Lambda^{1,1}}(x,q_1,q_2)=
P_{S^{2}}(x,q_1,q_2),
\nonumber \\
%P_{\Lambda^{2}}(x,q_1,q_2)&=&\frac{p_1^2}{2}-\frac{p_2}{2},
%\quad
%P_{\Lambda^{1,1}}(x,q_1,q_2)=\frac{(1+q_1)(1-q_2)}{1-q_1q_2}\frac{p_1^2}{2}+\frac{(1-q_1)(1+q_2)}{1-q_1q_2}\frac{p_2}{2},
%\nonumber \\
P_{\Lambda^{3}}(x;q_1,q_2)
&=&\frac{p_1^3}{6}-\frac{p_1p_2}{2}+\frac{p_3}{3},
\nonumber \\
%P_{\Lambda^{2,1}}(x;q_1,q_2)
%&=&\frac{(1-q_2)(2 + q_1 + q_2 + 2 q_1 q_2)}{1-q_1q_2^2}\frac{p_1^3}{6}
%+\frac{(q_2-q_1)(1+q_2)}{1-q_1q_2^2}\frac{p_1p_2}{2}
%\nonumber \\
%&&
%-\frac{(1-q_1)(1-q_2^3)}{(1-q_2)(1-q_1q_2^2)}\frac{p_3}{3}.
P_{\Lambda^{2,1}}(x;q_1,q_2)&=&P_{S^{2,1}}(x;q_1,q_2)
\end{eqnarray}
Using {\tt SF},\footnote{
SF is package of Maple program created by J. Stembridge, which is available
from\\
\url{http://www.math.lsa.umich.edu/~jrs/maple.html#SF/}
.}  we can generate the explicit
expression of the Macdonald polynomials for more examples.

\noindent{\underline{Specialization $M_R(q_2^{\varrho};q_1,q_2)$}}\\
Specializing $x=q_2^{\varrho}$ ($\varrho_j:=(N+1)/2-j$, $j=1,\cdots
,N$), we find the polynomial
$M_R(q_2^{\varrho};q_1,q_2):=P_R(x=q_2^{\varrho};q_1,q_2)$ as (\ref{Mac_sp}).
Up to $r\le 3$, the Macdonald functions
$M_{S^{r+\ell,r-\ell}}(q_2^{\varrho};q_1,q_2)$ and
$M_{\Lambda^{r+\ell,r-\ell}}(q_2^{\varrho};q_1,q_2)$ are listed as:
\begin{eqnarray}
M_{S^1}(q_2^{\varrho};q_1,q_2)&=&\frac{A^{\frac{1}{2}}-A^{-\frac{1}{2}}}{q_2^{\frac{1}{2}}-q_2^{-\frac{1}{2}}},
\nonumber \\
M_{S^2}(q_2^{\varrho};q_1,q_2)&=&\frac{\left(A^{\frac{1}{2}}-A^{-\frac{1}{2}}\right)\left(
A^{\frac{1}{2}}q_1^{\frac{1}{2}}-A^{-\frac{1}{2}}q_1^{-\frac{1}{2}}\right)}
{\left(q_2^{\frac{1}{2}}-q_2^{-\frac{1}{2}}\right)\left(q_1^{\frac{1}{2}}q_2^{\frac{1}{2}}-q_1^{-\frac{1}{2}}q_2^{-\frac{1}{2}}\right)},
\nonumber \\
M_{S^{1,1}}(q_2^{\varrho};q_1,q_2)&=&\frac{\left(A^{\frac{1}{2}}-A^{-\frac{1}{2}}\right)\left(
A^{\frac{1}{2}}q_2^{\frac{1}{2}}-A^{-\frac{1}{2}}q_2^{-\frac{1}{2}}\right)}
{\left(q_2^{\frac{1}{2}}-q_2^{-\frac{1}{2}}\right)\left(q_2-q_2^{-1}\right)},
\nonumber \\
M_{S^3}(q_2^{\varrho};q_1,q_2)&=&\frac{\left(A^{\frac{1}{2}}-A^{-\frac{1}{2}}\right)\left(
A^{\frac{1}{2}}q_1^{\frac{1}{2}}-A^{-\frac{1}{2}}q_1^{-\frac{1}{2}}\right)
\left(
A^{\frac{1}{2}}q_1-A^{-\frac{1}{2}}q_1^{-1}\right)}
{\left(q_2^{\frac{1}{2}}-q_2^{-\frac{1}{2}}\right)\left(q_1^{\frac{1}{2}}q_2^{\frac{1}{2}}-q_1^{-\frac{1}{2}}q_2^{-\frac{1}{2}}\right)\left(q_1q_2^{\frac{1}{2}}-q_1^{-1}q_2^{-\frac{1}{2}}\right)},
\nonumber \\
M_{S^{2,1}}(q_2^{\varrho};q_1,q_2)&=&\frac{\left(A^{\frac{1}{2}}-A^{-\frac{1}{2}}\right)\left(A^{\frac{1}{2}}q_1^{\frac{1}{2}}-A^{-\frac{1}{2}}q_1^{-\frac{1}{2}}\right)\left(A^{\frac{1}{2}}q_2^{-\frac{1}{2}}-A^{-\frac{1}{2}}q_2^{\frac{1}{2}}\right)}{\left(q_2^{\frac{1}{2}}-q_2^{-\frac{1}{2}}\right)^2\left(q_1^{\frac{1}{2}}q_2-q_1^{-\frac{1}{2}}q_2\right)},
\nonumber \\
M_{\Lambda^1}(q_2^{\varrho};q_1,q_2)&=&M_{S^1}(q_2^{\varrho};q_1,q_2),
\nonumber \\
M_{\Lambda^2}(q_2^{\varrho};q_1,q_2)&=&M_{S^{1,1}}(q_2^{\varrho};q_1,q_2),
\quad M_{\Lambda^{1,1}}(q_2^{\varrho};q_1,q_2)=M_{S^{2}}(q_2^{\varrho};q_1,q_2),
\nonumber \\
%M_{\Lambda^1}(q_2^{\varrho};q_1,q_2)&=&\frac{A^{\frac{1}{2}}-A^{-\frac{1}{2}}}{q_2^{\frac{1}{2}}-q_2^{-\frac{1}{2}}},
%\nonumber \\
%M_{\Lambda^{2}}(q_2^{\varrho};q_1,q_2)&=&\frac{\left(A^{\frac{1}{2}}-A^{-\frac{%1}{2}}\right)\left(
%A^{\frac{1}{2}}q_2^{\frac{1}{2}}-A^{-\frac{1}{2}}q_2^{-\frac{1}{2}}\right)}
%{\left(q_2^{\frac{1}{2}}-q_2^{-\frac{1}{2}}\right)\left(q_2-q_2^{-1}\right)},
%\nonumber \\
%M_{\Lambda^{1,1}}(q_2^{\varrho};q_1,q_2)&=&\frac{\left(A^{\frac{1}{2}}-A^{-\frac{1}{2}}\right)\left(
%A^{\frac{1}{2}}q_1^{\frac{1}{2}}-A^{-\frac{1}{2}}q_1^{-\frac{1}{2}}\right)}
%{\left(q_2^{\frac{1}{2}}-q_2^{-\frac{1}{2}}\right)\left(q_1^{\frac{1}{2}}q_2^{\frac{1}{2}}-q_1^{-\frac{1}{2}}q_2^{-\frac{1}{2}}\right)},
%\nonumber \\
M_{\Lambda^3}(q_2^{\varrho};q_1,q_2)&=&\frac{\left(A^{\frac{1}{2}}-A^{-\frac{1}{2}}\right)\left(
A^{\frac{1}{2}}q_2^{\frac{-1}{2}}-A^{-\frac{1}{2}}q_2^{\frac{1}{2}}\right)
\left(
A^{\frac{1}{2}}q_2^{-1}-A^{-\frac{1}{2}}q_2\right)}
{\left(q_2^{\frac{1}{2}}-q_2^{-\frac{1}{2}}\right)
\left(q_2-q_2^{-1}\right)
\left(q_2^{\frac{3}{2}}-q_2^{-\frac{3}{2}}\right)
},\nonumber \\
%M_{\Lambda^{2,1}}(q_2^{\varrho};q_1,q_2)&=&\frac{\left(A^{\frac{1}{2}}-A^{-\frac{1}{2}}\right)\left(A^{\frac{1}{2}}q_1^{\frac{1}{2}}-A^{-\frac{1}{2}}q_1^{-\frac{1}{2}}\right)\left(A^{\frac{1}{2}}q_2^{-\frac{1}{2}}-A^{-\frac{1}{2}}q_2^{\frac{1}{2}}\right)}{\left(q_2^{\frac{1}{2}}-q_2^{-\frac{1}{2}}\right)^2\left(q_1^{\frac{1}{2}}q_2-q_1^{-\frac{1}{2}}q_2\right)},
M_{\Lambda^{2,1}}(q_2^{\varrho};q_1,q_2)&=&M_{S^{2,1}}(q_2^{\varrho};q_1,q_2),
\end{eqnarray}
where $A:=q_2^N$, and the power sum symmetric function $p_n$ is
\begin{eqnarray}
 p_n=\frac{A^{\frac{n}{2}}-A^{-\frac{n}{2}}}{q_2^{\frac{n}{2}}-q_2^{-\frac{n}{2}}}.
\end{eqnarray}

\subsection{Consistency check of the gamma factor (\protect\ref{gamma_sym})
   and (\protect\ref{gamma_anti-sym})}
\label{sec:gamma_proof2}
Here we check the consistency of the gamma factors (\ref{gamma_sym})
   and (\ref{gamma_anti-sym})  by applying
the identity of the q-hypergeometric function.
%Plugging the gamma factor into the refined amplitude for $T^{(2,2p+1)}$,
%one finds
%\begin{eqnarray}
%Z_{S^r}(T^{2,2p+1})&=&\sum_{\ell=0}^r
%\frac{(q_2;q_1)_{\ell}(q_1;q_1)_r(A^2;q_1)_{r+\ell}(q_2^{-1}A^2;q_1)_{r-\ell}}
%{(q_1;q_1)_{\ell}(q_2;q_1)_r(q_2;q_1)_{r+\ell+1}(q_1;q_1)_{r-\ell}}(1-q_2q_1^{2%\ell})(q_1q_2^{-1})^{\frac{r-\ell}{2}}
%q_2^{2n-\ell}A^{-2n}
%\nonumber \\
%&&\quad\quad\times
%\biggl((-1)^{r-\ell}q^{\frac{r^2-\ell^2}{2}}q_2^{-\frac{\ell}{2}}A^r\biggr)^{2p+1}.
%\end{eqnarray}
Plugging $\gamma_{S^rS^r}^{S^{r+\ell,r-\ell}}$ of (\ref{gamma_sym}) into
the consistency condition for
$Z_{SU(N)}^{\text{ref}}({\bf S}^3,T^{2,1}_{S^r};q_1,q_2)=Z_{SU(N)}^{\text{ref}}({\bf
S}^3,\unknot_{S^r};q_1,q_2)$ in
(\ref{consistency_gamma}), we expect the following relation:
\begin{eqnarray}
&&P_{S^r}(q_2^{\varrho};q_1,q_2)
=\frac{(A;q_1)_r}{(q_2;q_1)_r}A^{-\frac{r}{2}}q_2^{\frac{r}{2}}
\nonumber \\
&&
=\sum_{\ell=0}^r
\frac{(q_2;q_1)_{\ell}(q_1;q_1)_r(A;q_1)_{r+\ell}(q_2^{-1}A;q_1)_{r-\ell}}
{(q_1;q_1)_{\ell}(q_2;q_1)_r(q_1q_2;q_1)_{r+\ell}(q_1;q_1)_{r-\ell}}\frac{(1-q_2q_1^{2\ell})}{(1-q_2)}
%\nonumber \\
%&&\quad\quad\quad\times
(-1)^{r-\ell}A^{-\frac{r}{2}}q_1^{\frac{(r-\ell)(r+\ell+1)}{2}}q_2^{\frac{3r}{2}-\ell}.
\nonumber \\
&&
\label{unknot_consis}
\end{eqnarray}
Here we quote an identity  eq.(2.8.1) of \cite{Gasper_Rahman}:
\begin{eqnarray}
&&\frac{(a;q)_r(b;q)_r(c;q)_r}{(q;q)_r(aq/b;q)_r(aq/c;q)_r}
\nonumber \\
&&=\frac{(\lambda bc/a;q)_r}{(qa^2/\lambda bc;q)_r}
\sum_{\ell=0}^r\frac{(\lambda;q)_{\ell}(1-\lambda q^{2\ell})(\lambda
b/a;q)_{\ell}(\lambda c/a;q)_{\ell}(aq/bc;q)_{\ell}}{(q;q)_{\ell}(1-\lambda)(aq/b;q)_{\ell}(aq/c;q)_{\ell}(\lambda bc/a;q)_{\ell}}
%\nonumber \\
%&&\quad\quad\times
\frac{(a;q)_{r+{\ell}}(a/\lambda;q)_{r-{\ell}}}{(\lambda q;q)_{r+\ell}(q;q)_{r-\ell}}
\left(\frac{a}{\lambda}\right)^{\ell}.
\nonumber \\
&&
\label{2.8.1}
\end{eqnarray}
Choosing $q=q_1$, $\lambda=q_2$ and $a=A$ in the above identity and
taking $b,c\to 0$ limit, we find
\begin{eqnarray}
&&\frac{(A;q_1)_r}{(q_1;q_1)_r}(-1)^rA^{-2r}q_1^{-r(r+1)}(bc)^{-r}
\nonumber \\
&&=(-1)^rA^{-2r}q_1^{-\frac{r(r+1)}{2}}q_2^{r}(bc)^{-r}
\sum_{\ell=0}^r\frac{(q_2;q_1)_{\ell}(1-q_2q_1^{2\ell})}{(q_1;q_1)_{\ell}(1-q_2)}
\frac{(A;q_1)_{r+\ell}(q_2^{-1}A;q_1)_{r-\ell}}{(q_2q_1;q_1)_{r+\ell}(q_1;q_1)_{r-\ell}}
(-1)^{\ell}q_1^{-\frac{\ell(\ell+1)}{2}}q_2^{-\ell},
\nonumber
\end{eqnarray}
where $(x/b;q)_n\to (-x)^nq^{\frac{n(n-1)}{2}}b^{-n}$ and $(xb;q)_n\to 1$
under $b\to 0$ limit.
This coincides with (\ref{unknot_consis}) and proves the gamma factor
for $R=S^r$.

For the $\gamma$ factor for the anti-symmetric representation $\Lambda^r$,
we expect the following relation:
\begin{eqnarray}
&&P_{\Lambda^r}(q_2^{\varrho};q_1,q_2)=
(-1)^rA^{r/2}q_2^{r/2}\frac{(A^{-1};q_2)_r}{(q_2;q_2)_r}
\nonumber \\
&&
=\sum_{\ell=0}^r\frac{(q_1;q_2)_{\ell}(q_2;q_2)_{r+\ell}(A^{-1};q_2)_{r+\ell}(q_1^{-1}A^{-1};q_2)_{r-\ell}}
{(q_2;q_2)_{\ell}(q_1q_2;q_2)_{r+\ell}(q_2;q_2)_{r+\ell}(q_2;q_2)_{r-\ell}}\frac{(1-q_1q_2^{2\ell})}{(1-q_1)}
%\nonumber \\
%&&\quad\quad\quad \times
(-1)^{\ell}A^{\frac{3r}{2}}q_1^{r}q_2^{\frac{\ell(\ell-1)-r(r-2)}{2}}.
\nonumber \\
&&
\label{gamma_id_anti-sym}
\end{eqnarray}
Choosing $q=q_2$, $\lambda=q_1$, and $a=A^{-1}$ and taking $b,c\to\infty$
in (\ref{2.8.1}), we also find the same identity as (\ref{gamma_id_anti-sym}).

%As an observation, the gamma factors are also rewritten as:
%\begin{eqnarray}
%&&\gamma_{S^rS^r}^{S^{r+\ell,r-\ell}}
%=\prod_{s\in C_{S^{r+\ell,r-\ell}/S^r}}b_{S^r}(s),
%\\
%&&\gamma_{\Lambda^r\Lambda^r}^{\Lambda^{r+\ell,r-\ell}}
%=\prod_{s\in C_{S^{r+\ell,r-\ell}/S^r}-R_{S^{r+\ell,r-\ell}/S^r}}b_{S^r}(s),
%\end{eqnarray}
%where
%\begin{eqnarray}
%b_R(s)=\left\{
%\begin{array}{ll}
%\frac{1-q^{a_R(s)}t^{l_R(s)+1}}{1-q^{a_R(s)+1}t^{l_R(s)}} & {\rm if}\;
% s\in R,\\
%1 & {\rm otherwise},
%\end{array}
%\right.
%\end{eqnarray}
%and $C_{Q/R}$ (resp. $R_{Q/R}$) denote the union of columns (resp. rows)
%that intersect $Q-R$.

\subsection{WKB analysis}
\label{sec:WKB_app}
\subsubsection{Another expression for $P_n(T^{2,2p+1};q,t)$}
\label{sec:Another}
As an analogy with an expression for the colored Jones polynomial for
$T^{2,2p+1}$,we find a conjecture for yet another expression of $P_{n}(T^{2,2p+1};q,t)$ which is consistent with (\ref{refined_SU(2)}).

In \cite{Habiro,Le,Masbaum,Hikami_AJ},
the expression for the colored Jones polynomial for $(2,2p+1)$ torus knots
is given as:
\begin{eqnarray}
J_n(T^{2,2p+1};q)
&=&q^{p(n^2-1)}\sum_{k_p\ge \cdots \ge k_2\ge k_1\ge 0}
(-1)^{k_p}q^{-\frac{k_p(k_p+1)}{2}}
\frac{(q^{1-n};q)_{k_p}(q^{1+n};q)_{k_p}}{(q;q)_{k_p}}
\nonumber \\
&&\quad\quad\quad\quad\quad\times
\prod_{a=1}^{p-1}q^{-(k_a-k_p)(k_a-k_p-1)+k_a(k_a-k_{a+1})}
\left[\begin{array}{c}
k_{a+1} \\
k_a
\end{array}
\right]_q,
\label{Jones_2p+1}
\end{eqnarray}
where the q-binomial coefficient is defined by
\begin{eqnarray}
\left[\begin{array}{c}
n \\
m
\end{array}
\right]_q:=\frac{(q;q)_{n}}{(q;q)_{n-m}(q;q)_{m}}.
\end{eqnarray}

By comparison with (\ref{refined_SU(2)})
for some lower orders in $r$ and $p$,
we can introduce the $t$ deformation as follows:
\begin{eqnarray}
&&P_{n=r+1}(T^{2,2p+1};q,t)
\nonumber \\
&=&q^{(p-1)r(r+2)}\sum_{i+j\le r}
\Biggl[
\sum_{r-i=k_p\ge k_{p-1}\ge\cdots\ge k_2\ge
 k_1\ge 0}
\Biggl(
q^{r+(r+1)(i+j)+\frac{j(j+1)}{2}}t^{2pi+3j}
\frac{[r]^{\prime}!}{[i]^{\prime}![j]^{\prime}![r-i-j]^{\prime}!}
\nonumber \\
&&
\times\prod_{a=1}^{p-1}
t^{2k_a}q^{-(k_a-k_p)(k_a-k_p-1)}q^{k_a(k_a-k_{a+1})}
\frac{[k_{a+1}]^{\prime}!}{[k_a]^{\prime}![k_{a+1}-k_{a}]^{\prime}!}
\Biggr)
\Biggr].     \label{Pn-torus-another}
\end{eqnarray}
where
$$
[n]'! = [1]' [2]' \cdots [n-1]' [n]',\qquad\qquad [n]' = \frac{q^n - 1}{q-1} = 1 + q + \ldots + q^{n-1}.
$$

Furthermore, we can simplify the expression using the Cauchy's
q-binomial identity:
\begin{eqnarray}
(-xq;q)_k=\sum_{j=0}^kx^jq^{\frac{j(j+1)}{2}}\frac{(q;q)_k}{(q;q)_j(q;q)_{k-j}}.
\end{eqnarray}
Specializing $x=q^{r+1}t^3$ and $k=r-i$ in the above identity, we find
\begin{eqnarray}
&&P_{n=r+1}(T^{2,2p+1};q,t)
\nonumber \\
&=&q^{(p-1)r(r+2)}\sum_{r=k_p\ge k_{p-1}\ge\cdots\ge k_2\ge
 k_1\ge i\ge 0}
\Biggl[
q^{r+(r+1)i}t^{2pi}\frac{(-t^3q^{r+2};q)_{r-i}(q;q)_r}{(q;q)_i(q;q)_{k_1-i}}
\nonumber \\
&&
\times \prod_{a=1}^{p-1}t^{2(k_a-i)}q^{-(k_a-k_p)(k_a-k_p-1)}q^{(k_a-i)(k_a-k_{a+1})}\frac{1}{(q;q)_{k_{a+1}-k_{a}}}\Biggr].
\label{conj_2p+1}
\end{eqnarray}

Now let us study the asymptotic behavior of (\ref{conj_2p+1}).
In $\hbar\to 0$ limit, we find the following expansion:
\begin{eqnarray}
&& P_{n=r+1}(T^{2,2p+1};q,t)\sim \int \prod_{\alpha=0}^{p-1}dz_{\alpha}\;
e^{\frac{1}{\hbar}(V_{(2,2p+1)}(z_{\alpha},x,t))+{\cal O}(\hbar)}, \\
&& V_{(2,2p+1)}(\vec{z},x,t)=
(p-1)(\log x)^2+2p(\log t)\cdot(\log z_0)
+(\log x)\cdot(\log z_0)
\nonumber \\
&&\quad\quad\quad\quad\quad\quad\quad\quad
+\sum_{a=1}^{p-1}\Bigl[
2(\log t)\cdot (\log (z_az_0^{-1}))
-(\log z_ax^{-1})^2+(\log z_az_0^{-1})\cdot(\log z_az_{a+1}^{-1})
\Bigr]
\nonumber \\
&&\quad\quad\quad\quad\quad\quad\quad\quad
+{\rm Li}_2(-t^3x)-{\rm Li}_2(-t^3x^2z_0^{-1})-{\rm Li}_2(x)
+{\rm Li}_2(z_0)+{\rm Li}_2(z_1z_0^{-1})
\nonumber \\
&&\quad\quad\quad\quad\quad\quad\quad\quad
-p{\rm Li}_2(1)+\sum_{a=1}^{p-1}{\rm Li}_2(z_{a+1}z_{a}^{-1}),
\end{eqnarray}
where $k_a=\frac{1}{\hbar}\log z_a$ for $a=1,\cdots p$,
$i=\frac{1}{\hbar}\log z_0$, and $r=\frac{1}{\hbar}\log x$.

The critical point of this potential is determined by
\begin{eqnarray}
1=\exp\left[z_{\alpha}\frac{\partial  V_{(2,2p+1)}(\vec{z},x,t)}{\partial z_{\alpha}}\right],\quad (\alpha=0,\cdots,p-1).
\end{eqnarray}
Combining a condition $y=\exp\left(x\partial V_{(2,2p+1)} /\partial x\right)$,
we find the following set of algebraic equations for $z_{\alpha}$:
\begin{eqnarray}
&&1=\frac{t^2 x^2 (z_{1} - z_{0})}{z_{1} (z_{0}-1) (t^3 x^2 +
 z_0)},
\quad
1=\frac{t^2 x (x - z_{p-1})}{z_{p-1} (z_{p-1} - z_{p-2})},
\nonumber \\
&&1=\frac{t^2 x^2 (z_{p-2}-z_{p-1})}{z_{p-1}z_{p-2}(z_{p-3}-z_{p-2})},
 \quad\cdots,\quad
1=\frac{t^2 x^2 (z_a - z_{a+1})}{z_a z_{a+1}(z_{a-1}-z_{a})},
\quad \cdots,\quad
1=\frac{t^2 x^2 (z_{1} - z_{2})}{z_{1} z_{2}(z_{0}-z_{1})},
\nonumber \\
&&
\nonumber \\
&&y=\frac{(-1 + x) z_1^2 z_2^2\cdots z_{p-1}^2 (t^3 x^2 + z_0)^2}{(1 + t^3 x) (x - z_{p-1})}.
\end{eqnarray}
After eliminating $z_{\alpha}$ ($\alpha=0,\cdots,p-1$), we obtain
the same refined A-polynomials $A_{T^{2,2p+1}}(x,y;t)$ as given in Table \ref{table_A-poly4},
which are computed on the basis of (\ref{Superpolynomial_sym_refined}) for $p=1,2,3,4,5$.

\subsubsection{Asymptotic limit of  $\P^{S^{n-1}}_{\rm{DGR}}(T^{2,2p+1};a,q,t)$ with $a=q^2 t^2$}
\label{sec:DGR_another}

As we discussed in section \ref{sec:saddle_DGR},
for the (colored) superpolynomials in grading conventions of \cite{DGR}, the right limit to consider is
\be
q^2 t^2 \; \to \; 1 \,,
\ee
with $x = q^n$ and $t$ kept fixed. In other words, in this grading conventions,
the combination $\log (qt)$ plays the role of a small expansion parameter $\hbar \to 0$.
On the other hand, if we think of the HOMFLY variable $a$ as $a = e^{N \hbar}$ (as opposed to $a=q^N$),
then it might be natural to consider a specialization of the superpolynomial to $a = (e^{\hbar})^2 = q^2 t^2$ (instead of $a=q^2$).
With this motivation in mind, here we consider the ``large color'' asymptotics of the superpolynomial
specialized to $a = q^2 t^2$.

\begin{figure}[H]
\centering
\includegraphics[width=0.6\textwidth]{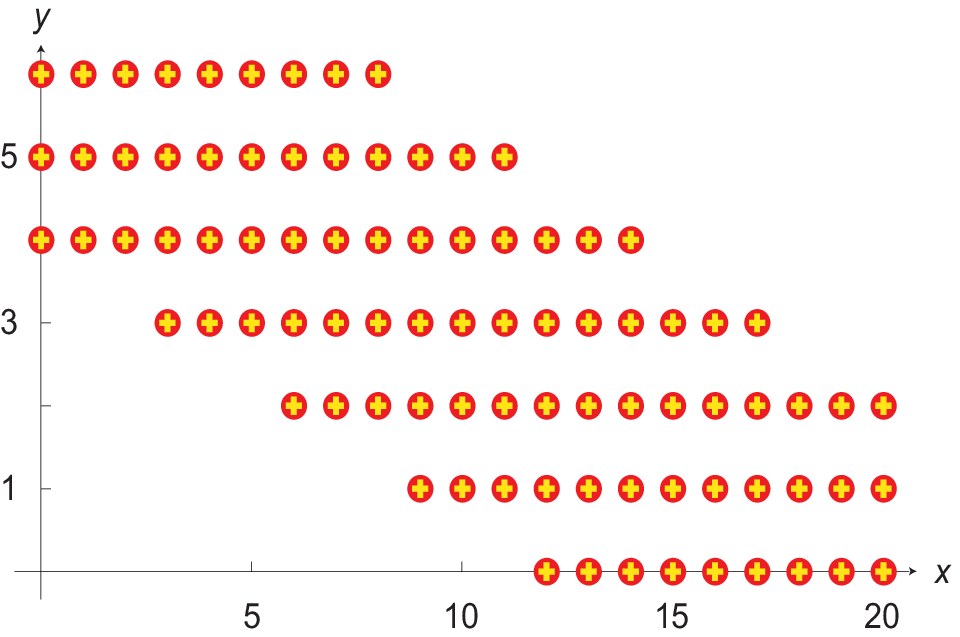}
\caption{Newton polygons for trefoil knot in DGR (see section \protect\ref{sec:saddle_DGR}) and DGR' (see section \protect\ref{sec:DGR_another}) gradings: refined (red circles) and unrefined (yellow crosses).}  \label{fig-At31dgr-Newton}
%\label{fig:AAAA}
\end{figure}

\begin{figure}[H]
\centering
\includegraphics[width=\textheight,angle=90]{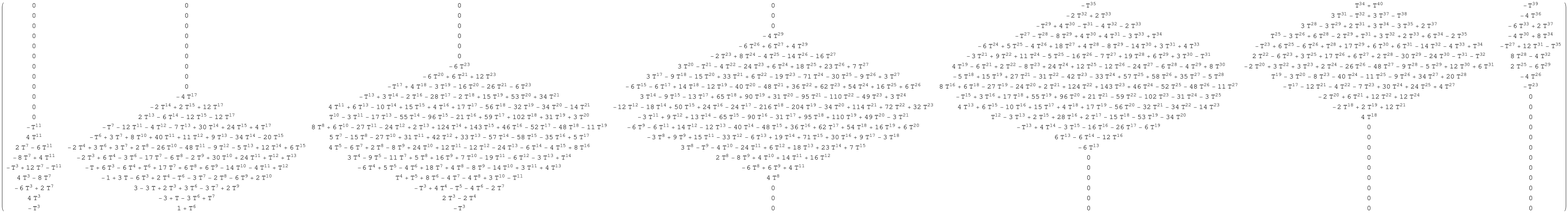}
$\qquad \qquad $
\includegraphics[width=\textheight,angle=90]{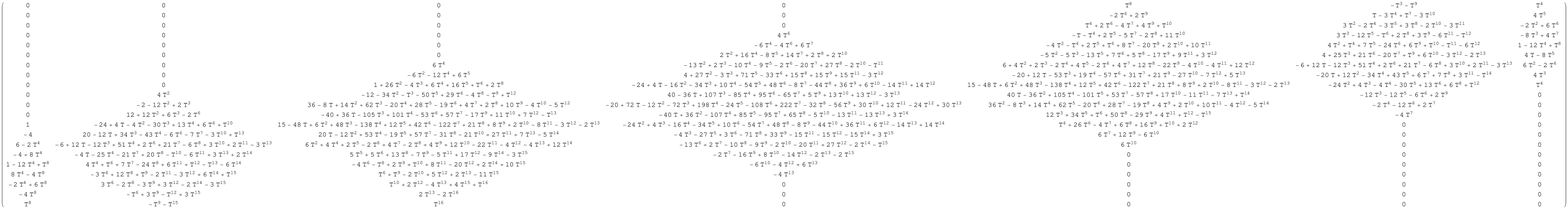}
\caption{Matrix form of A-polynomial for trefoil knot in DGR (see section \protect\ref{sec:saddle_DGR}) and DGR' (see section \protect\ref{sec:DGR_another}) gradings.}    \label{fig-At31dgr-matrix}
%\label{fig:AAAA}
\end{figure}

Much as in \eqref{Pn-saddle}, we find
\be
\P_{\text{DGR}}^{S^{n-1}}(T^{2,2p+1};a=q^2 t^2,q,t)
\sim \int dz\; e^{\frac{1}{\hbar}\left(V_{(2,2p+1)}^{\text{DGR}'}(z,x;t)
+{\cal O}(\hbar)\right)} \,,
\ee
with the potential function
\begin{eqnarray}
&&V_{(2,2p+1)}^{\text{DGR}'}(z,x;t)= {\rm Li}_2(-t^3) - {\rm Li}_2(x) + {\rm Li}_2(-t x) +
  {\rm Li}_2(xz^{-1}) -  {\rm Li}_2(-t^3 xz^{-1})] + {\rm Li}_2(z)
\nonumber \\
&&\quad\quad\quad\quad\quad\quad\quad
-
 {\rm Li}_2(zt^{-2}) +  {\rm Li}_2(x zt^{-2}) - {\rm Li}_2(-t x z)-\frac{\pi^2}{6}
\nonumber \\
&&\quad\quad\quad\quad\quad\quad\quad
+  (\log t)\cdot (\log zx^{-2}) +
 (2 p+1)(\log xz^{-1})\cdot (\log (-x^{\frac{1}{2}}z^{\frac{1}{2}})).
\end{eqnarray}
{}From this potential function, eqs. (\ref{saddle_point}) and (\ref{braid_saddle2}) yield
the equations for the saddle point that dominates the integral:
\begin{eqnarray}
&&1=\frac{t (t^2 - z_0) z_0^{- 2 p-1} (-x + z_0) (1 + t x z_0)}{(-1 + z_0) (t^3 x +
 z_0) (t^2 - x z_0)},
\nonumber \\
&&y(x,t)=\frac{(-1 + x) x^{2 p+1} (t^3 x + z_0) (1 + t x z_0)}{(1 + t x) (x - z_0) (-t^2 + x z_0)}.
\label{Saddle_DGR1}
\end{eqnarray}
Eliminating $z_0$, we obtain the refined A-polynomial.
In Appendix \ref{sec:AppendixB} the refined A-polynomial for $p=1$ is described explicitly,
with Newton polygon presented in figure \ref{fig-At31dgr-Newton}, and matrix form given in figure \ref{fig-At31dgr-matrix} (right).

The saddle point equations (\ref{Saddle_DGR1}) can be easily solved near $t=-1$.
Up to the second order, we find the approximate solution for $y(x,t)$ in the form:
\begin{eqnarray}
y(x,t)&=&-x^{2 p + 1} \Biggl(
1-\frac{x + 2 \xi_j - 3 x \xi_j - 4 x^2 \xi_j + 3 x^3 \xi_j + x \xi_j^2} {(-1 + x) (x - \xi_j) (-1 + x \xi_j)}(t+1)
\Biggr)
\nonumber \\
&&+{\cal O}(1+t)^2 \,.
\end{eqnarray}
Note that the leading order term in this equation encodes the unrefined curve $y+x^{2p+1}=0$, which agrees with $t=-1$ specialization of the second equation in (\ref{Saddle_DGR1}).
{}From (\ref{S0ref}), one also finds the refined classical action $S_0(u,t)$:
\begin{eqnarray}
S_0(u,t)&=&S_0(u)+\sum_{a=1}^{\infty}(1+t)^aS_0^{(a)}(u),
\nonumber \\
S_0(u)&=&\frac{1}{2(2p+1)}\log(-x^{2p+1})^2,
\nonumber \\
S_0^{(1)}(u)&=&\log\frac{(1-x)x^2}{(x-\xi_j)^3(1-x\xi_j)^3},
\nonumber \\
S_0^{(2)}(u)&=&\frac{1}{2}S_0^{(1)}
\nonumber \\
&&
+
\bigl(
-7 x^2 - 8 p x^2 + 6 x^3 + 6 p x^3 - 7 x \xi_j - 2 p x \xi_j +
 21 x^2 \xi_j + 6 p x^2 \xi_j - 13 x^3 \xi_j
\nonumber \\
&&\quad
+ 10 p x^3 \xi_j +
 3 x^4 \xi_j - 6 p x^4 \xi_j + 5 \xi_j^2 + 10 p \xi_j^2 -
 18 x \xi_j^2 - 12 p x \xi_j^2 + 50 x^2 \xi_j^2
\nonumber \\
&&\quad
+ 16 p x^2 \xi_j^2 -
 54 x^3 \xi_j^2 - 24 p x^3 \xi_j^2 + 11 x^4 \xi_j^2 -
 2 p x^4 \xi_j^2 - 19 x \xi_j^3 - 26 p x \xi_j^3
\nonumber \\
&&\quad
+ 33 x^2 \xi_j^3 +
 30 p x^2 \xi_j^3 - 25 x^3 \xi_j^3 - 14 p x^3 \xi_j^3 +
 15 x^4 \xi_j^3 + 18 p x^4 \xi_j^3 + 5 x^2 \xi_j^4
\nonumber \\
&&\quad
+  16 p x^2 \xi_j^4
- 6 x^3 \xi_j^4 - 18 p x^3 \xi_j^4
\bigr)
%\nonumber \\
%&&\quad
/\bigl(2(1 + 2 p) (-1 + x) (x - \xi_j)^2 (-1 + x \xi_j)^2\bigr).
\nonumber \\
&&
\end{eqnarray}
and the general form of the classical action:
\begin{eqnarray}
S_0(u,t)&=&\frac{1}{2(2p+1)}\left(\log(-x^{2p+1})\right)^2-\log(-t)
 \log\left(\frac{(1-x)x^2}{(x-\xi_j)^3(1-x\xi_j)^3}\right)+R(x,t).
\nonumber \\
&&
\end{eqnarray}

%*******************************************************************
%*******************************************************************
%*******************************************************************
%*******************************************************************

\newpage

\section{Proof of the gamma factor (\protect\ref{gamma_sym}) and
 (\protect\ref{gamma_anti-sym}), by Hidetoshi Awata}
\label{sec:gamma_proof}
%\begin{center}
%{\it Hidetoshi Awata}
%\end{center}
%\subsubsection{Proof for the gamma factor}
%\label{sec:gamma_proof1}
Let $P_R\left[ c \frac{1-L}{1-q_2} \right]$
be the Macdonald function $P_R(x;q_1,q_2)$
with the specialization
$p_n := \sum_{i=1}^\infty x_i^n =  c^n \frac{1-L^n}{1-q_2^{n}}$.
%Let $R^t$ be the conjugate of the Young diagram $R=(R_1,R_2,\cdots)$.
%
%
%%%%%%%%%%%%%%%%%%% Pieri %%%%%%%%%%%%%%%%%%%%%%%%
%
The Pieri formula
\cite{McD}(Ch.\ VI.6) %(6.24) p.\ 340
gives
\begin{eqnarray}
P_{S^r}(x;q_1,q_2) P_{S^r}(x;q_1,q_2)
&=&
\sum_{\ell=0}^r N_{S^rS^r}^{S^{r+\ell,r-\ell}} P_{S^{r+\ell,r-\ell}} (x;q_1,q_2)
,
\\
P_{\Lambda^r}(x;q_1,q_2) P_{\Lambda^r}(x;q_1,q_2)
&=&
\sum_{\ell=0}^r N_{\Lambda^r\Lambda^r}^{\Lambda^{r+\ell,r-\ell}} P_{\Lambda^{r+\ell,r-\ell}} (x;q_1,q_2)
\end{eqnarray}
with
\begin{eqnarray}
N_{S^rS^r}^{S^{r+\ell,r-\ell}}
&:=&
\prod_{j=1}^{r-\ell}
\frac{ 1-q_1^{j-1} q_2} {1-q_1^j }
\cdot
\prod_{j=2\ell+1}^{r+\ell}
\frac{ 1-q_1^{j-1} q_2^2}{ 1-q_1^j q_2 }
\cdot
\prod_{j=\ell+1}^{r}
\left(
\frac{ 1-q_1^{j}}{1-q_1^{j-1} q_2 }
\right)^2
,
\\
N_{\Lambda^r\Lambda^r}^{\Lambda^{r+\ell,r-\ell}}
&:=&
\prod_{i=1}^{\ell}
\frac{ 1-q_1 q_2^{i-1}}{ 1-q_2^i }
\cdot
\prod_{i=\ell+1}^{2\ell}
\frac{ 1-q_2^{i}}{1-q_1q_2^{i-1} }
.
\end{eqnarray}
Note that above $N_{S^rS^r}^{S^{r+\ell,r-\ell}}$ and $N_{\Lambda^r\Lambda^r}^{\Lambda^{r+\ell,r-\ell}}$ are
invariant under the transformation
$q_1\rightarrow q_1^{-1}$ and $q_2\rightarrow q_2^{-1}$.
%
%%%%%%%%%%%%%%%%%%%% Specialization %%%%%%%%%%%%%%%%%%%%%%%
%
The specialization formula
\cite{McD}(Ch.\ VI.6) %(6.11') p.\ 337
\begin{eqnarray}
P_R\left[ c\frac{1-L}{1-q_2} \right]
=
c^r
\prod_{(i,j)\in R}
\frac{
q_2^{i-1} - q_1^{j-1} L
}{
1-q_1^{R_i-j} q_2^{R^t_j-i+1}
}
\end{eqnarray}
yields
\begin{eqnarray}
\frac{
P_{S^{r+\ell,r-\ell}}\left[ c\frac{1-L}{1-q_2} \right]
}
{
P_{S^r}\left[ c\frac{1-L}{1-q_2} \right]
}
&=&
c^r
\frac{
\prod_{j=1}^{r-\ell} ( q_2-q_1^{j-1}L )
\cdot
\prod_{j=r+1}^{r+\ell} ( 1-q_1^{j-1}L )
\cdot
\prod_{j=r-\ell+1}^{r}( 1-q_1^{j-1}q_2 )
}{
\prod_{j=1}^{2\ell} ( 1-q_1^{j-1}q_2 )
\cdot
\prod_{j=2\ell+1}^{r+\ell}( 1-q_1^{j-1}q_2^2 )
}
,
\nonumber \\
\frac{
P_{\Lambda^{r+\ell,n-\ell}}\left[ c\frac{1-L}{1-q_2} \right]
}{
P_{\Lambda^r}\left[ c\frac{1-L}{1-q_2} \right]
}
&=&
c^r
\frac{
\prod_{i=1}^{r-\ell} ( q_2^{i-1}-q_1L )
\cdot
\prod_{i=r+1}^{r+\ell} ( q_2^{i-1}-L )
\cdot
\prod_{i=r-\ell+1}^{r}( 1-q_2^i )
}{
\prod_{i=1}^{2\ell} ( 1-q_2^i )
\cdot
\prod_{i=2\ell+1}^{r+\ell}( 1-q_1q_2^i )
}
.
\nonumber
\end{eqnarray}
%
%%%%%%%%%%%%%%%%%%%% coefficient %%%%%%%%%%%%%%%%%%%%%%%
%
Let
\begin{eqnarray}
g_\ell
&:=&
c^{-r} \prod_{j=\ell+1}^{r}\frac{1-q_1^{j-1}q_2}{1-q_1^{-j} }
,
\\
g'_\ell
&:=&
c^{-r} \prod_{i=1}^r q_2^{1-i}
\cdot
\prod_{i=1}^\ell\frac{ 1-q_2^{i}}{ 1-q_1^{-1}q_2^{1-i} }
,
\end{eqnarray}
then we have
%%%%%%%%%%%%%% proposition %%%%%%%%%%%%%%
\\
\noindent{\bf Proposition.}
\begin{eqnarray}
P_{S^{r}}\left[ c\frac{1-L}{1-q_2} \right]
&=&
\sum_{\ell=0}^r
g_{\ell}
N^{S^{r+\ell,r-\ell}}_{S^rS^r}
P_{S^{r+\ell,r-\ell}}\left[ c\frac{1-L}{1-q_2} \right]
,
\label{eq:PropRow}%
\\
P_{\Lambda^r}\left[ c\frac{1-L}{1-q_2} \right]
&=&
\sum_{\ell=0}^r
g'_{\ell}
N^{\Lambda^{r+\ell,r-\ell}}_{\Lambda^r\Lambda^r}
P_{\Lambda^{r+\ell,r-\ell}}\left[ c\frac{1-L}{1-q_2} \right]
.
\label{eq:PropColumn}%
\end{eqnarray}

%%%%%%%%%%%%%% end of proposition %%%%%%%%%%%%%%

%%%%%%%%%%%%% proof %%%%%%%%%%%%%%%%%%
\noindent{\it Proof.}
First, let
\begin{eqnarray}
f_\ell
&:=&
g_{\ell}
N_{S^rS^r}^{S^{r+\ell,r-\ell}}
\frac{
P_{S^{r+\ell,r-\ell}}\left[ c\frac{1-L}{1-q_2} \right]
}{
P_{S^r}\left[ c\frac{1-L}{1-q_2} \right]
}
\nonumber \\
&=&
( 1-q_1^{2\ell}q_2 )
\frac{
\prod_{j=\ell+1}^{r}(-q_1^{j})( 1-q_1^{j} )
\cdot
\prod_{j=1}^{r-\ell} ( q_2-q_1^{j-1}L )
\cdot
\prod_{j=r+1}^{r+\ell} ( 1-q_1^{j-1}L )
}{
\prod_{j=1}^{r-\ell}( 1-q_1^{j} )
\cdot
\prod_{j=\ell}^{r+\ell} ( 1-q_1^{j}q_2 )
}
,
\\
f'_\ell
&:=&
g'_{\ell}
N_{\Lambda^r\Lambda^r}^{\Lambda^{r+\ell,r-\ell}}
\frac{
P_{\Lambda^{r+\ell,r-\ell}}\left[ c\frac{1-L}{1-q_2} \right]
}{
P_{\Lambda^r}\left[ c\frac{1-L}{1-q_2} \right]
}
\nonumber \\
&=&
(-q_1)^{\ell}
( 1-q_1q_2^{2\ell} )
\prod_{i=\ell+1}^{r}q_2^{1-i}
\nonumber \\
&&\hskip24pt
\times
\frac{
\prod_{i=r-\ell+1}^{r}( 1-q_2^{i} )
\cdot
\prod_{i=1}^{r-\ell} ( q_2^{i-1}-q_1L )
\cdot
\prod_{i=r+1}^{r+\ell} ( q_2^{i-1}-L )
}{
\prod_{i=1}^{\ell}( 1-q_2^{i} )
\cdot
\prod_{i=\ell }^{r+\ell} ( 1-q_1q_2^{i} )
}
.
\end{eqnarray}
%
%%%%%%%%%%%%%%%%%%%% residue %%%%%%%%%%%%%%%%%%%%%%%
%
The residues of $f_\ell$ and $f'_\ell$
at $q_2=q_1^{-k}$ and $q_1=q_2^{-k}$
($\ell\leq k\leq \ell+r$),
%($k=\ell,\ell+1,\cdots,\ell+n$),
respectively, are
\begin{eqnarray}
-q_1^{k} {\rm Res}_{q_2=q_1^{k}} f_\ell
&=&
%q^{k(\ell-n)}( 1-q^{2\ell-k} )
q_1^{-kr}
( q_1^{-\ell}-q_1^{\ell-k} )
q_1^{(k+1)\ell}
\prod_{j=1}^{\ell} (-q_1^{-j})\cdot\prod_{j=1}^{k-\ell} (-q_1^{j})
\nonumber \\
&&\times
\frac{
\prod_{j=1}^{r}(-q_1^{j})( 1-q_1^{j} )
}{
\prod_{j=1}^{\ell} ( 1-q_1^{j} )
\cdot
\prod_{j=1}^{k-\ell} ( 1-q_1^{j} )
}
\cdot
\frac{
\prod_{j=k+1}^{r+k-\ell} ( 1-q_1^{j-1}L )
\cdot
\prod_{j=r+1}^{r+\ell} ( 1-q_1^{j-1}L )
}{
\prod_{j=1}^{r-\ell}( 1-q_1^{j} )
\cdot
\prod_{j=1}^{r+\ell-k} ( 1-q_1^{j} )
},
\nonumber \\
&&
\label{eq:resRow}%
\\
-q_2^{k} {\rm Res}_{q_1=q_2^{k}} f'_\ell
&=&
%(-1)^{n+k-\ell} t^{k(\ell-n)}
q_2^{-kr} ( q_2^{-\ell}-q_2^{\ell-k} )
\prod_{i=1}^{\ell} (-q_2^{i}) \cdot\prod_{i=1}^{k-\ell} (-q_2^{i})
%\prod_{i=1}^{k-\ell} t^{i} \cdot\prod_{i=\ell+1}^{n} t^{1-i}
\nonumber \\
&&\times
\frac{
\prod_{i=1}^{r}q_2^{1-i}( 1-q_2^{i} )
}{
\prod_{i=1}^{\ell} ( 1-q_2^{i} )
\cdot
\prod_{i=1}^{k-\ell} ( 1-q_2^{i} )
}
\cdot
\frac{
\prod_{i=k+1}^{r+k-\ell} ( q_2^{i-1}-L )
\cdot
\prod_{i=r+1}^{r+\ell} ( q_2^{i-1}-L )
}{
\prod_{i=1}^{r-\ell}( 1-q_2^{i} )
\cdot
\prod_{i=1 }^{r+\ell-k} ( 1-q_2^{i} )
}
.
\nonumber \\
&&
\label{eq:resColumn}%
\end{eqnarray}
Note that each second line
(\emph{i.e.} (\ref{eq:resRow}) and (\ref{eq:resColumn}))
of the above two equations
is symmetric under the replacement $\ell \leftrightarrow k-\ell$.
Thus one can show that
\begin{eqnarray}
{\rm Res}_{q_2=q_1^{-k}} ( f_\ell + f_{k-\ell} )
=
{\rm Res}_{q_1=q_2^{-k}} ( f'_\ell + f'_{k-\ell} )
=0.
\end{eqnarray}
%\ba{\rm Res}_{t=q^{-k}} f_{k-\ell}&=&-{\rm Res}_{t=q^{-k}} f_\ell , \\
%{\rm Res}_{q=t^{-k}} f'_{k-\ell}&=&-{\rm Res}_{q=t^{-k}} f'_\ell . \ea
%
%%%%%%%%%%%%%%%%%%%% f and f' %%%%%%%%%%%%%%%%%%%%%%%
%
Next, let
\begin{eqnarray}
f:=\sum_{\ell=0}^{r} f_\ell
,
\qquad
f':=\sum_{\ell=0}^{r} f'_\ell
.
\end{eqnarray}
The residue of $f$ at $q_2=q_1^{-k}$ is
\begin{eqnarray}
{\rm Res}_{q_2=q_1^{-k}} f
&=&
\sum_{\ell={\rm max}(0,k-r)}^{{\rm min}(k,r)}
{\rm Res}_{q_2=q_1^{-k}} f_\ell
\nonumber \\
&=&
\left[
\sum_{\ell={\rm max}(0,k-r)}^{\lfloor\frac{k}{2}\rfloor}
+
\sum_{\ell=\lceil\frac{k}{2}\rceil}^{{\rm min}(k,r)}
\right]
{\rm Res}_{q_2=q_1^{-k}} f_\ell
\nonumber \\
&=&
\sum_{\ell={\rm max}(0,k-r)}^{\lfloor\frac{k}{2}\rfloor}
{\rm Res}_{q_2=q_1^{-k}} ( f_\ell + f_{k-\ell} )
=0,
\end{eqnarray}
and also
${\rm Res}_{q=t^{-k}} f' = 0 $.
Here the floor function $\lfloor x\rfloor$
denotes the largest integer not greater than $x$
and the ceiling function $\lceil x\rceil$
denotes the smallest integer not less than $x$.
Furthermore
$\lim_{q_2\rightarrow\infty}f$,
$\lim_{q_1\rightarrow\infty}f'<\infty$.
Therefore
$f$ and $f'$ are constant in $q_2$ and $q_1$, respectively.
But
\begin{eqnarray}
\lim_{q_2\rightarrow 0} f
=
\lim_{q_2\rightarrow 0} f_0
=1
,
\qquad
\lim_{q_1\rightarrow \infty} f'
=
\lim_{q_1\rightarrow \infty} f'_0
=1
.
\end{eqnarray}
Therefore $f=f'=1$.
%So is $f'$.
\hfill\fbox{}
%%%%%%%%%%%%% end of proof %%%%%%%%%%%%%%%%%%

%%%%%%%%%%%%% proof of gamma formula %%%%%%%%%%%%%%%%%%

Substituting $c:=(q_2/L)^{\frac{1}{2}}$ into
(\ref{eq:PropRow}) and
(\ref{eq:PropColumn})
yields %, we have
\begin{eqnarray}
P_{S^r}
\left[ \frac{L^{\frac{1}{2}}-L^{-\frac{1}{2}}}{q_2^{\frac{1}{2}}-q_2^{-\frac{1}{2}}} \right]
&=&
\sum_{\ell=0}^r
g_{\ell}
N_{S^rS^r}^{S^{r+\ell,r-\ell}}
P_{S^{r+\ell,r-\ell}}
\left[ \frac{L^{\frac{1}{2}}-L^{-\frac{1}{2}}}{q_2^{\frac{1}{2}}-q_2^{-\frac{1}{2}}} \right]
,
\\
P_{\Lambda^r}
\left[ \frac{L^{\frac{1}{2}}-L^{-\frac{1}{2}}}{q_2^{\frac{1}{2}}-q_2^{-\frac{1}{2}}} \right]
&=&
\sum_{\ell=0}^r
g'_{\ell}
N_{\Lambda^r\Lambda^r}^{\Lambda^{r+\ell,r-\ell}}
P_{\Lambda^{r+\ell,r-\ell}}
\left[ \frac{L^{\frac{1}{2}}-L^{-\frac{1}{2}}}{q_2^{\frac{1}{2}}-q_2^{-\frac{1}{2}}} \right]
,
\end{eqnarray}
where $g_\ell = \gamma_\ell \lambda_\ell$
and
$g'_\ell = \gamma'_\ell \lambda'_\ell $
with
\begin{eqnarray}
\lambda_\ell
&:=&
(L/q_2)^{\frac{r}{2}}
\prod_{j=\ell+1}^{r}
(-q_1^{j-\frac{1}{2}}q_2^{\frac{1}{2}})
,
\qquad
~
\gamma_\ell
:=
\prod_{j=\ell+1}^{r}
\frac{ q_1^{\frac{j-1}{2}}q_2^{\frac{1}{2}} - q_1^{\frac{1-j}{2}}q_2^{-\frac{1}{2}}}{ q_1^{\frac{j}{2}}-q_1^{-\frac{j}{2}}}
,
\\
\lambda'_\ell
&:=&
(L/q_2^r)^{\frac{r}{2}}
\prod_{i=1}^\ell
(-q_2^{i-\frac{1}{2}}q_1^{\frac{1}{2}})
,
\qquad~~
\gamma'_\ell
:=
\prod_{i=1}^\ell
\frac{
q_2^{\frac{i}{2}}-q_2^{-\frac{i}{2}}
}{
q_2^{\frac{i-1}{2}}q_1^{\frac{1}{2}} - q_2^{\frac{1-i}{2}}q_1^{-\frac{1}{2}}
}
.
\end{eqnarray}
Note that
$\gamma_\ell =
\gamma_{S^rS^r}^{S^{r+\ell,r-\ell}}$,
$\gamma'_\ell =
\gamma_{\Lambda^r\Lambda^r}^{\Lambda^{r+\ell,r-\ell}}$.
%$N_{S^rS^r}^{S^{r+\ell,r-\ell}}=
%N_{(n)(n)}^{(n+\ell,n-\ell)}$ and
%$N'_\ell=
%N_{(1^n)(1^n)}^{(2^{n-\ell}1^{2\ell})}$.
If we set $L:=q_2^{N}$, then
$P_R
\left[ \frac{L^{\frac{1}{2}}-L^{-\frac{1}{2}}}{q_2^{\frac{1}{2}}-q_2^{-\frac{1}{2}}} \right]
=
M_R(q_2^{\varrho};q_1,q_2)$,
$\lambda_\ell=\lambda^{(+)}_{S^{r+\ell,r-\ell}}(S^r,S^r)$ and
$\lambda'_\ell=\lambda^{(+)}_{\Lambda^{r+\ell,r-\ell}}(\Lambda^r,\Lambda^r)$.
Hence we obtain %the formulas (...) and (...).
\begin{eqnarray}
M_{S^r}(q_2^{\varrho})
&=&
\sum_{\ell=0}^r
\gamma_{S^rS^r}^{S^{r+\ell,r-\ell}}
\lambda^{(+)}_{S^{r+\ell,r-\ell}}(S^r,S^r)
N_{S^rS^r}^{S^{r+\ell,r-\ell}}
M_{S^{r+\ell,r-\ell}}(q_2^{\varrho};q_1,q_2)
,
\\
M_{\Lambda^r}(q_2^{\varrho};q_1,q_2)
&=&
\sum_{\ell=0}^r
\gamma_{\Lambda^r\Lambda^r}^{\Lambda^{r+\ell,r-\ell}}
\lambda^{(+)}_{\Lambda^{r+\ell,r-\ell}}(\Lambda^r,\Lambda^r)
N_{\Lambda^r\Lambda^r}^{\Lambda^{r+\ell,r-\ell}}
M_{\Lambda^{r+\ell,r-\ell}}(q_2^{\varrho};q_1,q_2)
.
\end{eqnarray}

%*******************************************************************
%*******************************************************************
%*******************************************************************
%*******************************************************************

\newpage

\section{Refined A-polynomials}     \label{sec:AppendixB}

In this appendix we present detailed structure of refined $A$-polynomials $A^{\text{ref}}(x,y;t)$ in various examples which we found in this paper: the unknot, $(2,2p+1)$ torus knots, and the trefoil in DGR grading. In section \ref{ssec-appBnewton} we present Newton polygons for refined polynomials, as well as for their $t=-1$ limit. A red circle at position $(i,j)$ in such a polygon represents a monomial of the form $c_{i,j}x^i y^j$ in the refined $A$-polynomial. Smaller yellow crosses represent such monomials in the unrefined $A$-polynomial $A^{\text{ref}}(x,y;-1)$.

In section \ref{ssec-appBmatrix} we present matrix form of refined $A$-polynomials. The entry $(i+1,j+1)$ of such a matrix represents the coefficient $c_{i,j}$ in $A^{\text{ref}}(x,y;t)$. Note that the role of rows and columns in Newton polygons and matrices is exchanged. All these conventions are the same as in figure \ref{fig-unknotNewtonMatrix}.

%*******************************************************************
%*******************************************************************

\subsection{Newton polygons}        \label{ssec-appBnewton}

\begin{figure}[H]
\centering
\includegraphics[width=0.25\textwidth]{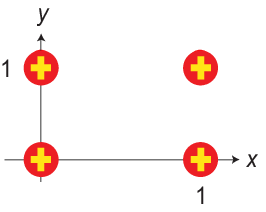}
\caption{Newton polygons for the unknot: refined (red circles) and unrefined (yellow crosses).}
%\label{fig:AAAA}
\end{figure}

\begin{figure}[H]
\centering
\includegraphics[width=0.4\textwidth]{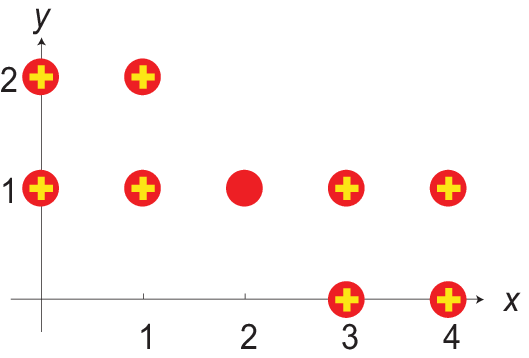}
\caption{Newton polygons for trefoil, \emph{i.e.} $T^{2,3}$ torus knot: refined (red circles) and unrefined (yellow crosses).}
%\label{fig:AAAA}
\end{figure}

\begin{figure}[H]
\centering
\includegraphics[width=0.6\textwidth]{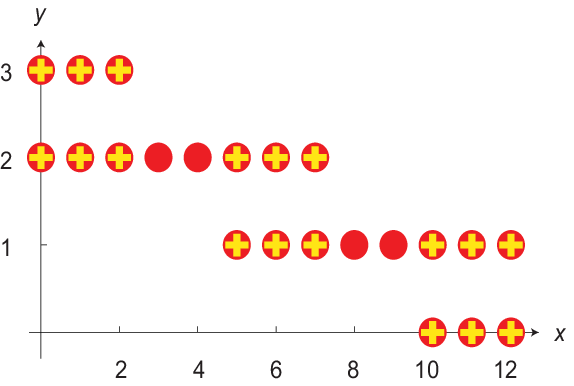}
\caption{Newton polygons for $T^{2,5}$ torus knot: refined (red circles) and unrefined (yellow crosses).}
%\label{fig:AAAA}
\end{figure}

\begin{figure}[H]
\centering
\includegraphics[width=0.6\textwidth]{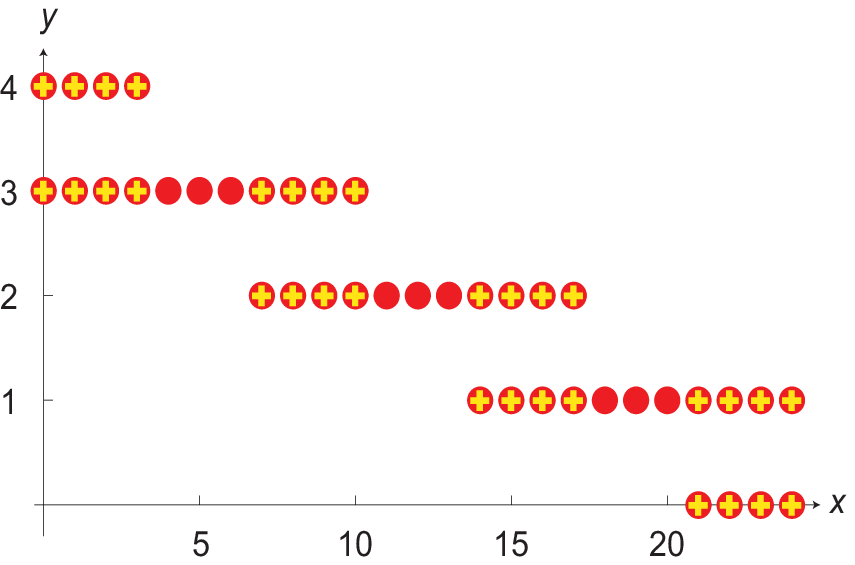}
\caption{Newton polygons for $T^{2,7}$ torus knot: refined (red circles) and unrefined (yellow crosses).}
%\label{fig:AAAA}
\end{figure}

\begin{figure}[H]
\centering
\includegraphics[width=0.8\textwidth]{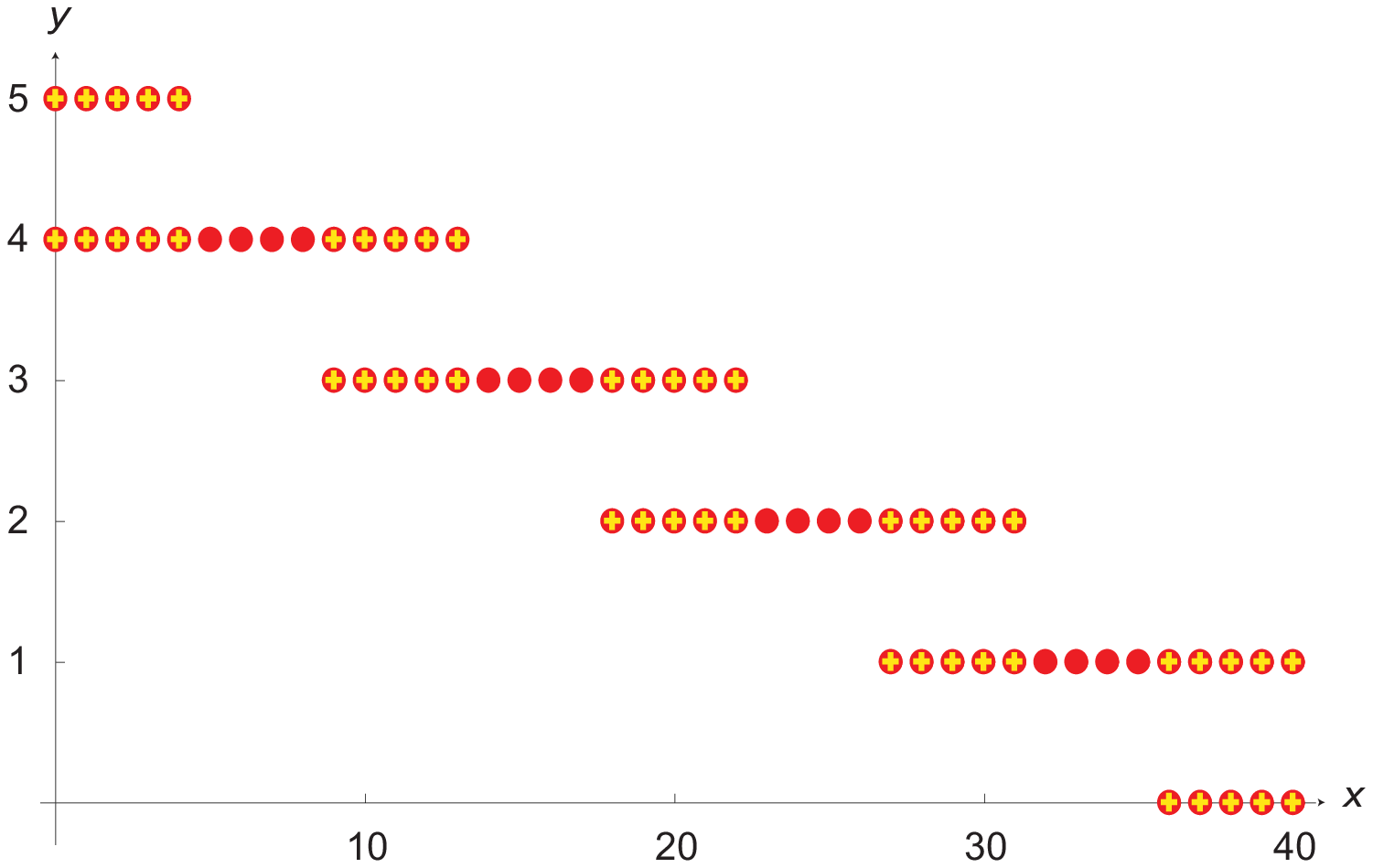}
\caption{Newton polygons for $T^{2,9}$ torus knot: refined (red circles) and unrefined (yellow crosses).}
%\label{fig:AAAA}
\end{figure}

\begin{figure}[H]
\centering
\includegraphics[width=0.8\textwidth]{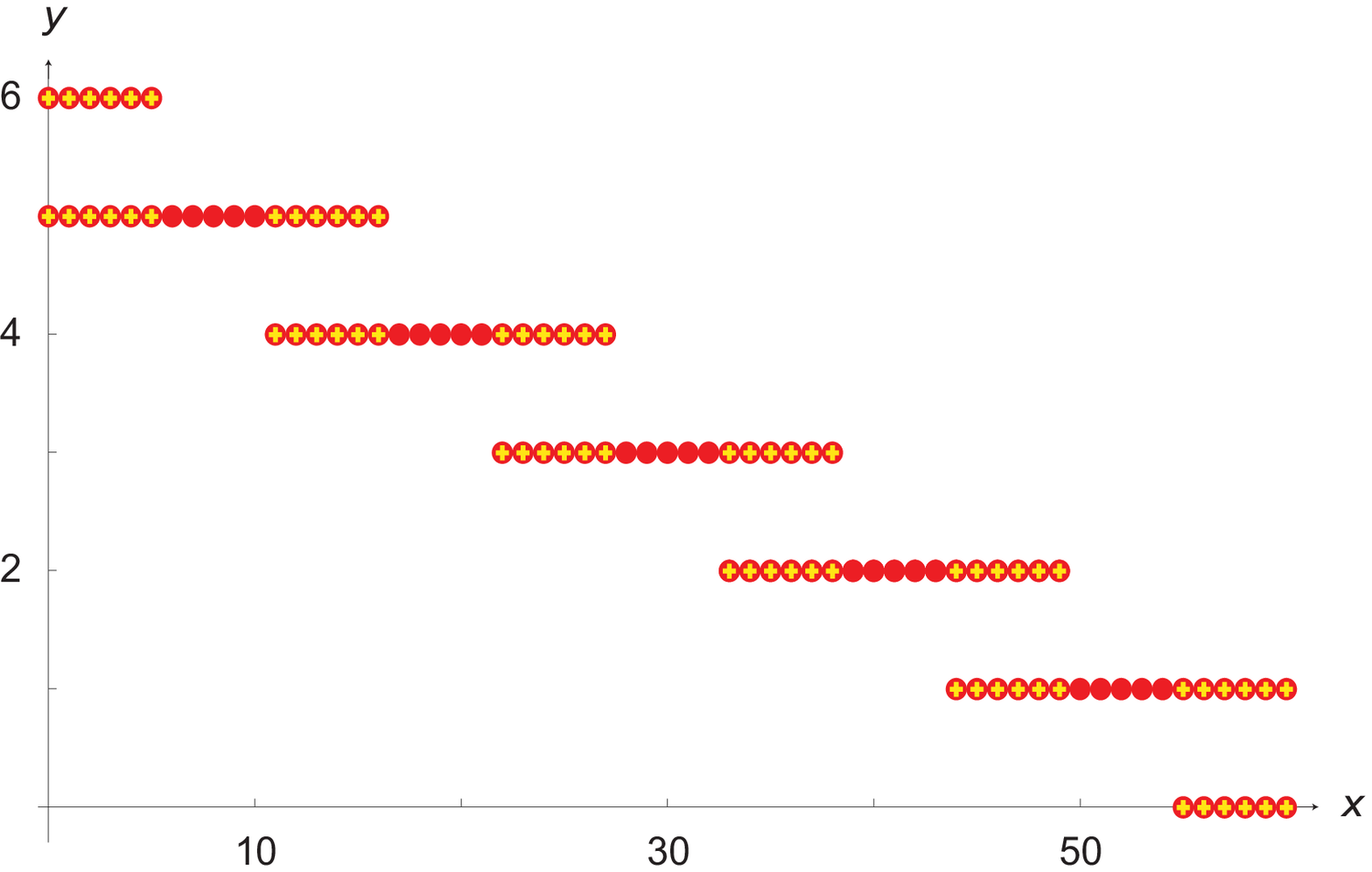}
\caption{Newton polygons for $T^{2,11}$ torus knot: refined (red circles) and unrefined (yellow crosses).}
%\label{fig:AAAA}
\end{figure}

%*******************************************************************
%*******************************************************************

\subsection{Matrix forms of A-polynomials}   \label{ssec-appBmatrix}

\begin{figure}[H]
\centering
\includegraphics[width=0.15\textwidth]{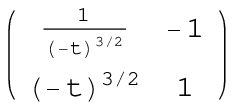}
\caption{Matrix form of A-polynomial for the unknot.}
%\label{fig:AAAA}
\end{figure}

\begin{figure}[H]
\centering
\includegraphics[width=0.22\textwidth]{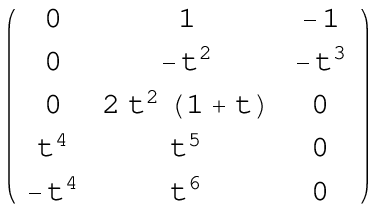}
\caption{Matrix form of A-polynomial for trefoil, \emph{i.e.} $T^{2,3}$ torus knot.}
%\label{fig:AAAA}
\end{figure}

\begin{figure}[H]
\centering
\includegraphics[width=0.65\textwidth]{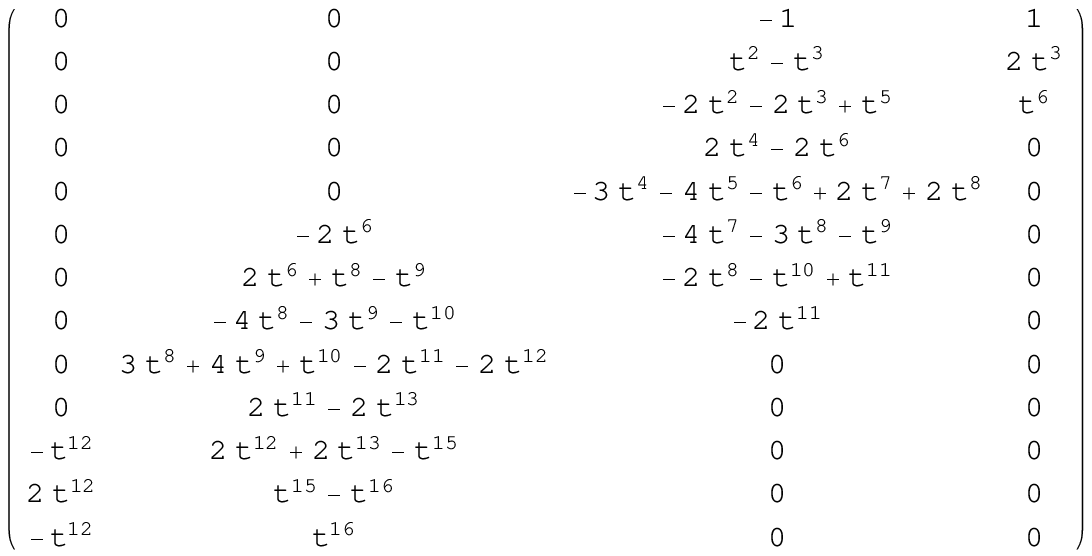}
\caption{Matrix form of A-polynomial for $T^{2,5}$ torus knot.}
%\label{fig:AAAA}
\end{figure}

\begin{figure}[H]
\centering
\includegraphics[width=0.95\textheight,angle=90]{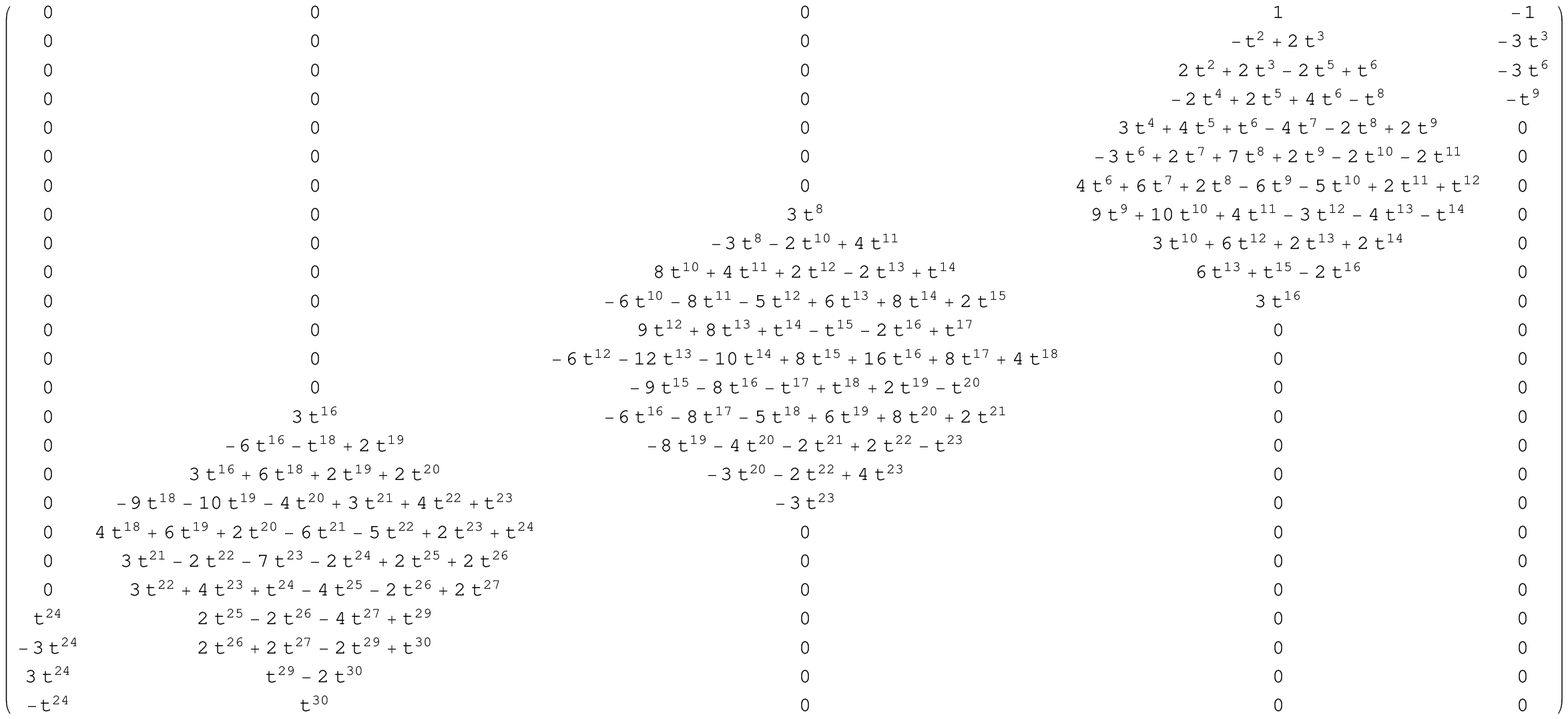}
\caption{Matrix form of A-polynomial for $T^{2,7}$ torus knot.}
%\label{fig:AAAA}
\end{figure}

\newpage

\begin{figure}[H]
\centering
\includegraphics[width=\textheight,angle=90]{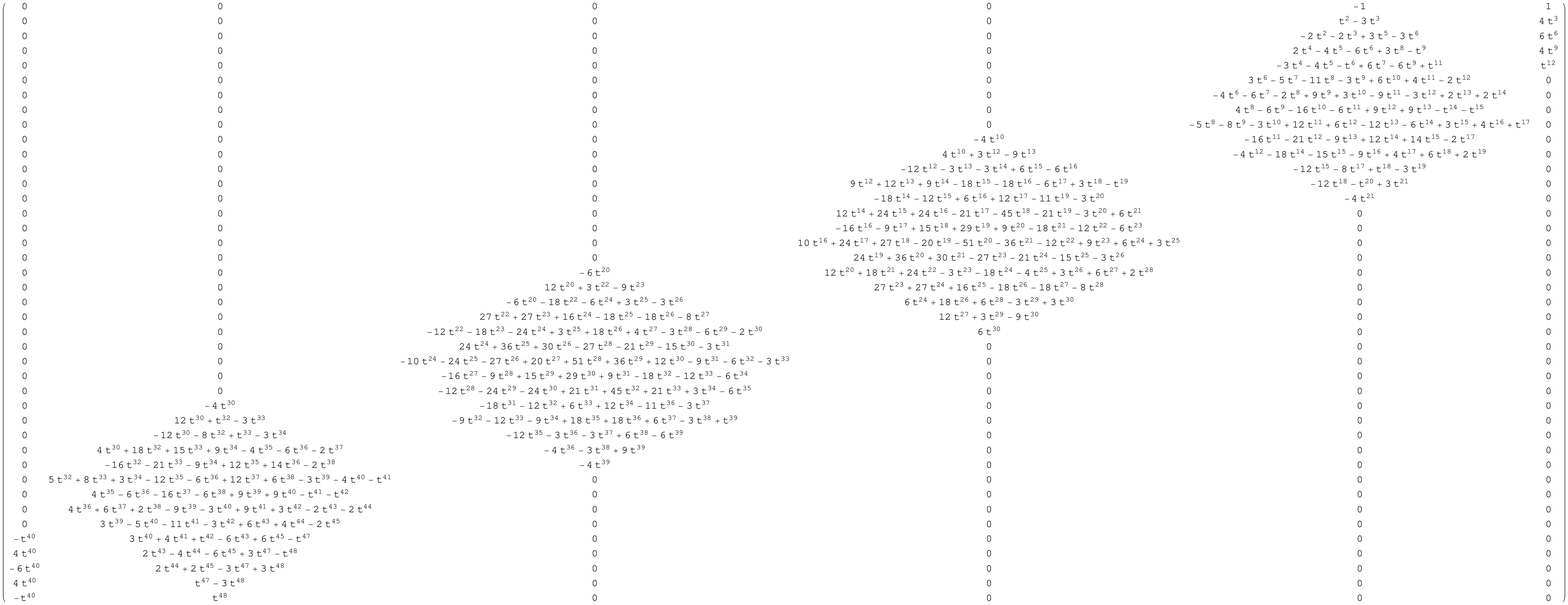}
\caption{Matrix form of A-polynomial for $T^{2,9}$ knot.}
%\label{fig:AAAA}
\end{figure}

\begin{figure}[H]
\centering
\includegraphics[width=\textheight,angle=90]{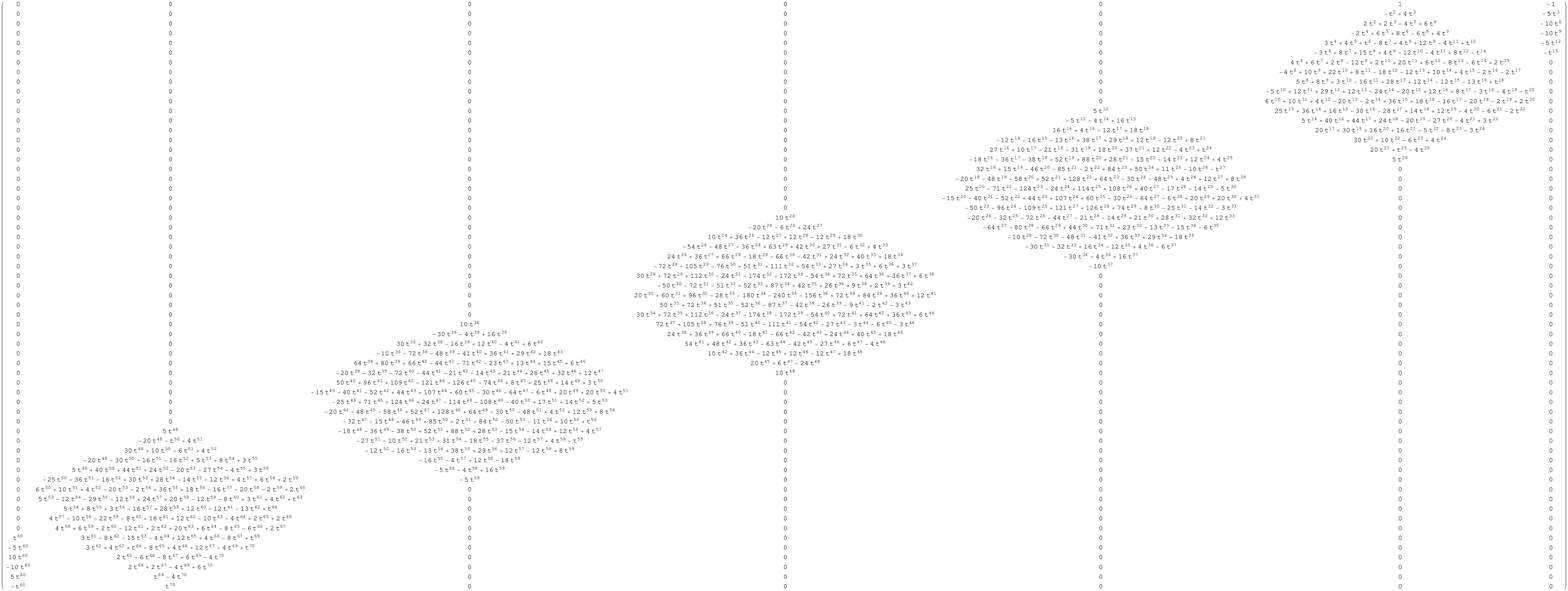}
\caption{Matrix form of A-polynomial for $T^{2,11}$ torus knot.}
%\label{fig:AAAA}
\end{figure}

%*******************************************************************
%*******************************************************************

%%%%%%%%%%%%%%%%%%%%%%%%%%%%%%%%%%%%%%%%%%%%%%%%%%%%%%%%%%%%%%%%%%%%%%%%%%%

\newpage

\bibliographystyle{JHEP_TD}
\bibliography{abmodel}

\end{document}